\documentclass[useAMS,usenatbib]{mn2e_mod}
\usepackage{natbib}
\usepackage{graphicx}
\usepackage{amsmath}
\usepackage{amssymb}
\usepackage{deluxetable}
\usepackage{longtable}
\usepackage{url}
\usepackage{array}
\usepackage{pdflscape, float,lscape}
\usepackage{graphicx,kantlipsum, rotating}
\usepackage{hyperref}

\let\ACMmaketitle=\maketitle
\renewcommand{\maketitle}{\begingroup\let\footnote=\thanks \ACMmaketitle\endgroup}

\title[SF and molecular gas in post-starbursts]{Star formation and molecular gas properties of post-starburst galaxies}

\author[Baron et al.]
{Dalya Baron$^{1}$\thanks{dalyabaron@gmail.com},
Hagai Netzer$^{1}$,
K. Decker French$^{2}$,
Dieter Lutz$^{3}$, 
\newauthor 
Ric I. Davies$^{3}$ \&
J. Xavier Prochaska$^{4}$
\\
\\
$^{1}$School of Physics and Astronomy, Tel-Aviv University, Tel Aviv 69978, Israel.\\
$^{2}$Department of Astronomy, University of Illinois, 1002 W. Green St., Urbana, IL 61801, USA.\\
$^{3}$Max-Planck-Institut f$\ddot{u}$r Extraterrestrische Physik, Giessenbachstrasse 1, 85748 Garching, Germany.\\
$^{4}$Department of Astronomy and Astrophysics, UCO/Lick Observatory, University of California, 1156 High Street, Santa Cruz, CA 95064, USA.
}

\begin{document}

\maketitle

\label{firstpage}
\begin{abstract}

Post-starburst galaxies are believed to be in a rapid transition between major merger starbursts and quiescent ellipticals. Their optical spectrum is dominated by A-type stars, suggesting a starburst that was quenched recently. While optical observations suggest little ongoing star formation, some have been shown to host significant molecular gas reservoirs. This led to the suggestion that gas depletion is not required to end the starburst, and that star formation is suppressed by other processes. We present NOEMA CO(1-0) observations of 15 post-starburst galaxies with emission lines consistent with Active Galactic Nucleus (AGN) photoionization. We collect post-starburst candidates with molecular gas measurements from the literature, with some classified as classical E+A, while others with line ratios consistent with AGN and/or shock ionization. Using far-infrared observations, we show that systems that were reported to host exceptionally-large molecular gas reservoirs host in fact obscured star formation, with some systems showing star formation rates comparable to ULIRGs. Among E+A galaxies with molecular gas measurements, 7 out of 26 (26\%) host obscured starbursts. Using far-infrared observations, post-starburst candidates show similar SFR-$M_{\mathrm{H_2}}$ and Kennicutt-Schmidt relations to those observed in star-forming and starburst galaxies. In particular, there is no need to hypothesize star formation quenching by processes other than the consumption of molecular gas by star formation. The combination of optical, far-infrared, and CO observations indicates that some regions within these galaxies have been recently quenched, while others are still forming stars in highly obscured regions. All this calls into question the traditional interpretation of such galaxies.

\end{abstract}


\vspace{1cm}
\section{Introduction}\label{s:intro}

Observations of the local galaxy population reveal a bimodal distribution in their properties, suggesting two broad classes of galaxies: star-forming and quiescent. These two classes differ in their morphology, color, stellar kinematics, star formation, and gas properties (see e.g., \citealt{baldry06} for $z < 0.1$ galaxies). Post-starburst galaxies (also called E+A or K+A) are rare objects that are located in the ``green valley" of the galaxy color-magnitude diagram \citep{wong12}. They are believed to represent a rapid transition between the star-forming and quiescent classes (see \citealt{french21} for a review). Their optical spectra are dominated by A-type stars, suggesting a recent burst of star formation that was terminated abruptly $\sim$1 Gyr ago \citep{dressler83, couch87, poggianti99}. Stellar population synthesis modeling of their ultraviolet-optical spectral energy distributions (SEDs) suggest peak star formation rates (SFRs) in the range 10--300 $M_{\odot}\,\mathrm{yr}^{-1}$ \citep{kaviraj07, french18}, with 10\%--80\% of their total stellar mass forming in the most recent star formation episode	 \citep{liu96, norton01, yang04, kaviraj07, french18}. Many E+A galaxies show disturbed morphologies with tidal features, suggesting that they are merger remnants  ($\sim$50\%; e.g., \citealt{zabludoff96, canalizo00, yang04, goto04, cales11}).

Studies suggest that some post-starburst galaxies are the evolutionary link between major merger ultra luminous infrared galaxies and quiescent ellipticals (e.g., \citealt{yang04, yang06, kaviraj07, wild09, yesuf14, cales15, wild16, almaini17, baron17b, baron18, french18}). According to the current paradigm, a major merger between gas-rich spirals triggers a starburst \citep{mihos94, mihos96, barnes96} which is highly-obscured by dust and is primarily visible in infrared and mm wavelengths. Soon after, gas in funneled to the vicinity of the supermassive black hole, triggering an Active Galactic Nucleus (AGN; e.g., \citealt{sanders88, springel05, hopkins06}). The molecular gas reservoir is then quickly depleted by the starburst and by stellar and AGN feedback, terminating additional star formation and black hole accretion (e.g., \citealt{tremonti07, coil11, diamond_stanic12, rowlands15, baron17b, baron18, maltby19, baron20}). The regions undergoing rapid quenching become optically-thin, and their observed spectrum is dominated by A-type stars (E+A spectrum). Having no molecular gas left, they are expected to transition into red and dead ellipticals in a few hundreds Myrs.

Some details of this simple picture have recently been challenged, when studies found significant molecular gas reservoirs in optically-selected post-starburst galaxies \citep{french15, rowlands15, alatalo16b, yesuf17, li19, yesuf20b, bezanson22, smercina22}. This led to the suggestion that complete gas consumption or expulsion are not required to end the starburst in these systems. In addition, post-starbursts were found to be offset from the Kennicutt--Schmidt (KS) relation \citep{kennicutt98_ks} observed in normal star-forming and starburst galaxies \citep{french15, smercina18, li19, smercina22}, with SFRs that are suppressed by a factor of $\sim$10 compared to galaxies with similar molecular gas surface densities. \citet{smercina22} suggested that the star formation in these systems is suppressed by significant turbulent heating of the molecular gas.

The traditional selection and classification scheme of post-starburst galaxies relies on their optical spectra (see review by \citealt{french21} and section \ref{s:other_samples} below). As a result, the star formation properties of post-starburst galaxies have been typically derived from optical observations, using for example, the H$\alpha$ emission line or the Balmer absorption equivalent width (EW). The optically-derived SFRs place most of these systems on or below the star-forming main sequence, consistent with the idea that they are completely quenched or transitioning to quiescence. However, using infrared and radio observations, several studies found significant obscured star formation in systems showing post-starburst signatures in their optical spectrum \citep{smail99, poggianti00, smercina18, baron22}. 

In \citet{baron22}, we used IRAS far-infrared observations to study the star formation properties of several large samples of post-starburst galaxies. We found that their far-infrared properties depend on their selection criteria. Post-starbursts selected with the traditional E+A criteria (i.e., weak emission lines; see additional details in section \ref{s:other_samples}) show IRAS far-infrared detection fraction of $4.5$\%, suggesting that only $\sim$4\% are $>0.3$ dex above the star-forming main sequence. On the other hand, post-starbursts selected with emission lines, where the emission line ratios are consistent with AGN and/or shocks, are found $>0.3$ dex above the star-forming main sequence in 13--45\% of the cases, and hosting starbursts ($> 0.6$ dex) in 7--32\% of the cases, where the exact fraction depends on the selection criteria being used (see section 4.1 there for confidence intervals). We found that post-starbursts with more luminous emission lines are more likely to be found above the main sequence, and among AGN hosts, post-starbursts are more likely to be found above the main sequence compared to typical Seyfert galaxies with similar redshifts and stellar masses.

Our finding that some post-starburst galaxies still host significant star-formation raises the question of whether systems that were previously reported to have significant molecular gas reservoirs host in fact obscured star formation. If found to be true, it may solve the challenges raised by \citet{french15} and \citet{smercina22} to the simple evolutionary picture described above. We therefore aim to revisit the recent results by \citet{french15} and \citet{smercina22} using IRAS far-infrared and molecular gas observations of post-starburst galaxies selected with different selection criteria.

In this paper we combine far-infrared and Carbon Monoxide (CO) observations to study the star formation and molecular gas properties of post-starburst galaxy candidates. We present new CO observations for a sample of 15 galaxies selected from our parent sample of post-starburst candidates with AGN and ionized outflows from \citet{baron22}. We combine this sample with several other samples of post-starburst candidates with available CO observations. This allows us to study the molecular gas properties of post-starburst candidates with different emission line properties. The paper is organized as follows. In Section \ref{s:data} we describe our observations and data analysis, and in section \ref{s:all_samples} we present all the literature samples we consider. We present our results in Section \ref{s:results}, discuss their implications in Section \ref{s:disc}, and summarize in Section \ref{s:conclusions}. We provide a description of the data availability in Section \ref{s:data_avail}. Throughout this paper we use a Chabrier initial mass function (IMF; \citealt{chabrier03}), and assume a standard $\Lambda$CDM cosmology with $\Omega_{\mathrm{M}}=0.3$, $\Omega_{\Lambda}=0.7$, and $h=0.7$. 

\section{Observations and data analysis}\label{s:data}

\subsection{Sample selection}\label{s:data:sample}

Our parent sample of post-starburst galaxy candidates with AGN and ionized outflows was drawn from the 14th data release of the Sloan Digital Sky Survey (SDSS; \citealt{york00}), and is described in detail in  \citet{baron22}. We fitted stellar population synthesis models to all the galaxies at $z < 1$, and measured the EW of the H$\delta$ absorption line. We define post-starburst galaxy candidates as systems for which $\mathrm{EW(H\delta) > 5}$\AA. We performed line profile decomposition into narrow and broad kinematic components, where the former traces stationary gas, while the latter originates from outflowing gas. We used the Balmer emission lines to filter out type I AGN. We selected systems with narrow line ratios that are consistent with AGN photoionization (including LINERs; \citealt{kewley01, cidfernandes10}). Systems with detected broad components in both Balmer and forbidden lines were defined as post-starburst galaxies hosting AGN and ionized outflows. We found a total of 215 post-starburst candidates with AGN and ionized outflows, out of which 144 show evidence for an ionized outflow in multiple lines: [OIII]~$\lambda \lambda$ 4959,5007\AA, $\mathrm{H\alpha}$~$\lambda$ 6563\AA, [NII]~$\lambda \lambda$ 6548,6584\AA, [OII]~$\lambda \lambda$ 3725,3727\AA, $\mathrm{H\beta}$~$\lambda$ 4861\AA, and [SII]~$\lambda \lambda$ 6717,6731\AA\,.

Using the ionized outflow properties, we selected a subset of 32 systems for follow-up observations. These galaxies show the highest signal-to-noise ratios (SNRs) in the broad kinematic components of the [OIII] and H$\alpha$ emission lines. Such a selection is biased towards sources with more luminous emission lines, and since emission line luminosity is correlated with AGN bolometric luminosity and SFR (\citealt{baron22}), this subset is biased towards sources with higher AGN luminosities and SFRs. Roughly half of the objects can be observed with the Northern Extended Millimeter Array. 

\begin{table*}
 \caption{\textbf{The NOEMA sample.}}\label{tab:NOEMA_observations}
 	\centering
	\footnotesize
	
\begin{tabular}{c c c c c c c c c}
\hline
\hline
Object ID & RA      & Dec.    & $z$ & $\mathrm{t_{obs}}$ & RMS   & $\theta_{a} \times \theta_{b}$ & P.A  \\
          & (J2000) & (J2000) &     &      (hr)          & (mJy) & (arcsec)                       & (deg) \\
   (1)    & (2)     & (3)     & (4) &      (5)           &  (6)  & (7)                            & (8) \\
\hline
0626-52057-0460 & 16:27:38.07 & 44:53:42.12 & 0.179 & 3.00 & 1.12 &  $2.5 \times 1.6$  & -0.01 \\
1008-52707-0017 & 10:28:37.36 & 50:23:50.93 & 0.092 & 2.13 & 2.35 &  $2.7 \times 1.7$  & 82.38 \\
1463-53063-0262 & 13:24:01.63 & 45:46:20.69 & 0.125 & 3.00 & 1.41 &  $2.7 \times 1.4$  & 40.72 \\
1605-53062-0122 & 11:19:43.84 & 10:50:37.17 & 0.177 & 3.50 & 0.65 &  $4.6 \times 4.3$  & -23.19\\
1609-53142-0564 & 11:52:59.31 & 13:07:18.76 & 0.104 & 2.49 & 1.00 &  $5.7 \times 3.5$  & 44.95 \\
1646-53498-0129 & 14:46:39.62 & 33:21:29.32 & 0.070 & 2.10 & 1.22 &  $2.9 \times 2.3$  & -11.85\\
1686-53472-0386 & 16:36:14.09 & 29:16:12.73 & 0.136 & 2.10 & 1.46 &  $2.8 \times 1.6$  & 6.32  \\
1766-53468-0172 & 12:16:22.73 & 14:17:53.02 & 0.082 & 2.16 & 1.31 &  $5.3 \times 3.6$  & 45.39 \\
1773-53112-0588 & 13:21:11.30 & 14:54:52.13 & 0.103 & 4.50 & 0.83 &  $4.5 \times 3.3$  & 26.90 \\
1870-53383-0446 & 07:59:27.49 & 53:37:48.23 & 0.084 & 2.10 & 1.63 &  $2.1 \times 1.7$  & 49.90 \\
2026-53711-0219 & 10:50:41.50 & 31:19:56.38 & 0.115 & 2.31 & 0.82 &  $4.1 \times 4.1$  & 50.63 \\
2268-53682-0278 & 08:04:27.84 & 13:29:30.73 & 0.134 & 2.46 & 0.83 &  $4.7 \times 3.6$  & 31.34 \\
2345-53757-0457 & 10:03:34.27 & 28:50:31.53 & 0.138 & 2.60 & 0.77 &  $4.3 \times 4.1$  & 49.64 \\
2494-54174-0129 & 11:20:23.91 & 15:43:54.63 & 0.159 & 2.46 & 0.93 &  $4.5 \times 4.1$  & -220.43\\
6281-56295-0012 & 00:34:43.68 & 25:10:20.96 & 0.118 & 2.25 & 2.41 &  $2.8 \times 1.3$  & 11.78 \\
\hline
\vspace{-0.3in}
\tablenotetext{} {(1): SDSS identifier PLATE-MJD-FIBER. (2) and (3): SDSS right ascension and declination. (4): SDSS redshift. (5): NOEMA integration time. (6) RMS noise in channel. (7) Beam FWHM along major and minor axes. (8) Beam position angle.}\\
\end{tabular}
\end{table*}

\subsection{NOEMA observations}\label{s:data:NOEMA}

We used the Northern Extended Millimeter Array (NOEMA) of the Institute of Radioastronomy in the Millimeter (IRAM)\footnote{IRAM is supported by INSU/CNRS (France), MPG (Germany) and IGN (Spain).} to observe the CO(1-0) line in 15 galaxies (project W19BS). The observations were carried out with 10 antennas in C or D configurations under good weather conditions. The NOEMA receivers provide two orthogonal linear polarizations, each delivering a bandwidth of 7.744 GHz in the lower sideband (LSB) and upper sideband (USB) simultaneously. These are fed into the wide-band correlator {\sc PolyFix}, which has a channel width of 2 MHz. We tuned the CO(1-0) line frequency in the USB. Out of the 15 targets, 9 were observed using separate tracks while 6 were observed in pairs in a track-sharing mode. In Table \ref{tab:NOEMA_observations} we summarize the observations. 

We performed data reduction, calibration, and imaging analysis using the CLIC and MAPPING softwares of GILDAS\footnote{\url{http://www.iram.fr/IRAMFR/GILDAS}}. We used standard pipeline reduction and calibration outputs for most of the tracks. In a few tracks, we flagged poor data scans, applied a baseline from a different track, or enforced a different calibrator. We produced uv-tables for every LSB and USB separately, with a frequency resolution of 20 MHz (corresponding to $\sim$60 km/sec at the rest-frame frequency of our galaxies). 

We identified the emission line channels in each USB uv-table and used the task \emph{uv\_baseline} to fit and subtract a constant baseline. We produced clean images using the task \emph{go clean}, and show in Figure \ref{f:CO_spectra} in the appendix the integrated spectra around the CO(1-0) transition. The CO line intensity was estimated by integrating over these spectra. The statistical uncertainty in the line intensity is defined as $\sigma_{l}^{2} = \Delta v^2 \sigma^2 N_{l} \big(1 + N_l/N_b \big)$, where $\Delta v$ is the channel velocity width, $\sigma$ is the channel RMS noise, $N_{l}$ is the number of channels used to integrate over the line, and $N_{b}$ is the number of channels used for the baseline fitting \citep{sage07, young11}. We detect CO emission in 14 out of 15 sources, with SNRs ranging from 3.1 to 25. For the undetected source, we define an upper limit of the line intensity as three times the statistical uncertainty, assuming $\Delta v = 200$ km/sec.

\begin{table*}
 \caption{\textbf{CO and mm continuum emission properties.}}\label{tab:CO_and_mm_properties}
 	\centering
	\footnotesize
\begin{tabular}{c c c c c c c c c}
\hline
\hline
Object ID & $f_{\mathrm{CO}}$        & $L^{'}_{\mathrm{CO}}$                 & $\mathrm{FWHM_{CO}}$ & $\nu_{\mathrm{LSB}}$ & $f_{\mathrm{LSB}}$ & $\nu_{\mathrm{USB}}$ & $f_{\mathrm{USB}}$ & $\mathrm{FWHM_{mm}}$ \\
          &   [$\mathrm{Jy\, km\,sec^{-1}}$]   &  [$10^8\, \mathrm{K\, km\,sec^{-1} pc^{-2}}$]  &  [arcsec]             &       [GHz]            & [$\mathrm{\mu Jy}$]  &  [GHz]                 &  [$\mathrm{\mu Jy}$] & 
		   [arcsec]        \\
   (1)    & (2)                      & (3)                            & (4)                  &      (5)             & (6)                &  (7)                 & (8)                &           (9)   \\
\hline

0626-52057-0460 & $0.64 \pm 0.10$    & $9.93 \pm 1.61$  & $< 0.407$         & $94.03$ & $65 \pm 15$   & $112.29$ & $65 \pm 15$   & --             \\
1008-52707-0017 & $7.53 \pm 0.82$    & $29.91 \pm 3.24$ & $1.017 \pm 0.046$ & $95.28$ & $513 \pm 27$  & $112.20$ & $617 \pm 33$  & $0.74 \pm 0.12$\\
1463-53063-0262 & $1.46 \pm 0.18$    & $10.90 \pm 1.35$ & $0.895 \pm 0.069$ & $94.79$ & $< 40 \pm 13$ & $112.21$ & $< 40 \pm 13$ & --             \\
1605-53062-0122 & $0.75 \pm 0.11$    & $11.40 \pm 1.72$ & $3.54 \pm 0.26$   & $95.64$ & $< 44 \pm 15$ & $113.87$ & $< 43 \pm 14$ & --             \\
1609-53142-0564 & $0.52 \pm 0.16$    & $2.66 \pm 0.84$  & $3.50 \pm 0.60$   & $97.43$ & $152 \pm 22$  & $114.53$ & $158 \pm 26$  & --             \\
1646-53498-0129 & $1.48 \pm 0.46$    & $3.37 \pm 1.05$  & $1.64 \pm 0.11$   & $96.04$ & $136 \pm 15$  & $112.61$ & $148 \pm 16$  & --             \\
1686-53472-0386 & $5.19 \pm 0.29$    & $45.56 \pm 2.57$ & $1.311 \pm 0.048$ & $95.67$ & $132 \pm 17$  & $113.26$ & $143 \pm 17$  & --             \\
1766-53468-0172 & $0.92 \pm 0.18$    & $2.89 \pm 0.56$  & $3.69 \pm 0.33$   & $95.48$ & $149 \pm 23$  & $112.23$ & $195 \pm 25$  & --             \\
1773-53112-0588 & $8.46 \pm 0.33$    & $42.34 \pm 1.67$ & $2.180 \pm 0.048$ & $95.14$ & $95 \pm 12$   & $112.22$ & $112 \pm 13$  & --             \\
1870-53383-0446 & $<0.93$            & $<3.06$          & --                & $95.13$ & $83 \pm 14$   & $111.92$ & $87 \pm 16$   & --             \\
2026-53711-0219 & $1.49 \pm 0.22$    & $9.37 \pm 1.40$  & $3.02 \pm 0.30$   & $95.06$ & $361 \pm 24$  & $112.33$ & $322 \pm 28$  & $2.00 \pm 0.37$\\
2268-53682-0278 & $0.249 \pm 0.065$  & $2.14 \pm 0.56$  & $2.88 \pm 0.77$   & $94.41$ & $48 \pm 18$   & $111.98$ & $84 \pm 19$   & --             \\
2345-53757-0457 & $0.68 \pm 0.10$    & $6.19 \pm 0.92$  & $1.56 \pm 0.49$   & $97.00$ & $56 \pm 15$   & $114.63$ & $63 \pm 15$   & --             \\
2494-54174-0129 & $3.28 \pm 0.43$    & $40.07 \pm 5.23$ & $4.01 \pm 0.20$   & $94.17$ & $207 \pm 47$  & $112.13$ & $169 \pm 44$  & $4.03 \pm 1.15$\\
6281-56295-0012 & $2.45 \pm 0.35$    & $16.19 \pm 2.31$ & $2.66 \pm 0.10$   & $95.31$ & $< 50 \pm 17$ & $112.63$ & $<50 \pm 17$  & --             \\
\hline
\vspace{-0.2in}
\tablenotetext{} {(1) SDSS identifier PLATE-MJD-FIBER. (2) Velocity-integrated CO flux. (3) CO line luminosity. (4) Deconvolved FWHM of the CO emission from a circular Gaussian fit. (5) Central frequency of the LSB. (6) Continuum mm flux in the LSB. (7) Central frequency of the USB. (8) Continuum mm flux in the USB. (9) Deconvolved FWHM of the mm continuum emission in the LSB and USB combined from a circular Gaussian fit. }
\tablenotetext{} {\textbf{Table notes:} The reported flux uncertainties the quadratic sum of the uncertainty in the fit and the absolute flux scale. For NOEMA in 3 mm, a conservative estimate of the absolute flux scale uncertainty is 10\% \citep{tacconi13}.}
\end{tabular}
\end{table*}

The CO line luminosity in $\mathrm{K\, km/sec\, pc^2}$ is given by (\citealt{solomon97}):
\begin{equation}\label{eq:1}
	{L^{'}_{\mathrm{CO}} = 3.25 \times 10^7 S_{\mathrm{CO}}\Delta v \, \nu_{\mathrm{obs}}^{-2} \, D^{2}_{\mathrm{L}} \, (1 + z)^{-3} },
\end{equation}
where $S_{\mathrm{CO}}\Delta v$ is the velocity-integrated flux in $\mathrm{Jy\, km/sec}$, $\nu_{obs}$ is the observed frequency in GHz, $D_{\mathrm{L}}$ is the luminosity distance in Mpc, and $z$ is the redshift. The molecular gas mass is given by $M_{\mathrm{H_2}} = \alpha_{\mathrm{CO}} L^{'}_{\mathrm{CO}}$, where $\alpha_{\mathrm{CO}}$ is the CO conversion factor. We assume a constant conversion factor similar to the Milky Way disc, $\alpha_{\mathrm{CO}} = 4.3\, \mathrm{M_{\odot}(K\, km\, sec^{-1} \, pc^2)^{-1}}$ (\citealt{bolatto13}). In section \ref{s:results:molecular_gas} we explain why other definitions for the conversion factor (bimodal or CO intensity-weighted $\alpha_{\mathrm{CO}}$; e.g., \citealt{narayanan12}) do not affect our conclusions. 

To study the spatial extent of the CO line emission we combined the baseline-subtracted line channels in each USB with the task \emph{uv\_continuum}. Using the graphical interface of MAPPING, we fitted each object with a point source and a circular Gaussian in the uv-plane. For the point source fits, we inspected the clean images of the residuals and found significant residuals in most of our sources, suggesting that the emission is spatially-resolved. For the circular Gaussian fits, the residuals in the clean images are consistent with those of the background. We verified that the integrated CO fluxes from the circular Gaussian fits are consistent with the fluxes estimated from the spatially-integrated CO spectra. Out of 14 sources with detected CO lines, the best-fitting full width at half maximum (FWHM) of the circular Gaussian fit is larger than three times its uncertainty in 13 of them. Therefore, we consider the CO line spatially-resolved in 13 out of 14 sources.

To estimate the mm continuum flux, we combined all the non-line channels in the LSB and USB separately. We fitted each source in each sideband with a point source and a circular Gaussian in the uv-plane. In the cases where the continuum was spatially-resolved, we used the flux obtained from the best fit circular Gaussian. In the cases where the continuum was not resolved, we used the flux obtained from the point source fit. For the sources where continuum emission was not detected, we used a polygon to estimate the RMS noise, and used it to place an upper limit on the continuum flux. To study the spatial extent of the continuum emission we combined the LSB and USB channels and fitted each source with a point source and a circular Gaussian in the uv-plane. We detect mm continuum emission in 11 sources, and spatially-resolve it in the three brightest sources. In Section \ref{s:data:FIR} below, we combine the mm fluxes with archival far-infrared observations to estimate the SFR or its upper limit for every source in our sample. Table \ref{tab:CO_and_mm_properties} summarizes the CO and mm continuum emission properties for our systems, and Table \ref{tab:integ_properties} provides the measured molecular masses.

\subsection{{\it IRAS} far-infrared data}\label{s:data:FIR}

We make use of publicly-available far infrared photometry from the \emph{Infrared Astronomical Satellite} ({\it IRAS}; \citealt{neugebauer84}). {\it IRAS} provides a full-sky coverage in four bands centered around 12, 15, 60, and 100 $\mathrm{\mu m}$, including several scans of individual fields. Our method to extract the 60 $\mathrm{\mu m}$ fluxes follows the approach described in \citet{baron22} which we briefly summarize here. We used {\sc scanpi} (\citealt{helou88}) to stack the calibrated surveys scans around the SDSS coordinates of our sources. We used the default settings of {\sc scanpi}, with `Source Fitting Range'=3.2', `Local Background Fitting Range'=30', and `Source Exclusion Range for Local Background Fitting'=4', and used median stacks since the {\it IRAS} data has non-Gaussian noise. The {\sc scanpi} output includes the best-fitting flux density ($f_{\nu}$), the RMS deviation of the residuals after the subtraction of the best-fitting template ($\sigma$), the offset of the peak of the best-fitting template from the galaxy coordinates ($\Delta$), and the correlation coefficient between the best-fitting template and the data ($\rho$). To avoid contamination by false matches and noise fluctuations, we consider a source detected if the following requirements are met: $\rho > 0.8$, $\Delta< 0.4'$, and $f_{\nu}/\sigma > 3$ (see details in \citealt{baron22}). We defined the flux uncertainty of the detected sources to be $\sigma$. For undetected sources, we defined their flux upper limit to be $3 \sigma$. Out of 15 sources, 5 are detected in 60 $\mathrm{\mu m}$ and the rest are considered upper limits. 

To estimate the SFR in each source, we combined the mm continuum fluxes (LSB and USB) from NOEMA and the 60 $\mathrm{\mu m}$ flux by {\it IRAS}. We used the \citet[hereafter CE01]{chary01} templates to search for the template that best fits the far-infrared and mm flux measurements or upper limits. Each CE01 template corresponds to a different star formation luminosity, $L_{\mathrm{SF}}$, allowing us to estimate $L_{\mathrm{SF}}$ in sources where at least one of the three flux measurements is available (60 $\mathrm{\mu m}$, LSB, and USB). For sources where both far-infrared and mm fluxes are measured, we found a good correspondence between the three flux measurements and the best-fitting template. For sources where only upper limits are available, we placed an upper limit on $L_{\mathrm{SF}}$. Following \citet{baron22}, we adopted a conservative uncertainty on $L_{\mathrm{SF}}$ of 0.2 dex. In Figure \ref{f:FIR_and_mm_fit} in the appendix we show the far-infrared and mm luminosities, along with the best-fitting star formation templates. We estimated the SFR from $L_{\mathrm{SF}}$ assuming a Chabrier initial mass function (IMF; \citealt{chabrier03}) which we rounded off slightly to give $\mathrm{SFR} = L_{\mathrm{SF}} / \, 10^{10}\, L_{\odot}\, [M_{\odot}\,\mathrm{yr}^{-1}]$. Out of the 15 sources in our sample, we estimated the SFR in 12 systems, and placed upper limits on the remaining 3. We list the SFRs in Table \ref{tab:integ_properties}. 

\begin{table*}
 \caption{\textbf{Integrated galaxy properties.}}\label{tab:integ_properties}
 	\centering
	\scriptsize
\begin{tabular}{c | c c c c | c c c c c}
\hline
\hline
          & \multicolumn{4}{c|}{Optical properties} & \multicolumn{5}{c}{FIR and mm properties} \\
\hline
Object ID & $r_{\mathrm{opt}}$ & $\log \mathrm{SFR(Dn4000\AA)}$ & $\log M_{*}$ & $t_{\mathrm{burst}}$ & $\log M_{\mathrm{H_2}}$ & $\log \mathrm{SFR(FIR+mm)}$ & $r_{\mathrm{CO}}$ & $\log \Sigma_{\mathrm{gas}}$ & $\log \Sigma_{\mathrm{SFR}}$ \\

          & [kpc]   & [$M_{\odot}\,yr^{-1}$] & [$M_{\odot}$]                  & [Myr]                  & [$M_{\odot}$]             & [$M_{\odot}\,yr^{-1}$]      & [kpc]               & [$M_{\odot}\, \mathrm{pc}^{-2}$]                        & [$M_{\odot}\, \mathrm{yr}^{-1} \mathrm{kpc}^{-2}$] \\
		  
 (1)& (2) & (3) & (4) & (5) & (6) & (7) & (8) & (9) & (10) \\

\hline

0626-52057-0460 & 5.89 & 0.89  & 10.35 & 90              & $9.636 \pm 0.071$    & $1.37 \pm 0.20^{*}$   & $< 0.61$          & --     & --      \\
1008-52707-0017 & 6.47 & 0.17  & 10.50 & 79              & $10.114 \pm 0.047$   & $1.55 \pm 0.20^{*}$   & $0.870 \pm 0.039$ & $3.74$ & $1.18$  \\
1463-53063-0262 & 8.65 & 0.56  & 10.49 & --              & $9.676 \pm 0.054$    & $< 0.67 \pm 0.20$ & $1.003 \pm 0.078$ & $3.18$ & $< 0.17$  \\
1605-53062-0122 & 8.17 & 0.81  & 10.70 & $-$88$^{\dagger}$ & $9.696 \pm 0.065$  & $< 1.03 \pm 0.20$ & $5.30 \pm 0.39$   & $1.75$ & $< -0.91$ \\
1609-53142-0564 & 4.63 & 0.13  & 10.47 & 69              & $9.06 \pm 0.14$      & $1.06 \pm 0.20$   & $3.34 \pm 0.57$   & $1.52$ & $-0.48$ \\
1646-53498-0129 & 4.81 & $-$1.19 & 10.62 & 279             & $9.17 \pm 0.13$    & $0.67 \pm 0.20^{*}$   & $1.100 \pm 0.077$ & $2.59$ & $0.09$  \\
1686-53472-0386 & 14.35 & 0.95  & 11.09 & 57             & $10.298 \pm 0.024$   & $1.48 \pm 0.20^{*}$   & $1.574 \pm 0.058$ & $3.41$ & $0.60$  \\
1766-53468-0172 & 3.22 & 0.03  & 10.36 & 20              & $9.100 \pm 0.085$    & $0.89 \pm 0.20$   & $2.84 \pm 0.25$   & $1.70$ & $-0.51$ \\
1773-53112-0588 & 4.77 & 0.64  & 10.20 & --              & $10.267 \pm 0.017$   & $0.89 \pm 0.20$   & $2.063 \pm 0.045$ & $3.14$ & $-0.23$ \\
1870-53383-0446 & 3.99 & $-$0.88 & 10.67 & 104             & $< 9.12 \pm 0.14$  & $0.63 \pm 0.20$   & --                & -- 	   & --      \\
2026-53711-0219 & 7.50 & $-$0.04 & 10.53 & 121             & $9.611 \pm 0.065$  & $1.29 \pm 0.20$   & $3.15 \pm 0.32$   & $2.12$ & $-0.20$ \\
2268-53682-0278 & 6.44 & 0.00  & 10.60 & 63              & $8.97 \pm 0.11$      & $0.99 \pm 0.20$   & $3.42 \pm 0.91$   & $1.40$ & $-0.57$ \\
2345-53757-0457 & 5.01 & $-$1.00 & 10.40 & 8               & $9.430 \pm 0.064$  & $0.93 \pm 0.20$   & $1.90 \pm 0.60$   & $2.38$ & $-0.11$ \\
2494-54174-0129 & 10.73 & 1.14  & 10.89 & 68             & $10.241 \pm 0.057$   & $1.55 \pm 0.20^{*}$   & $5.50 \pm 0.28$   & $2.26$ & $-0.42$ \\
6281-56295-0012 & 10.19 & 0.00  & 11.05 & --             & $9.848 \pm 0.062$    & $< 0.67 \pm 0.20$ & $2.84 \pm 0.11 $  & $2.44$ & $< -0.73$ \\
\hline 
\vspace{-0.2in}
\tablenotetext{} {\textbf{Columns.} (1): SDSS identifier PLATE-MJD-FIBER. (2): Petrosian radius measured in $r$-band from the SDSS. (3) Dn4000\AA-based SFR from the MPA-JHU catalogue. (4) Stellar mass from the MPA-JHU catalogue. (5) Post-burst age: the time elapsed since 90\% of the stars were formed in the recent burst(s). (6) Molecular gas mass from CO emission. (7) SFR derived from fitting the far-infrared fluxes by {\it IRAS} and mm fluxes by NOEMA. (8) Half-light radius of the CO emission. (9) Gas surface density. (10) SFR surface density.}
\tablenotetext{} {\textbf{Notes.} $^\dagger$: a negative post-burst age, suggesting that 90\% of the population has not yet been formed. $^{*}$: marks objects detected in IRAS $60\, \mathrm{\mu m}$.}

\end{tabular}
\end{table*}

\subsection{Galaxy properties from optical and ultraviolet observations}\label{s:data:optical-sdss}

We use several properties derived from the SDSS optical spectra and photometry. We use the Petrosian radius measured in $r$-band from the SDSS photometric catalogues (\citealt{petrosian76}). We use the stellar masses and the Dn4000\AA-based SFRs reported in the MPA-JHU value added catalogue (\citealt{b04, kauff03b, t04, salim07}). All these properties are listed in Table \ref{tab:integ_properties}.

The sources in our sample were selected to have strong H$\delta$ absorption ($\mathrm{EW(H\delta) > 5}$\AA), suggesting a recent starburst that was quenched abruptly. To derive their star formation history (SFH), we performed the stellar population synthesis modeling described in \citet{french18}. \citet{french18} combined \emph{Galaxy Evolution Explorer} (GALEX; \citealt{martin05_galex}) ultraviolet photometry and SDSS photometry and spectra to fit the SED of post-starburst galaxy candidates. They modeled the SEDs as a combination of old and young stellar populations, where the young population represents the recently-quenched starburst. The old stellar population is modeled as a linear-exponential SFR over time ($\propto t \times e^{-t/\tau}$). The young stellar population is modeled as either one or two declining exponents in the SFR over time, representing one or two recent bursts at $t < 2$ Gyr. 

The \citet{french18} SFH model has several free parameters. The first is the time elapsed since the first starburst began. In the case of a single recent burst, the burst duration is a free parameter of the model. In the case of two recent bursts, the burst duration is set to a characteristic timescale of 25 Myrs, while the time separation between the bursts is a free parameter of the model. In both cases, the mass fraction of the young stellar population with respect to the total stellar mass is a free parameter of the model. The total stellar mass is assumed to be the reported stellar mass in the MPA-JHU catalogue (see additional details there). 

Using the modeled SFH, \citet{french18} constructed model template spectra using the stellar population synthesis models by \citet{conroy09} and \citet{conroy10}. By fitting the templates to the GALEX and SDSS observations, they derived the best SFH model parameters. They also defined the "post-burst age" as the time elapsed since 90\% of the stars were formed in the more recent burst(s). Therefore, the "post-burst age" represents the age of the recently-quenched starburst, as seen in ultraviolet and optical wavelengths.

We applied the above age-dating technique to the sources in our sample. For three of the sources we found non-physical SFHs, where the mass fractions of the young stellar populations are unity, suggesting that all stars in these galaxies were formed in the recent ($t < 200$ Myr) bursts. The derived peak SFRs during the burst exceed 100 $M_{\odot}\,\mathrm{yr}^{-1}$, comparable to those observed in the most extreme ultra luminous infrared galaxies in the local universe. The ultraviolet and optical observations of these sources are completely dominated by the young stellar population, resulting in a young stellar mass fraction of unity. Near-infrared photometry (e.g., 2MASS $J$, $H$, and $K$ bands; \citealt{skrutskie06}) is required to constrain the old stellar population in these sources, which is outside the scope of this paper. We therefore omit these three sources from further optical age-dating analysis. For one source, we obtained a negative post-burst age, suggesting ongoing star formation. We list the post-burst ages in Table \ref{tab:integ_properties}, and the best-fitting SFH parameters are given in Table \ref{table:ages} in the appendix.

In this work we assumed that galaxies with EW(H$\delta$)$> 5$\AA\, are post-starbursts. However, \citet{wild20} applied the stellar population synthesis code {\sc bagpipes} to a sample of H$\delta$-strong galaxies and found that some of the sources show no evidence for a starburst that was quenched abruptly. By applying the age-dating technique by \citet{french18}, we explicitly assume that the systems we consider hosted a recent burst in the form of exponentially-declining SFR. However, when the more flexible {\sc bagpipes} code was applied to these sources, their star formation histories showed no evidence of recently-quenched burst in some of the cases (V. Wild private communication). The two codes also differ in their dust attenuation law, where \citet{french18} use the \citet{calzetti00} law while {\sc bagpipes} uses the \citet{charlot00} model. These dust models differ in particular in their attenuation of the young $t < 10^{7}$ yr stellar population, which may result in significantly-different star formation histories for the sources we consider. We treat these differences as uncertainties on our model and leave the comparison between the codes to a future work.

\begin{table*}
 \caption{\textbf{Properties of post-starburst candidates from the literature.} }\label{tab:other_samples}
 	\centering
	\scriptsize
\begin{tabular}{m{1.5cm} | m{2cm}| m{2cm}| m{1.5cm}| m{0.8cm} m{0.8cm} c| m{1.5cm} c | c c} 
\hline
\hline
Sample & Post-starburst selection & Emission line selection criteria  & Molecular subset selection &  $z$ & $\log M_{*}$            & $\mathrm{N_{obj}}$ & Telescope & RMS & CO  & FIR   \\
       &                          &                                   &                            &      & [$\mathrm{M_{\odot}}$]  &                    &           & [mJy]   & detections      &  detections\\
(1)	   &  (2)                     &   (3)                             &  (4)                       &  (5) &        (6)              &     (7)            & (8)       & (9)      &  (10) & (11)\\
\hline

\citet{french15}   & $\mathrm{H\delta_A - \sigma(H\delta_A)}$ $\mathrm{ > 4\AA}$ & $\mathrm{EW(H\alpha) < 3 \AA}$ & Half are mid-infrared-bright & $0.045 \pm 0.025$ & $10.49 \pm 0.33$ & 32 & IRAM 30m \& SMT & 6--25 & 17 (53\%) & 32 (100\%) \\
\hline

\citet{rowlands15} & PCA on 3175--4150\AA: selection according to  H$\delta$ and Dn4000\AA\, & No specific selection. Includes: SF, AGN, and composites. & No additional criteria & $0.0394 \pm	0.0067$ & $10.20 \pm	0.31$ & 11 & IRAM 30 m & 3--12 & 10 (90\%) & 9  (81\%)  \\

\hline

\citet{alatalo16b} & $\mathrm{H\delta_A > 5\AA}$                                                    & Line ratios consistent with shock models. Includes: SF, AGN, and composites. & Detected in mid-infrared & $0.104 \pm 0.049$ & $10.56 \pm 0.32$ & 52 & IRAM 30 m \& CARMA & 0.6--15 & 47 (90\%) & 35 (67\%)  \\

\hline

\citet{yesuf17} & $\mathrm{H\delta_F > 4\AA}$ \& cuts on (NUV-$g$) and Dn4000\AA\,  & Seyferts                       & H$\delta$ criterion relaxed to $\mathrm{H\delta_F > 0.8\AA}$ & $0.0305 \pm 0.0071$ & $10.32 \pm 0.20$ &  24 & SMT & 7--15 &  6 (25\%) & 12 (50\%)  \\

\hline 
\vspace{-2cm}
\tablenotetext{} {\footnotesize \textbf{Columns.} (1) Sample of post-starbursts with molecular gas measurements. (2) Selection criteria of a post-starburst galaxy. (3) Selection criteria of the emission lines. (4) Additional selection criteria for the subset observed with mm observations. (5) Median redshift and standard deviation. (6) Median stellar mass and standard deviation. (7) Number of galaxies in the sample. (8) The telescopes with which mm observations have been performed. (9) Minimum and maximum RMS per channel for the different galaxies reported by the studies. (10) Number and fraction of galaxies detected in mm observations. (11) Number and fraction of galaxies with detected far-infrared observations, including IRAS and Herschel observations.}

\end{tabular}
\end{table*} 

\subsection{Statistical methods}\label{s:data:stat}

Our sample includes a number of upper limits on both SFR and $M_{\mathrm{H_2}}$. To perform linear regression in the case of upper limits only in the dependent variable, we used the Python implementation\footnote{\url{https://github.com/jmeyers314/linmix}} of {\sc linmix} \citep{kelly07}. {\sc linmix} is a Bayesian method that accounts for measurement uncertainties in both the dependent ($y$) and independent ($x$) variables, but it is only suitable for datasets with upper limits on $y$. To perform linear regression in the case of upper limits in both $x$ and $y$, we used the analytic likelihoods derived by \citet{pihajoki17}. These likelihoods take into account both measurement uncertainties and upper limits in both of the variables. Using these likelihoods, we derived posterior probability distributions of our model parameters using the nested sampling Monte Carlo algorithm {\sc MLFriends} (\citealt{buchner14, buchner17}) with the {\sc UltraNest}\footnote{\url{https://johannesbuchner.github.io/UltraNest/}} package \citep{buchner21}. For the cases where the upper limits are only on $y$, we applied both methods and found consistent results.

In our analysis, we fit linear relations between different variables of the form: $y = a \times x + b + N$, where $N \sim \mathcal{N}(0,\,\sigma^{2})$. The free parameters of the fit are $a$, $b$, and $\sigma^{2}$. Throughout our paper, when presenting the best-fitting relations, we mark with a solid line the relation $y = \overline{a} \times x + \overline{b}$, where $\overline{a}$ and $\overline{b}$ are the mean values of the posterior probability distributions. We use 1\,000 samples from the distribution $y = \overline{a} \times x + \overline{b} + \overline{N}$, where $\overline{N} \sim \mathcal{N}(0,\,\overline{\sigma}^{2})$ and mark them as shaded lines. Therefore, the total shaded region around the best-fitting line represents the full distribution of the scatter, and not $\pm \sigma$.

\section{Samples from the literature}\label{s:all_samples}

\subsection{Post-starburst candidate samples from the literature}\label{s:other_samples}

To obtain a more global view, we collected additional published information about post-starburst galaxies with CO measurements. We used the samples presented by \citet{french15}, \citet{rowlands15}, \citet{alatalo16b}, and \citet{yesuf17}\footnote{\citet{li19} combined the samples from \citet{french15}, \citet{rowlands15}, and \citet{alatalo16a}, and used their SED-fitting-based dust masses to estimate $M_{\mathrm{H_2}}$ in post-starburst galaxies for which CO measurements were not available. While we do not include this sample in our general analysis, we discuss it in Section \ref{s:results:molecular_gas}.}. The properties of the different samples are summarized in table \ref{tab:other_samples}.

All these studies selected post-starburst candidates using their optical spectra, in particular using the H$\delta$ absorption and/or the Dn4000\AA\, index. The sources in the different samples differ in their emission line properties. In particular, the \citet{french15} sample is the only sample selected using the traditional E+A selection criteria, which require strong H$\delta$ absorption and weak line emission (e.g., H$\alpha$ EW $<$ 3 \AA), where the latter is used to ensure that the galaxies have little on-going star formation. However, such criteria select against systems in which the line luminosity is powered by processes other than star-formation, such as AGN or shocks. \citet{rowlands15} did not apply specific selection criteria for the emission lines, and thus their sample includes galaxies classified as star-forming, composite, and AGN, according to standard line diagnostic diagrams (\citealt{kewley06}; see table 2 there). \citet{alatalo16a} selected sources with emission line ratios consistent with shock excitation models. These emission line ratios, however, are also consistent with star-forming, composite, and AGN ionization, according to standard line diagnostic diagrams (see figure 1 in \citealt{alatalo16a}). \citet{yesuf17} selected galaxies with Seyfert-like emission line ratios. These different selection criteria result in populations with somewhat different properties, as we discuss below.

In \citet{baron22} we used {\it IRAS} far-infrared observations to study the star formation properties of several large samples of post-starburst candidates. These samples are: \citet{french18}, \citet{alatalo16a}, \citet{yesuf14}, and \citet{baron22}, which are the parent samples of the subsets studied here: \citet{french15}, \citet{alatalo16b}, \citet{yesuf17}, and this work, respectively. We found that their far-infrared properties depend on their selection criteria. Using the $60\,\mathrm{\mu m}$ luminosity to estimate the SFR, we found that post-starburst candidates with emission lines, where the emission line ratios are consistent with AGN and/or shocks, are found $>0.3$ dex above the star-forming main sequence in 13\%, 30\%, and 45\% of the cases, for \citet{alatalo16a}, \citet{yesuf14}, and \citet{baron22} respectively (see table 1 there for the confidence intervals), and hosting starbursts ($> 0.6$ dex above the main sequence) in 7\%, 18\%, and 32\% of the cases. On the other hand, post-starburst candidates selected with the traditional E+A selection criteria (\citealt{french18}) showed far-infrared detection fraction of 4.5\%, suggesting that only $\sim$4\% are $>0.3$ dex above the star-forming main sequence.

Here we concentrate on the subsets of post-starburst candidates with available $M_{\mathrm{H2}}$ measurements. To select post-starburst candidates for follow-up mm observations, studies employed additional selection criteria. For example, \citet{french15} and \citet{alatalo16b} selected mid-infrared-bright sources, biasing their samples towards sources with higher SFRs. Indeed, the {\it IRAS} far-infrared detection fraction increases dramatically for the subsets chosen for mm follow-up, from 4.5\% to 40\% for \citet{french15}, and from 13\% to 60\% for \citet{alatalo16b}. To ensure enough Seyferts in their sample, \citet{yesuf17} relaxed the EW(H$\delta$)$>4$\AA\, requirement, allowing EW(H$\delta$) as low as 0.8 \AA. As a result, these subsets no longer represent the star-formation and molecular gas properties of their parent populations.

We used the molecular gas masses reported by the different studies, and scaled them so that $\alpha_{\mathrm{CO}} = 4.3\, \mathrm{M_{\odot}(K\, km\, sec^{-1}\, pc^2)^{-1}}$. To estimate the far-infrared-based SFR in these sources, we used {\sc scanpi} to extract their 60 $\mathrm{\mu m}$ fluxes and the \citet{baron22} method assuming $L_{\mathrm{SF}} = 1.716 \times \nu L_{\nu}(60\,\mathrm{\mu m})$, which we then converted into SFRs using $\mathrm{SFR} = L_{\mathrm{SF}} / \, 10^{10}\, L_{\odot}\, [M_{\odot}\,\mathrm{yr}^{-1}]$ as in section \ref{s:data:FIR}. Following \citet{baron22}, we adopted a conservative uncertainty of 0.2 dex which also accounts for our use of a single monochromatic luminosity to estimate SFR. The sources from \citet{french15} were followed up with far-infrared \emph{Herschel} observations \citep{smercina18}. Since these are deeper than the archival {\it IRAS} scans, we used the total infrared luminosities reported in \citet{smercina18}, and converted them to far-infrared luminosities (8--1000 $\mathrm{\mu m}$) using $L_{\mathrm{SF}} = L_{\mathrm{TIR}}/1.06$, where $L_{\mathrm{TIR}}$ is the total infrared luminosity (3--1100 $\mathrm{\mu m}$). For the \citet{french15} sources that were detected in 60 $\mathrm{\mu m}$ by {\it IRAS}, we verified that the {\it IRAS}-based SFRs are consistent with those obtained with \emph{Herschel}.

We also collected the stellar masses and Dn4000\AA-based SFRs reported in the MPA-JHU catalogue for these sources. \citet{french18} applied their age-dating technique to the sources of \citet{french15}, \citet{rowlands15}, and \citet{alatalo16b}. We used the best-fitting SFH parameters presented there. We applied the age-dating technique to the sources of \citet{yesuf17}, and list the best-fitting SFH parameters in Table \ref{table:ages_yesuf} in the appendix. Most of the sources in the different samples have emission line ratios above the theoretical `maximum starburst' line by \citet{kewley01}, placing them in the LINER or Seyfert regions in standard line diagnostic diagrams \citep{baldwin81, veilleux87}. For such sources, most of the H$\alpha$ emission is not associated with star formation (e.g., \citealt{wild10}). As a result, we cannot use the H$\alpha$ luminosity to estimate the SFR.

We list the optical and far-infrared SFRs, stellar masses, and molecular gas masses of the galaxies from the different samples in tables \ref{table:french_meas}, \ref{table:rowlands_meas}, \ref{table:alatalo_meas}, and \ref{table:yesuf_meas} in appendix \ref{a:SF_and_mol_PSB_sample}.

\subsection{Comparison samples}\label{s:comp_samples}

To compare the star formation and molecular gas properties of post-starburst candidates to those of other galaxies, we used several comparison samples. We make use of the xCOLD GASS survey (\citealt{saintonge17}), which includes 532 galaxies with CO(1-0) measurements from the IRAM 30 m telescope. The galaxies in this sample were mass-selected in $0.01 < z < 0.05$ to be representative of the local galaxy population with $M_{*} > 10^9\, M_{\odot}$. Out of these, we select all the emission line galaxies (413 out of 532), out of which 226 are star-forming galaxies, 83 are composites, 91 are LINERs, and 13 are Seyferts (see additional details in \citealt{saintonge17}). We used {\sc scanpi} to extract their 60 $\mathrm{\mu m}$ fluxes and estimated their far-infrared SFRs. We extracted their SFR(Dn4000\AA) from SDSS. In figure \ref{f:SFR_FIR_versus_optical_xcold_gass} in appendix \ref{a:SF_and_mol_comp_sample}, we plot SFR(far-infrared) versus SFR(Dn4000\AA), and present the best-fitting linear relation between them, including upper limits. 

We also use two samples of (ultra)luminous infrared galaxies. We make use of the far-infrared SFRs and molecular gas masses reported by \citet{gao04}, where we scale their molecular gas masses according to our selected conversion factor $\alpha_{\mathrm{CO}} = 4.3\, \mathrm{M_{\odot}(K\, km\, sec^{-1} \, pc^2)^{-1}}$, and their SFRs according to our conversion $\mathrm{SFR} = L_{\mathrm{SF}} / \, 10^{10}\, L_{\odot}\, [M_{\odot}\,\mathrm{yr}^{-1}]$. We also use the SFRs and stellar masses of the (ultra)luminous infrared galaxies by \citet{u12}, which were estimated using radio to X-ray SED fits. In figure \ref{f:SFR_versus_MH2_comparison_samples} in appendix \ref{a:SF_and_mol_comp_sample}, we plot SFR versus $M_{\mathrm{H_2}}$ for the xCOLD GASS and \citet{gao04} samples, and the best-fitting linear relation between them, including upper limits.

\section{Results}\label{s:results} 

\begin{figure}
\includegraphics[width=3.5in]{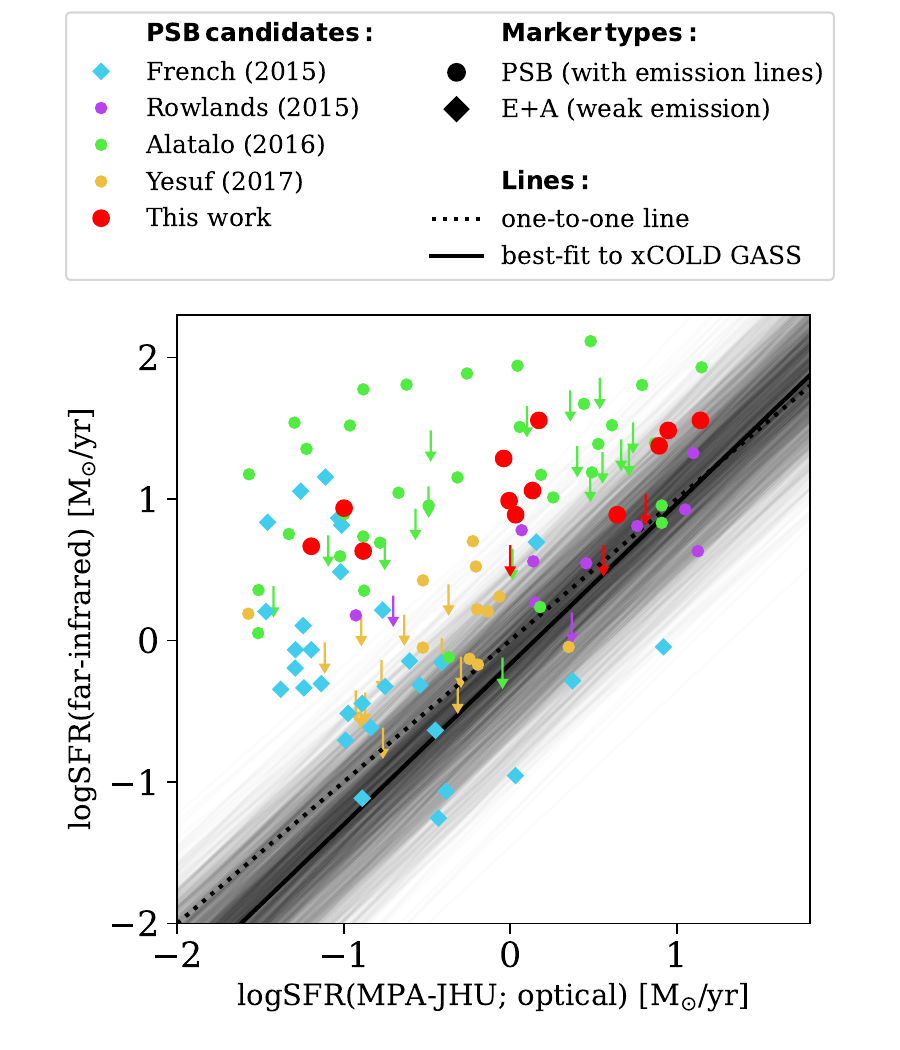}
\caption{\textbf{Comparison of far-infrared and optical SFRs in post-starburst candidates.} 60 $\mathrm{\mu m}$-based SFR versus the Dn4000\AA-based SFR for different post-starburst galaxy samples, where points represent measurements and arrows represent upper limits. Each color corresponds to a different sample, as indicated in the legend at the top. The different marker types separate classical E+A galaxies, selected with no/weak emission lines, from post-starbursts with emission lines. The dotted black line is a 1:1 relation. The black solid line is the best-fitting relation obtained for the comparison galaxies from xCOLD GASS (see figure \ref{f:SFR_FIR_versus_optical_xcold_gass} and section \ref{s:comp_samples}), and the shaded area represents the full posterior distribution of the best-fit. The figure shows significant differences between SFR(far-infrared) and SFR(optical) for post-starburst candidates.} 
\label{f:SFR_FIR_versus_Dn4000}
\end{figure} 

\begin{figure*}
	\centering
\includegraphics[width=1\textwidth]{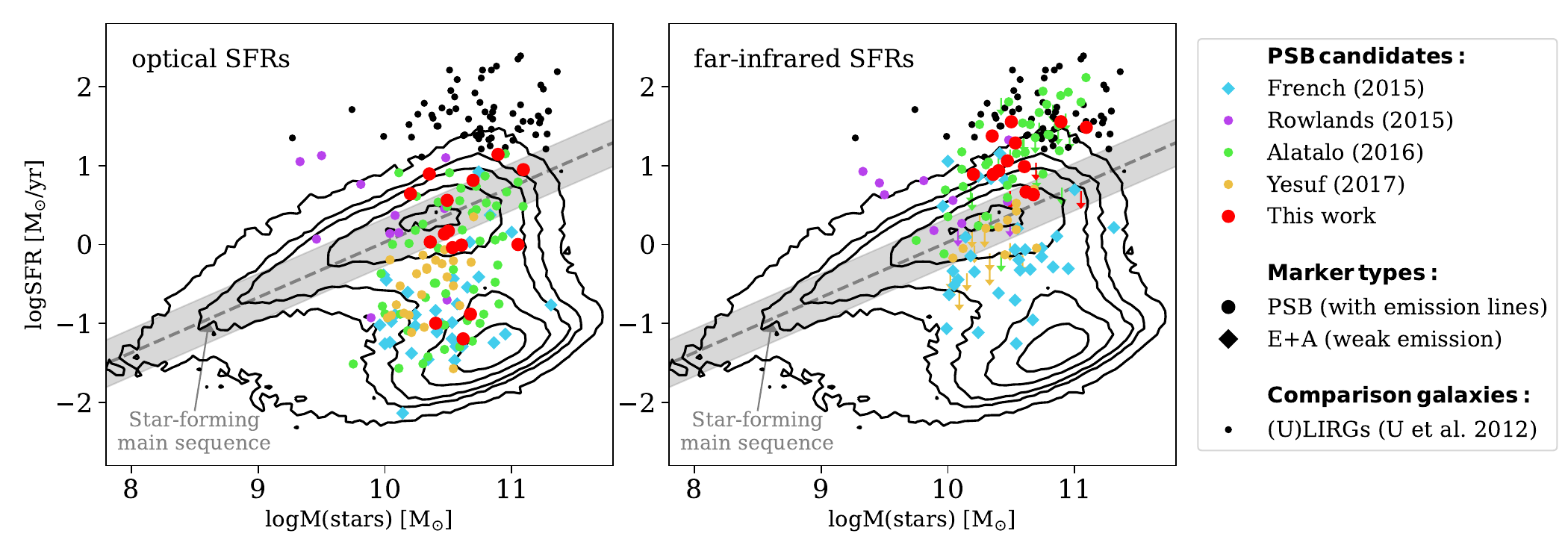}
\caption{\textbf{SFR versus stellar mass using optical and far-infrared SFRs.} 
The points represent measurements and arrows represent upper limits, where each color corresponds to a different sample, as indicated in the legend on the right. The different marker types separate classical E+A galaxies, selected with no/weak emission lines, from post-starbursts with emission lines. The black contours represent all SDSS galaxies at $z \leq 0.15$ from the MPA-JHU catalog. The grey dotted line represents the star-forming main sequence at $z=0$ (\citealt{whitaker12}), and the light-grey band a $\pm$0.3 dex interval around it. For comparison, we also show the (ultra)luminous infrared galaxies by \citet{u12}. For the post-starburst galaxy candidates, the left panel shows the optical SFR derived from the Dn4000\AA\, index, and the right panel shows the far-infrared SFR derived from 60 $\mathrm{\mu m}$ {\it IRAS} observations. Different SFR tracers place post-starburst galaxies in different locations with respect to the main sequence, with optical SFRs placing post-starbursts on or below the main sequence, and far-infrared SFRs above or on the main sequence.}\label{f:SFR_versus_m}
\end{figure*}

\subsection{Star formation properties}\label{s:results:obscured_SF}

The SFR of a galaxy can be derived using different SFR diagnostics which span the full electromagnetic spectrum, from UV to radio wavelengths, and using either continuum or line emission (e.g., \citealt{kennicutt98, calzetti13}). Different SFR tracers have different uncertainties, biases, and regimes of applicability. In this section we use optical and far-infrared tracers to estimate the SFR in post-starburst candidates, and find them to be significantly different from each other. We therefore start by discussing the details of the SFR diagnostics we used, paying particular attention to the limitations of the methods when applied to post-starburst galaxies.

The optical SFRs were extracted from the MPA-JHU value added catalog, and are based on a combination of SDSS optical spectroscopy and photometry \citep{b04}. 
For galaxies with emission line ratios consistent with star formation, the SFR was estimated using the observed emission lines (SFR$_{e}$). The same procedure cannot be applied to composite galaxies or AGN hosts, as their emission lines are not primarily powered by star formation ionization. Therefore, for composites and AGN hosts, \citet{b04} estimated the SFR using the measured Dn4000\AA\, index and the relationship between SFR$_{e}$ and Dn4000\AA\, for the star-forming galaxies. The relation between SFR$_{e}$ and Dn4000\AA\, has a scatter of $\sim 0.4$ dex (see figure 11 in \citealt{b04}), making SFR(Dn4000\AA) less reliable than SFR$_{e}$ (see discussion in \citealt{b04}). Most of the post-starburst galaxies we consider have emission line ratios consistent with composites or AGN, and thus their optical SFR were derived indirectly using the Dn4000\AA\, index. The major limitation of optical SFR diagnostics is that they are susceptible to dust attenuation and thus may miss obscured star formation. For example, \citet{poggianti00} found that 50\% of their luminous infrared galaxies show strong H$\delta$ absorption and weak optical line emission which suggest low SFR(optical), where in fact these systems host obscured starbursts. In \citet{baron22} we showed that many galaxies selected to be post-starburst using their optical spectra host in fact obscured star formation.

The far-infrared SFRs are based either on the 60 $\mathrm{\mu m}$ monochromatic luminosities from IRAS (for \citealt{rowlands15}, \citealt{alatalo16b}, and \citealt{yesuf17}), the bolometric far-infrared luminosity (\citealt{french15}; based on \citealt{smercina18}), or on a combination of the 60 $\mathrm{\mu m}$ monochromatic luminosity from IRAS with mm continuum flux from NOEMA (this work). In all cases, the monochromatic luminosities have been converted into bolometric far-infrared luminosities assuming the CE01 templates. While the far-infrared bolometric luminosity can trace obscured star formation, it has its own limitations. First, for highly-obscured systems with accreting supermassive black holes, the far-infrared luminosity could be contaminated by the AGN. In addition, the bolometric far-infrared luminosity has an averaging time scale of $>$100 Myr (e.g., \citealt{kennicutt98, calzetti13}), which may lead to strong systematic uncertainties in galaxies with rapidly-changing SFRs, such as the post-starbursts considered here (see simple model in section \ref{s:FIR_to_SFR}). For example, \citet{hayward14} used hydrodynamical simulations of galaxy mergers and showed that ultraviolet emission from the post-burst population (e.g., A-type stars) can power infrared emission which is unrelated to ongoing star formation. \citet{smercina18} found that SFR(far-infrared) is a factor of a few larger than the SFR derived using infrared emission lines, where the latter have averaging time scales of $\sim$10 Myr and they do not suffer from dust extinction. Thus, they suggested that the far-infrared bolometric luminosity overestimates the instantaneous SFR in post-starburst galaxies.

In Figure \ref{f:SFR_FIR_versus_Dn4000} we compare between the SFRs derived from optical and far-infrared observations for the post-starbursts we consider. For comparison, we also show the best-fitting relation between SFR(far-infrared) and SFR(Dn4000\AA) obtained for typical galaxies from the xCOLD GASS sample, which includes both star-forming and AGN host galaxies (see figure \ref{f:SFR_FIR_versus_optical_xcold_gass} and section \ref{s:comp_samples}). The shaded lines represent the full posterior distribution for the xCOLD GASS best fit. Figure \ref{f:SFR_FIR_versus_Dn4000} shows significant differences between SFR(far-infrared) and SFR(Dn4000\AA) for post-starburst galaxies, with most of the galaxies being above the one-to-one relation, and some sources showing SFR(far-infrared) which is two orders of magnitude larger than that derived using the Dn4000\AA\, index.

We quantify the significance of the deviation of each post-starburst sample from the best-fitting relation obtained from the xCOLD GASS sample. The scatter of the best-fitting relation is $\sigma_{\mathrm{xCOLD}} = 0.33$ dex. We use the Kaplan-Meier Estimator to estimate the median $(\log \mathrm{SFR(far-infrared)} - \log \mathrm{SFR(optical)})/\sigma_{\mathrm{xCOLD}}$ for each of the samples, including upper limits in the analysis \citep{feigelson85, davidson_pilon2020}. These median values represent the median distance in $\sigma$ of every post-starburst sample from the xCOLD GASS relation. Ordered in a descending order, the median (and the 16th and 84th percentiles) is $3.0\sigma$ ($0.2\sigma$, $5.7\sigma$) for \citet{alatalo16b}, $2.6\sigma$ ($0.7\sigma$, $4.6\sigma$) for this work, $1.6\sigma$ ($-0.7\sigma$, $5.5\sigma$) for \citet{french15}, $1.0\sigma$ ($-1.2\sigma$, $2.2\sigma$) for \citet{yesuf17}, and $0.3\sigma$ ($-1.5\sigma$, $2.1\sigma$) for \citet{rowlands15}.

In Figure \ref{f:SFR_versus_m} we show the SFR versus stellar mass for post-starburst galaxy candidates, where the left panel shows the optical SFR and the right panel the far-infrared SFR. For comparison, the contours show all SDSS galaxies at $z \leq 0.15$, where the SFR and stellar mass were extracted from the MPA-JHU catalog, similarly to the post-starbursts. We also present the sample of luminous and ultraluminous infrared galaxies by \citet{u12}, where the SFRs and stellar masses were estimated using radio to X-ray SED fits. Since in this case the SFRs and stellar masses were derived using other methods, one cannot directly compare between the post-starbursts and the (ultra)luminous infrared galaxies. Nevertheless, we use them to mark the region in the SFR-stellar mass plane occupied by extreme starburst systems. Using the \citet{whitaker12} star-forming main sequence as a reference, figure \ref{f:SFR_versus_m} qualitatively suggests that optical observations place post-starburst galaxies on or below the star-forming main sequence. On the other hand, far-infrared SFRs place many systems $>0.3$ dex above the star-forming main sequence, with some galaxies showing SFRs which are comparable to those of (ultra)luminous infrared galaxies.

\begin{figure}
\includegraphics[width=3.5in]{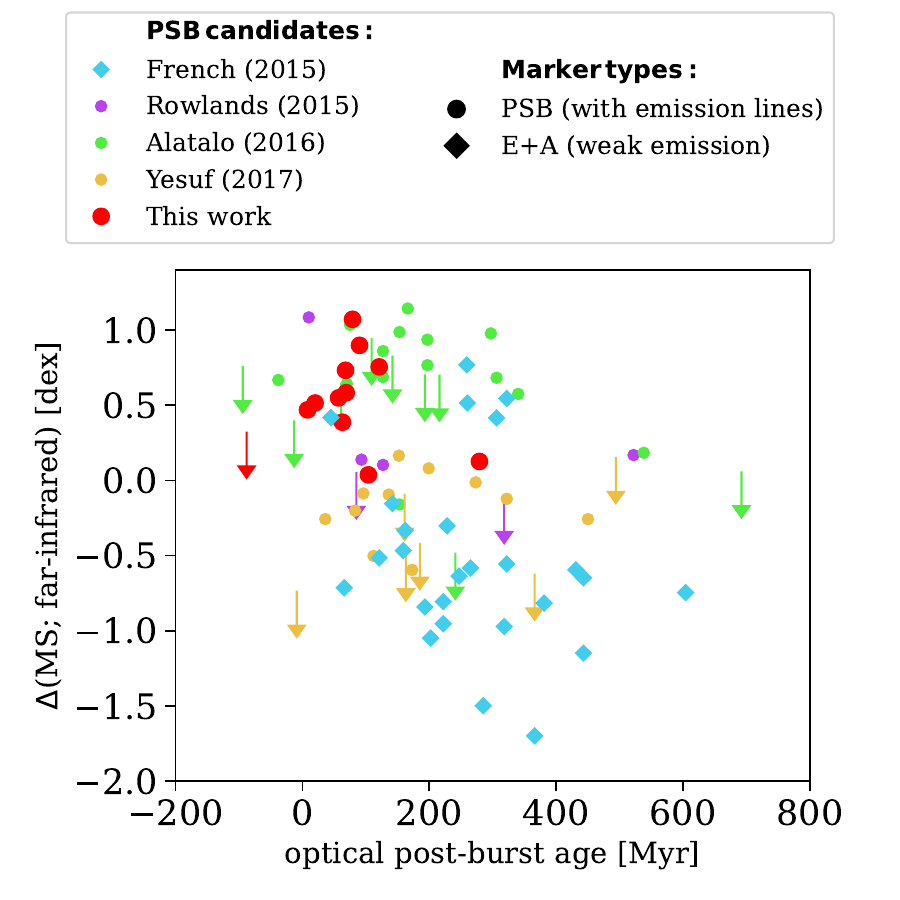}
\caption{\textbf{The location of a galaxy with respect to the star-forming main sequence versus the post-burst age.} The points represent measurements and arrows represent upper limits, where each color corresponds to a different sample, as indicated in the legend at the top. $\mathrm{\Delta(MS)}$ is estimated using the far-infrared-based SFR, while the post-burst age is estimated from stellar population synthesis modeling of the optical spectrum. The significant scatter in this relation suggests that while the presence of obscured star formation decreases on average as the post-burst age increases, there is no strong evidence that the optical and FIR trace the same stellar population.} \label{f:Delta_MS_versus_pb_age}
\end{figure}

In Figure \ref{f:Delta_MS_versus_pb_age} we show the location of the galaxy with respect to the star-forming main sequence, $\mathrm{\Delta(MS)}$, as a function of the optical post-burst age. $\mathrm{\Delta(MS)}$ was estimated using far-infrared SFRs, assuming the \citet{whitaker12} star-forming main sequence. The post-burst age was estimated from stellar population synthesis modeling of the optical spectra. Figure \ref{f:Delta_MS_versus_pb_age} shows a weak decreasing trend (generalized Kendall's $\tau=-0.22$ and p-value of 0.002) with significant scatter, with, for example, systems with post-burst ages of 200 Myrs showing SFRs 1 dex above and 1 dex below the main sequence. Therefore, we do not find strong evidence that the far-infrared and optical trace the same stellar population.

As discussed at the beginning of this section, optical and far-infrared SFR diagnostics have different limitations and biases. Given that, the significant difference between SFR(far-infrared) and SFR(optical) for many of the post-starburst galaxies (figures \ref{f:SFR_FIR_versus_Dn4000} and \ref{f:SFR_versus_m}) may be due to the following: (i) the far-infrared luminosity includes AGN contribution we did not account for, (ii) the far-infrared luminosity, which has an averaging time scale of $>$100 Myrs, is powered by the post-burst stellar population, and thus overestimates the instantaneous SFR in these rapidly-transitioning objects, and (iii) the far-infrared luminosity includes a contribution from obscured star-forming regions, in which case the optical underestimates the true SFR. In the rest of this section we explore these different options.

First, we demonstrate that the AGN contribution to the far-infrared emission of the galaxies we consider is negligible. The intrinsic AGN SED at infrared wavelengths has been the subject of extensive research (see \citealt{lani17} and references therein). Both observations and clumpy torus models show that the intrinsic AGN SED drops significantly above $20\,\mathrm{\mu m}$, with the luminosity ratio $\nu L_{\nu}(60\,\mathrm{\mu m}) / \nu L_{\nu}(22\,\mathrm{\mu m})$ ranging from 0.17 to 0.3 (e.g., \citealt{mor09, stalevski12, lani17}). In contrast, the post-starburst galaxies we consider show $\nu L_{\nu}(60\,\mathrm{\mu m}) / \nu L_{\nu}(22\,\mathrm{\mu m}) \sim 3$, consistent with the ratio observed in star-forming galaxies  \citep{chary01}, where star formation dominates the dust heating. In \citet{baron22} we used the near-infrared to far-infrared SED of post-starbursts to rule out the option that the far-infrared luminosity is powered by the AGN. We showed that even under the most extreme assumption that the AGN dominates the $22\,\mathrm{\mu m}$ completely, its contribution to the $60\,\mathrm{\mu m}$ flux is $<$10\% (see section 4.1.1 in \citealt{baron22}).

\begin{figure}
\includegraphics[width=3.5in]{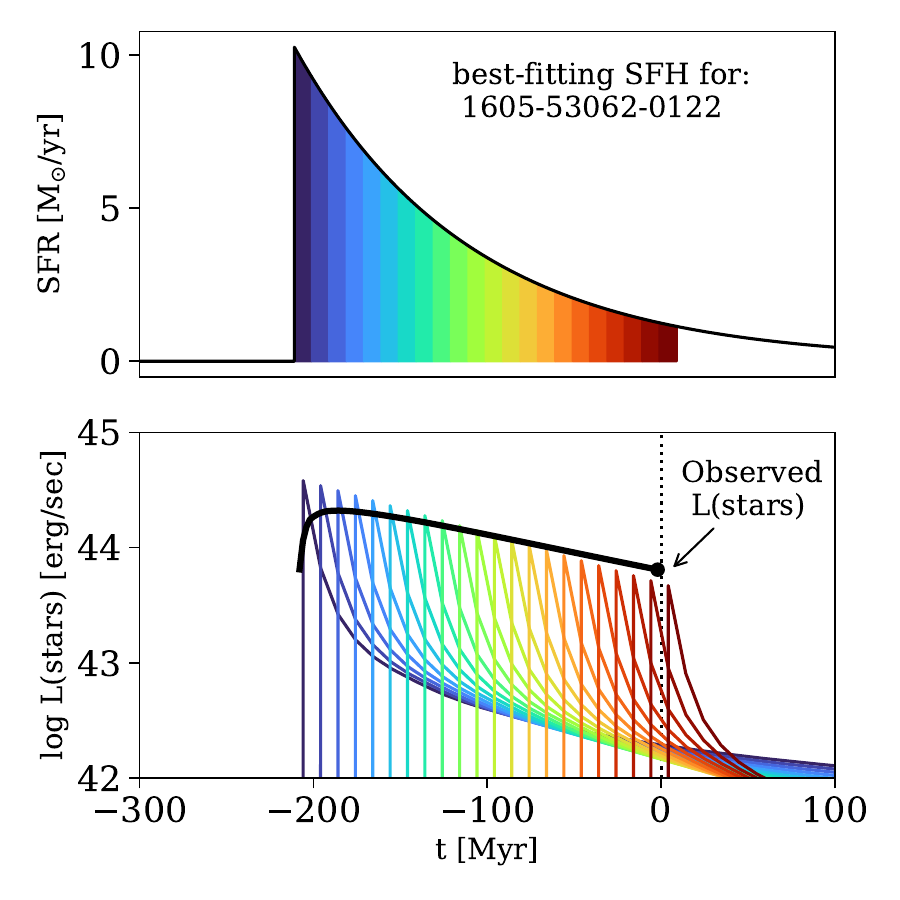}
\caption{\textbf{The stellar population luminosity following a starburst.} The top panel shows an example the best-fitting SFH obtained with the age-dating technique by \citet{french18} for one of the galaxies from our sample (ID: 1605-53062-0122; see table \ref{tab:integ_properties}). The best-fitting SFH also includes a contribution from an older ($>$ 2 Gyr) stellar population, but only the recent burst is shown. The bottom panel shows how the total bolometric luminosity of the stellar population changes as the burst ages. Each colored curve represents the bolometric luminosity of a stellar population formed at different times, as indicated in the top panel. The black curve represents the total bolometric luminosity of all the stars as a function of time. The larger black point represents the expected stellar bolometric luminosity today (t=0), which we use to estimate the expected dust far-infrared emission from the recent burst.} \label{f:SB99_visualization}
\end{figure}

Next, we test the possibility that the observed far-infrared emission is powered by the recently-quenched starburst seen in the optical (i.e., the A-type stars; \citealt{hayward14}). We followed the same approach we presented in \citet{baron22}. The approach is visualized in Figure \ref{f:SB99_visualization} using the best-fitting SFH obtained for the galaxy 1605-53062-0122 from our sample (see table \ref{tab:integ_properties}). We used the optically-derived best-fitting SFHs of the post-starburst candidates, and estimated the expected far-infrared emission at t=0 (present day). The best-fitting SFHs were obtained from the age-dating technique by \citet{french18}, and they are normalized to have the same total stellar mass as given in the MPA-JHU catalogue. We used {\sc starburst99} \citep{leitherer99} to simulate the SED of a stellar population following an instantaneous starburst that produced $M = 10^{7}\,\mathrm{M_{\odot}}$. We used a \citet{kroupa01} IMF, and the original Padova tracks \citep{girardi00}, assuming 0.4, 1, and 2.5 solar metallicity. The {\sc starburst99} output includes the SED of the stellar population following the starburst, from 0 to 1 Gyr, in steps of 10 Myr, where each SED covers the wavelength range $100$--$10^6$\AA. We integrated each {\sc starburst99} SED to obtain the total bolometric luminosity of the stellar population in each time step. For each post-starburst candidate, we used its best-fitting SFH to calculate the stellar mass that was formed in each time step of 10 Myrs from t=1 Gyr to t=0 (present time). We then used the {\sc starburst99} output to estimate the bolometric luminosity contributed by this stellar mass at t=0. By summing up all the contributions from the stellar populations from 1 Gyr to now, we obtained the total bolometric luminosity of the post-burst population today.

To estimate the far-infrared emission driven by the post-burst population, we have to assume the fraction of the stellar radiation that is absorbed by dust. As shown by \citet{hirashita03}, in the most extreme case of dust-obscured starbursts, dust covers the stellar population entirely and absorbs all its radiation. In this case, the dust far-infrared emission will be equal to the total bolometric luminosity. In the more realistic case of partial obscuration, the far-infrared emission will be some fraction of the total bolometric luminosity: $L_{\mathrm{FIR}} = k \times L_{\mathrm{bol}}^{\mathrm{SF}}$. \citet{hirashita03} showed that $k \sim 0.5$ in star-forming galaxies. That is, dust absorbs $\sim$50\% of the stellar radiated energy is star-forming galaxies. We examined a broader range of $k =$0.3--1, where the latter represents the case of full obscuration in starbursts. We also examined the \citet{charlot00} dust model, and found consistent results. The dust models by \citet{hirashita03} and \citet{charlot00} have been constrained using observations of star-forming and starburst galaxies, and their application to post-starburst candidates is valid only for systems on or above the star-forming main sequence. Systems below the main sequence are expected to be dust poor, where the dust is expected to absorb less than the minimum assumed 30\% of the stellar radiation. Therefore, in such sources, the expected far-infrared emission we estimated should be treated as an upper limit.

\begin{figure}
\includegraphics[width=3.2in]{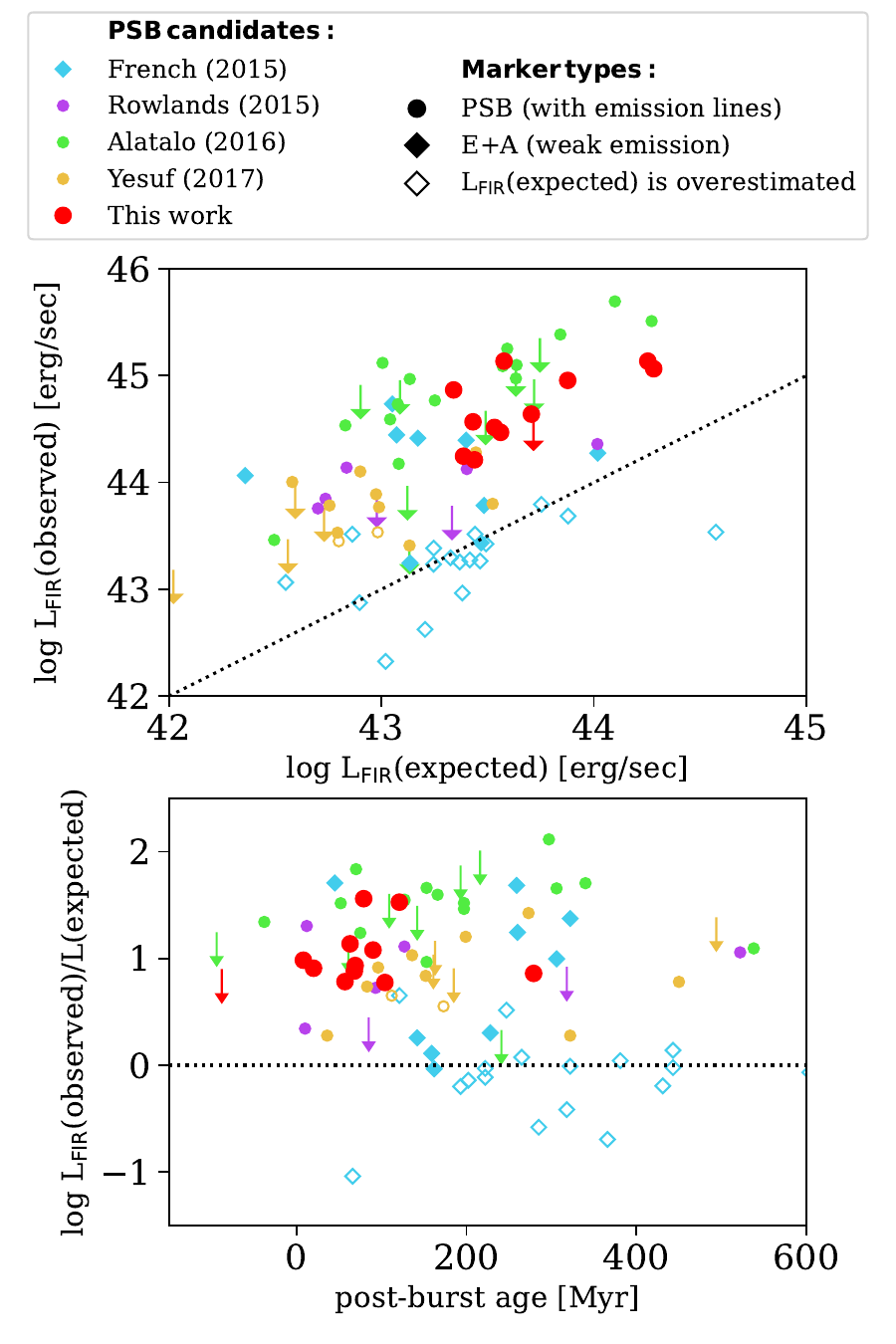}
\caption{\textbf{Comparison of observed far-infrared luminosity with that expected from the recently-quenched starburst observed in optical wavelengths.} Top panel: the points represent measurements and arrows represent upper limits, where each color corresponds to a different sample, as indicated in the legend at the top. The different marker types separate classical E+A galaxies, selected with no/weak emission lines, from post-starbursts with emission lines. The expected far-infrared luminosity was calculated using the best-fitting SFH and using {\sc starburst99}. The black dotted line represents a 1:1 relation. Most of the sources are above the 1:1 line, suggesting that the observed far-infrared emission is larger than that expected from the post-burst population (section \ref{s:results:obscured_SF}). The empty points represent sources which are $<$-0.5 dex below the star-forming main sequence. These sources are most-likely dust poor, and thus their expected far-infrared emission is overestimated in our model by a larger factor. Bottom panel: L$_{\mathrm{FIR}}$(observed)/L(expected) versus the post-burst age, where no relation is detected.} \label{f:LSF_FIR_observed_versus_predicted}
\end{figure}

In Figure \ref{f:LSF_FIR_observed_versus_predicted} we compare between the observed far-infrared emission and that expected from the recently-quenched starburst seen in optical. Here we use Padova tracks with solar metallicity, and use $k = 0.5$. We first discuss the post-starburst galaxies samples selected \emph{with} emission lines (\citealt{rowlands15, alatalo16b, yesuf17}; and this work). Figure \ref{f:LSF_FIR_observed_versus_predicted} shows that the observed far-infrared emission is 10--100 times brighter than that expected from the optical information. This suggests that the observed far-infrared emission is not powered by the remaining A-type stars, and thus the difference between SFR(far-infrared) and SFR(optical) cannot be explained by the long averaging timescale of the far-infrared emission. The most probable alternative is that the far-infrared emission is powered by ongoing obscured star formation in post-starburst galaxies selected \emph{with} emission lines. We reach similar conclusions when using Padova tracks with 0.4 or 2.5 solar metallicity, and for the entire range of $k$ observed in star-forming galaxies, from roughly 0.3 to 1. Finally, we find no relation between L$_{\mathrm{FIR}}$(observed)/L(expected) and the post-burst age. That is, we find no evidence that younger systems show larger difference between the observed far-infrared and that expected from the stellar population.

We now focus on the classical E+A galaxies, selected with weak/no emission lines (\citealt{french15}). Figure \ref{f:LSF_FIR_observed_versus_predicted} shows that 7 (26\%) out of the 26 E+A candidates selected with the traditional criteria have observed far-infrared emission $> 5$ times larger than that expected from the optical spectra, even when assuming that dust absorbs and reemits 100\% of the stellar radiation. A factor of 5 difference cannot be attributed to various uncertainties and free parameters in our {\sc starburst99} modeling, and thus for these sources, we favor the obscured starburst scenario. These are also the sources for which significant molecular gas reservoirs were reported by \citet{french15}. Roughly 50\% of the E+A galaxies lie on or below the one-to-one line, favoring the explanation that the far-infrared emission is powered by the aging post burst population. All the sources below the one-to-one line are sources which are 0.5--1 dex below the star forming main sequence, and are most-likely dust poor. For such sources, our assumption that the dust absorbs 50\% of the stellar radiation is invalid, and thus the expected far-infrared luminosities are overestimated by a large factor.

\underline{To summarize:} in this section we used optical and far-infrared diagnostics to trace the SFR in post-starburst galaxies, and found significant differences between them. In particular, many of the post-starburst candidates show SFR(far-infrared) 10--100 times larger than SFR(optical), placing them $>0.3$ dex above the star-forming main sequence. We examined different options that may cause such a difference. For most of the post-starburst galaxies selected \emph{with} emission lines, we favor the scenario in which the far-infrared emission is powered by ongoing obscured star formation. In this case, we argue that the optical SFR underestimates the true SFR. For classical E+A galaxies selected with no/weak emission lines, we suggest that 26\% host significant obscured star formation, while the rest show far-infrared luminosity that is consistent with being powered by the post-burst population.

\begin{figure*}
	\centering
\includegraphics[width=1\textwidth]{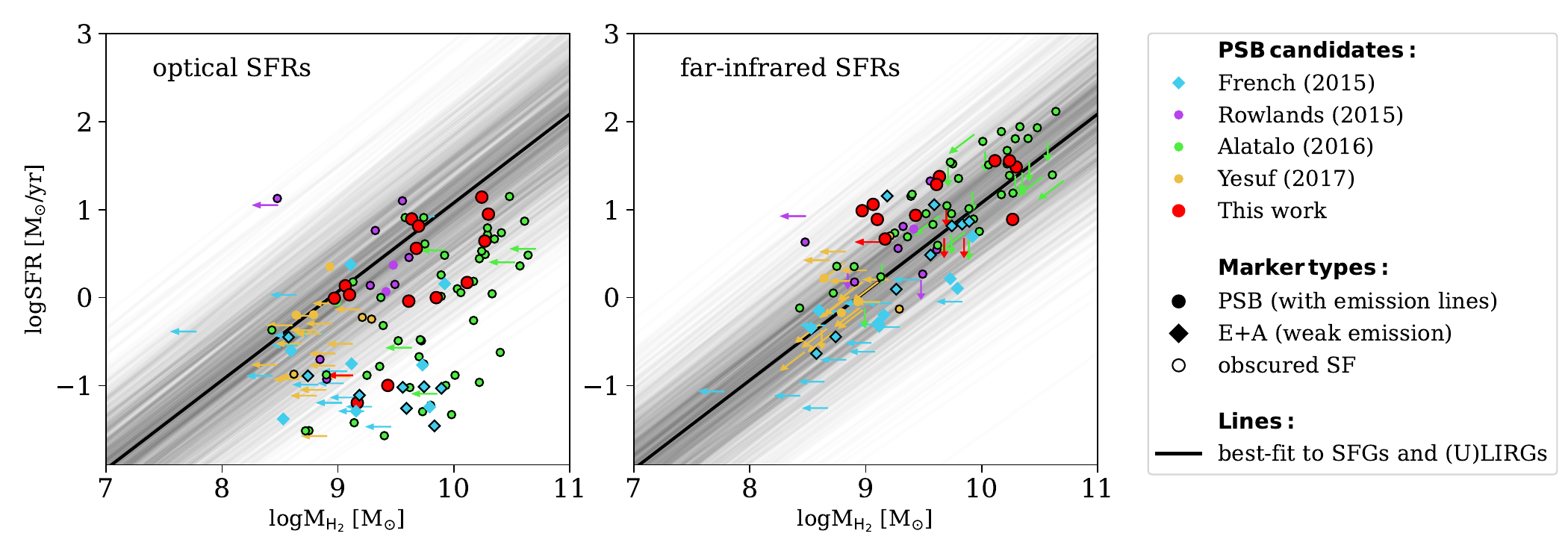}
\caption{\textbf{SFR versus $M_{\mathrm{H_2}}$ using optical and far-infrared-based SFRs.} 
The points represent measurements and arrows represent upper limits, where arrows pointing down are upper limits on SFR, arrows to the left on $M_{\mathrm{H_2}}$, and diagonal arrows to the bottom left on both SFR and $M_{\mathrm{H_2}}$. Each color corresponds to a different sample, as indicated in the legend on the right. The different marker types separate classical E+A galaxies, selected with no/weak emission lines, from post-starbursts with emission lines. The left panel shows the optical SFRs, and the right panel shows the far-infrared SFRs. We mark with black edges the galaxies for which we favor the obscured starburst scenario and argue that optical observations underestimate the true SFR (section \ref{s:results:obscured_SF}). The black solid line represents the best-fit SFR-$M_{\mathrm{H_2}}$ relation obtained for the comparison galaxies (section \ref{s:comp_samples} and figure \ref{f:SFR_versus_MH2_comparison_samples}). The shaded lines represent the full posterior distribution of the best-fit.}\label{f:SFR_versus_MH2}
\end{figure*}


\subsection{Molecular gas properties}\label{s:results:molecular_gas}

Standard galaxy evolution scenario suggests that the termination of star formation is the result of the exhaustion of the molecular gas reservoir, which can be depleted by star formation, stellar feedback, or AGN feedback (e.g., \citealt{hopkins06}). Recent studies challenged this picture, by finding significant molecular gas reservoirs in systems that were selected to have post-starburst signatures in optical wavelengths (e.g., \citealt{french15, rowlands15, alatalo16b, yesuf17, yesuf20b}). For example, \citet{french15} detected CO emission in 53\% of their post-starburst galaxies, finding molecular gas fractions that are consistent with star-forming galaxies. \citet{alatalo16b} detected CO emission in 90\% of their post-starbursts, and found molecular gas fractions that are even larger than those found by \citet{french15}. More recent studies combined the molecular gas measurements with optically-derived SFHs to study the evolution of the molecular gas reservoir as the starburst ages. In particular, \citet{french18} found a significant decline in the molecular gas fraction as a function of the post-burst age, which they attributed to AGN feedback (see also \citealt{bezanson22}). \citet{li19} found a significant decline in the star formation efficiency (SFR/$M_{\mathrm{H_2}}$) as a function of the post-burst age, suggesting that SFR is decoupled from $M_{\mathrm{H_2}}$ and declines faster, possibly also due to AGN feedback. Finally, several studies found that post-starburst galaxies are offset from the Kennicutt--Schmidt (KS) relation \citep{kennicutt98_ks}, with SFRs that are suppressed by a factor of $\sim$10 compared to galaxies with similar gas surface densities \citep{french15, smercina18, li19, smercina22}. 

In light of our discovery that many systems selected as post-starburst host in fact obscured star formation (\citealt{baron22} and section \ref{s:results:obscured_SF}), in this section we revisit previous results regarding the molecular gas properties of post-starburst galaxies. In section \ref{s:results:SFR_MH2} we study the SFR-$M_{\mathrm{H_2}}$ relation. In section \ref{s:results:SFE_PBage} we examine the star formation efficiency and its relation to the post-burst age, and in section \ref{s:results:KS} we present the Kennicutt--Schmidt (KS) relation.

Throughout this section we assume a constant conversion factor $\alpha_{\mathrm{CO}} = 4.3\, \mathrm{M_{\odot}(K\, km\, sec^{-1}\, pc^2)^{-1}}$ for both post-starbursts and comparison galaxies, where the latter includes star-forming and starburst galaxies. However, some of the sources in the different comparison samples are CO-bright galaxies, for which a lower conversion factor is more appropriate (e.g., \citealt{narayanan12, tacconi20} and references therein). Rescaling the molecular gas masses in these sources according to a lower $\alpha_{\mathrm{CO}}$ value will also require us to rescale the molecular gas masses of the CO-bright post-starbursts. Therefore, since the scaling is CO luminosity-dependent, both post-starburst and (ultra)luminous infrared galaxies will be rescaled by the same factor, which will not affect our conclusions.

\subsubsection{SFR--$M_{\mathrm{H_2}}$ relation}\label{s:results:SFR_MH2}

In Figure \ref{f:SFR_versus_MH2} we show SFR versus $M_{\mathrm{H_2}}$ for the post-starburst galaxy candidates, where the left panel shows the optical SFR and the right panel the far-infrared SFR. For comparison, we show the best-fitting SFR-$M_{\mathrm{H_2}}$ relation obtained for star-forming and starburst galaxies, where the shaded region represents the full distribution of the scatter around the relation (see section \ref{s:comp_samples} and figure \ref{f:SFR_versus_MH2_comparison_samples}). According to the left panel of Figure \ref{f:SFR_versus_MH2}, when using optically-derived SFRs, some post-starburst candidates deviate from the SFR-$M_{\mathrm{H_2}}$ relation observed in other galaxies, showing significantly larger molecular gas masses (our sample, most of the \citealt{alatalo16b} sample, and half of the \citealt{french15} sample). To quantify the significance of the deviation, we calculate $(\log\mathrm{SFR_{expected}} - \log\mathrm{SFR_{optical}})/\sigma_{\mathrm{SF}}$, where $\log\mathrm{SFR_{expected}}$ is the SFR expected from the best-fitting relation for star-forming and starburst galaxies and $\sigma_{\mathrm{SF}} = 0.31$ dex is the best-fitting scatter. The median (16th, 84th) deviation is 4.8$\sigma$ (1.6$\sigma$, 6.4$\sigma$) for \citet{french15}, 3.4$\sigma$ (1.5$\sigma$, 5.9$\sigma$) for \citet{alatalo16b}, and 1.0$\sigma$ (0.0$\sigma$, 3.2$\sigma$) for our work\footnote{\citet{rowlands15} and \citet{yesuf17} are not included in this analysis as these studies did not present evidence that their post-starbursts deviate from the SFR-$M_{\mathrm{H_2}}$ relation observed in other galaxies.}. That is, an E+A galaxy from the \citet{french15} sample has a median deviation of 4.8$\sigma$ from the best-fitting relation observed in other galaxies.

In section \ref{s:results:obscured_SF} we found that some of the post-starburst galaxies we consider host obscured star formation, and thus optical SFR diagnostics underestimate their true SFR. In figure \ref{f:SFR_versus_MH2} we mark these sources with black edges. The right panel of the figure shows that if far-infrared SFRs are used instead, neither post-starbursts nor classical E+A galaxies deviate significantly from the SFR-$M_{\mathrm{H_2}}$ relation observed in other galaxies. Using far-infrared SFRs, the median deviation $(\log\mathrm{SFR_{expected}} - \log\mathrm{SFR_{FIR}})/\sigma_{\mathrm{SF}}$ from the best-fitting relation (and the 16th, 84th percentiles) is 0.7$\sigma$ (-0.4$\sigma$, 1.5$\sigma$) for \citet{french15}, -0.8$\sigma$ (-1.7$\sigma$, 0.3$\sigma$) for \citet{alatalo16b}, and -1.2$\sigma$, (-2.2$\sigma$, 0.2$\sigma$) for our work. We therefore suggest that post-starburst candidates that were reported to have significant molecular gas reservoirs with respect to their optical SFR also have obscured star formation. Once the obscured star formation is taken into account, they are no longer significantly offset from the SFR-$M_{\mathrm{H_2}}$ relation.

\begin{figure*}
	\centering
\includegraphics[width=1\textwidth]{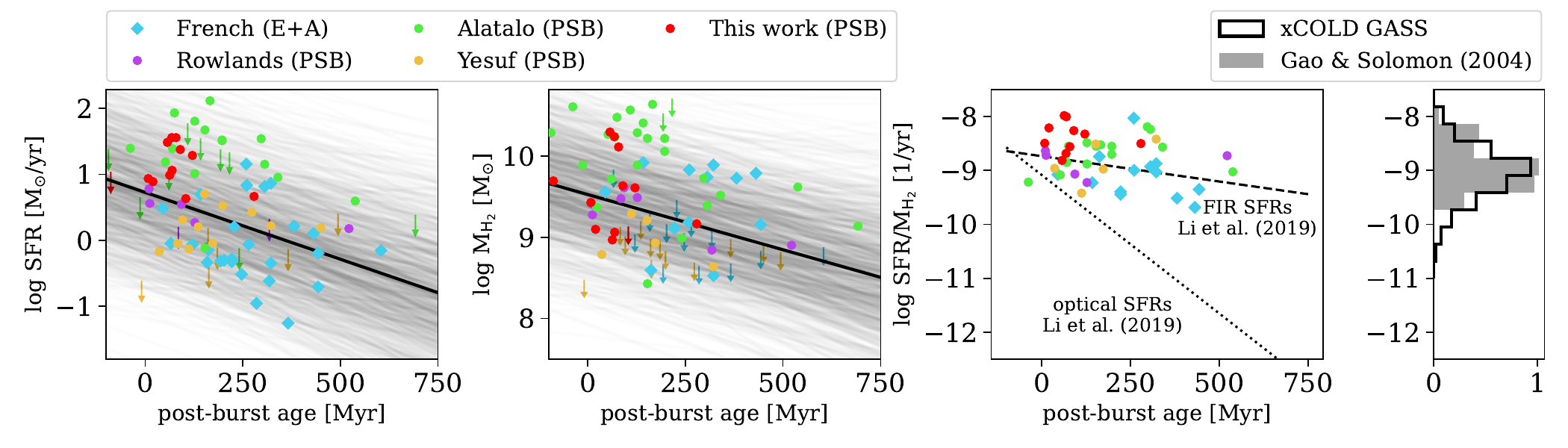}
\caption{\textbf{SFR and $M_{\mathrm{H_2}}$ versus post-burst age.} 
From left to right: first and second panels: far-infrared SFR and $M_{\mathrm{H_2}}$ versus the post-burst age, where measurements are marked with points and upper limits with arrows. The different colors correspond to the different post-starburst samples, as indicated in the legend at the top. The solid black lines represent the best-fitting relations obtained with {\sc linmix}, taking into account upper limits. The grey shaded lines represent the best-fitting scatter in the relations, which is quite high. Third panel: SFR/$M_{\mathrm{H_2}}$ versus post-burst age, where only sources with detected SFR and $M_{\mathrm{H_2}}$ are plotted. The y-axis limits are set to the minimum and maximum star formation efficiencies reported by \citet[see section \ref{s:results:SFE_PBage} for additional details]{li19}. We also show the best-fitting relations we obtain when reanalyzing the \citet{li19} sources using optical and far-infrared SFRs (see appendix \ref{a:li_et_al_reanalysis}). Fourth panel: the distribution of the star formation efficiency in the xCOLD GASS and \citet{gao04} galaxies. The Figure shows that once far-infrared SFRs are used, the trend between the star formation efficiency and age is much flatter than what \citet{li19} found, with star formation efficiencies consistent with those observed in other types of galaxies.}\label{f:SFE_versus_pb_age}
\end{figure*}

\subsubsection{Star formation efficiency and post-burst age}\label{s:results:SFE_PBage}

We now revisit the recently-discovered trends between SFR, $M_{\mathrm{H_2}}$ and post-burst age \citep{french18, li19}. In Figure \ref{f:SFE_versus_pb_age} we show SFR and $M_{\mathrm{H_2}}$ versus the post-burst age. Both SFR and $M_{\mathrm{H_2}}$ decline as a function of post-burst age, with a significant scatter around the best-fitting relations. The decline of $M_{\mathrm{H_2}}$ as a function of post-burst age is consistent with that reported by \citet{french18}, which they attributed to depletion of molecular gas by AGN feedback. We suggest that the decline is driven by the tight correlation between SFR(far-infrared) and $M_{\mathrm{H_2}}$, and the weak decline of SFR(far-infrared) as a function of post-burst age. Indeed, the ratio of the SFR to $M_{\mathrm{H_2}}$ is nearly constant with the post-burst age, suggesting that the reported decline in $M_{\mathrm{H_2}}$ is due to depletion of molecular gas by the starburst. The resulting depletion times, around $10^9$ yrs, are consistent with those observed in other non E+A galaxies.

\citet{li19} found a significant decline in star formation efficiency as a function of post-burst age, where the star formation efficiency drops from $\log \mathrm{SFE}\sim -8$ at post-burst age of $\sim 0$ Myrs to $\log \mathrm{SFE}\sim -12$ at 500 Myrs (dotted line in figure \ref{f:SFE_versus_pb_age}). In contrast, Figure \ref{f:SFE_versus_pb_age} shows a nearly constant star formation efficiency, with $\log \mathrm{SFE}\sim -9$ for post-burst ages 0--750 Myrs. In appendix \ref{a:li_et_al_reanalysis} we repeat the analysis presented in \citet{li19} and show that the significant decline they detect is due to their use of optical SFRs. Once far-infrared SFRs are used, we find that the star formation efficiency does not evolve with the post-burst age, consistent with Figure \ref{f:SFE_versus_pb_age}. For comparison, we show in figure \ref{f:SFE_versus_pb_age} the best-fitting relations we obtain for the \citet{li19} sample when using optical and far-infrared SFRs.

\underline{To summarize:} once we account for the presence of obscured star formation, we find no evidence for an evolution of the molecular gas mass or the star formation efficiency with the post-burst age. Instead, the star formation efficiency changes only weakly with post-burst age, and is consistent with those observed in other types of galaxies. Therefore, There is no need to hypothesize molecular gas depletion by AGN feedback. 

\begin{figure*}
\includegraphics[width=1\textwidth]{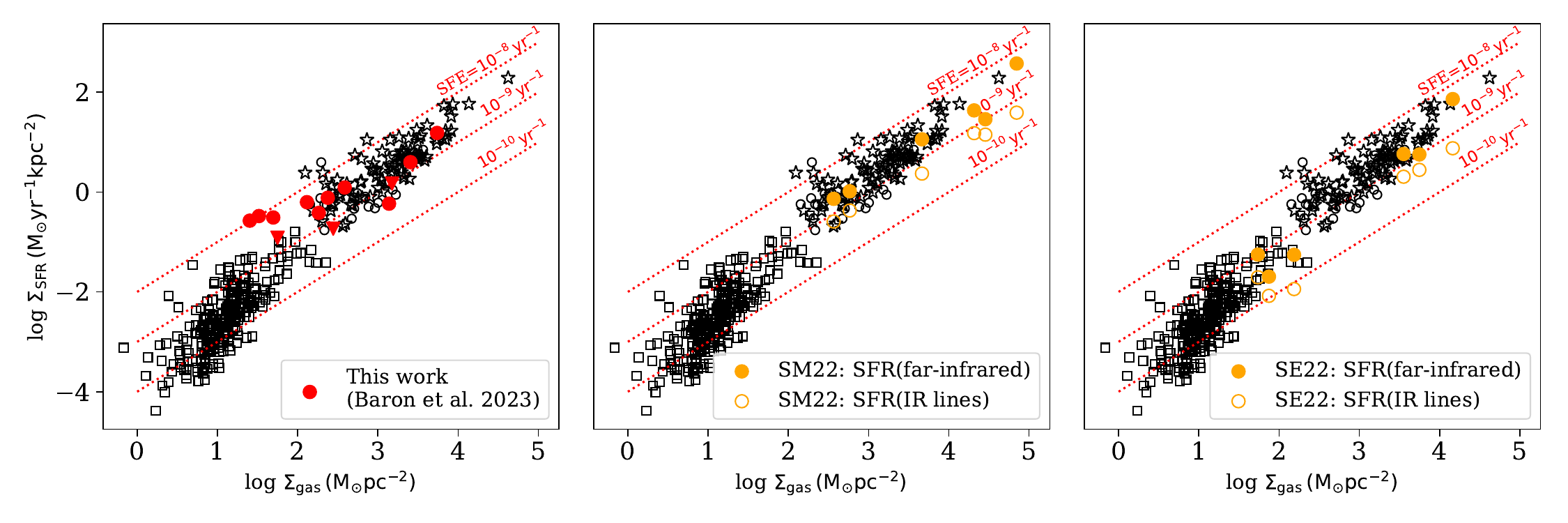}
\caption{\textbf{Kennicutt--Schmidt relation of the post-starburst galaxy candidates we consider.} The colored points represent measurements and red triangles represent upper limits. The comparison galaxies include the dwarfs, spirals, circumnuclear disks, and luminous infrared galaxies from \citet{de_los_reyes19} and \citet{kennicutt21}, and are marked with black empty symbols. For the dwarfs and spirals, $\mathrm{\Sigma(HI+H_{2})}$ is plotted, while for the infrared luminous galaxies, $\mathrm{\Sigma(H_{2})}$ is plotted. The left panel shows our NOEMA sample. The middle panel shows our reanalysis of the \citet{smercina22} sources, where we show both SFR(far-infrared) and SFR(infrared lines). The right panels shows the updated estimates by \citet{sun22} of the \citet{smercina22} sources.} \label{f:KS_relation}
\end{figure*}

\subsubsection{Kennicutt--Schmidt relation}\label{s:results:KS}

We now revisit the recent results that post-starbursts are significantly offset from the KS relation (\citealt{french15, li19, smercina22}). The earlier studies by \citet{french15} and \citet{li19} were based on optical SFRs and unresolved CO observations of post-starburst galaxies. However, in section \ref{s:results:obscured_SF} we showed that some post-starbursts host obscured star formation, and thus optical diagnostics underestimate their true SFR. In appendix \ref{a:li_et_al_reanalysis} we examine the KS relation of the \citet{li19} post-starbursts, which include all the sources presented in \citet{french15}. We find that sources that were reported to be significantly offset from the KS relation host in fact obscured star formation, and if far-infrared SFRs are used, they are no longer offset from the relation of other galaxies (figure \ref{f:reanalysis_of_KS_li_et_al}).

In the left panel of Figure \ref{f:KS_relation} we show the KS relation for the post-starburst candidates in our NOEMA sample. We used the CO half-light radii to convert the SFR and $M_{\mathrm{H_2}}$ to surface densities (see Table \ref{tab:integ_properties}). As a comparison, we show the dwarfs, spirals, circumnuclear disks, and luminous infrared galaxies from \citet{de_los_reyes19} and \citet{kennicutt21}, assuming a constant $\alpha_{\mathrm{CO}} = 4.3\, \mathrm{M_{\odot}(K\, km\, sec^{-1}\, pc^2)^{-1}}$ for all systems. We rescale their reported SFRs to match our conversion from far-infrared luminosity to SFR described in Section \ref{s:other_samples}. Figure \ref{f:KS_relation} shows that if far-infrared luminosity is used to trace star formation, the post-starburst galaxies we consider are not offset from the KS relation of other galaxies. 

\citet{smercina22} used ALMA observations to resolve the star formation and CO emitting regions in six post-starburst candidates. To estimate the SFR in their sources, they used their total infrared (TIR) luminosities and infrared emission lines such as [CII] and [NeII]+[NeIII]. They found that the line-based SFRs are lower than the TIR-based SFRs. Following \citet{hayward14}, they suggested that due to the long averaging timescales of the infrared luminosity, the TIR luminosity may overestimate the instantaneous SFR in these rapidly transitioning sources. Instead, they argued that the Neon-based SFRs are the most accurate, since they have short averaging timescales and are not affected by reddening like the H$\alpha$ line. However, as noted by \citet{smercina18}, a major assumption in their analysis is that the [Ne II] and [Ne III] lines are powered by star formation ionization. The optical line ratios of their sources are all consistent with LINERs rather than with star-forming galaxies. It is therefore unclear whether star formation ionization powers the Neon emission lines. Nevertheless, using these observations, they concluded that post-starburst galaxy candidates are offset from the KS relation, forming stars only 10\% as efficiently as starburst galaxies with similar gas surface densities. 

We performed an independent analysis of the \citet{smercina22} sources and reached different conclusions (see middle panel of figure \ref{f:KS_relation}). We used their minimal and maximal SFR surface densities, where the former is based on infrared emission lines ([CII] or [NeII]+[NeIII]) while the latter is based on the total infrared luminosity. For the gas surface densities, we used their CO luminosities and sizes, and estimated the gas surface density assuming $\alpha_{\mathrm{CO}} = 4.3\, \mathrm{M_{\odot}(K\, km\, sec^{-1}\, pc^2)^{-1}}$. If TIR SFRs are used, we find no offset from the KS relation. If infrared emission lines are used instead, 2 out of 6 sources show no offset, while 4 show an offset of a factor $\sim$2. This is in contrast with the factor of 10 reported by \citet{smercina22}, and the offsets observed in their figure 6. We found that this difference is because \citet{smercina22} did not correct the molecular gas masses in the comparison sample to include Helium, while the estimates for the post-starbursts have been corrected. Once we correct the molecular gas masses in both the comparison sample and the post-starburst samples, we find consistent results, with only a factor of $\sim$2 offset for 4 out of 6 of the sources.

\citet{sun22} recently re-analyzed the ALMA observations of the \citet{smercina22} galaxies. In contrast to \citet{smercina22} who reported compact unresolved CO cores in all their sources, the visibility-plane analysis by \citet{sun22} resolves the CO emission in all sources. Their derived CO sizes are 2--14 times larger than those reported by \citet{smercina22}, and the derived CO luminosities are 1--6 times larger. Furthermore, when using the total infrared luminosity, \citet{sun22} also find no offset of the \citet{smercina22} sources from the KS relation. Using the CO sizes and luminosities reported by \citet{sun22} and the infrared line SFRs from \citet{smercina22}, we find an offset of a factor of $\sim$2 for 4 out of 6 of the sources. In the right panel of figure \ref{f:KS_relation}, we present the KS relation using the updated estimates by \citet{sun22}, using both the far-infrared and infrared line SFRs.

We quantify the deviation of the different samples of post-starburst galaxies from the KS relation observed in other galaxies. The samples are: (i) our NOEMA sample, (ii) our re-analysis of the \citet{smercina22} sample, once using SFR(far-infrared) and once using SFR(infrared lines), and (iii) the updated estimates by \citet{sun22} of the \citet{smercina22} sample, once using SFR(far-infrared) and once using SFR(infrared lines). We use {\sc linmix} to fit a linear relation between $\log \Sigma_{\mathrm{SFR}}$ and $\log \Sigma_{\mathrm{gas}}$ for the comparison galaxies, once for the dwarfs and spirals and once for the circumnuclear disks and luminous infrared galaxies (see \citealt{kennicutt21} for a justification to perform separate fits). The best-fitting scatter parameters are $\sigma_{1} = 0.43$ dex and $\sigma_{2} = 0.33$ dex respectively. Then, for each galaxy in each sample, we estimate the deviation of $\log \Sigma_{\mathrm{SFR}}$ from the best-fitting relation, in units of $\sigma_{1}$ or $\sigma_{2}$, depending on whether the galaxy is in the dwarfs/spirals or circumnuclear disks/luminous infrared galaxies region. The median (16th, 84th percentile) deviation for our sample is -0.3$\sigma$ (-1.3$\sigma$, 1.1$\sigma$). For the \citet{smercina22} sample, the median deviation is 0.1$\sigma$ (-0.34$\sigma$, 0.5$\sigma$) when using SFR(far-infrared) and 1.6$\sigma$ (1.5$\sigma$, 1.9$\sigma$) when using SFR(infrared lines). Using the updated estimates by \citet{sun22}, the median deviation is 0.5$\sigma$ (-0.9$\sigma$, 0.7$\sigma$) using SFR(far-infrared) and 1.8$\sigma$ (1.2$\sigma$, 2.0$\sigma$) using SFR(IR lines). That is, using infrared emission lines, the median galaxy from the \citet{smercina22} sample is offset by 1.6$\sigma$ below the KS relation observed in other galaxies.

In this section we assumed that the molecular gas and star formation extents are similar and equal to the CO extent we derived. A similar assumption was made by \citet{smercina22} and for the infrared-bright comparison galaxies by \citealt{kennicutt21}. For the three brightest sources where we resolved both the CO and mm emission, we found that the mm extent is $0.6 R_{\mathrm{CO}}$, $0.74 R_{\mathrm{CO}}$, and $R_{\mathrm{CO}}$ respectively. However, \citet{bellocchi22} studied the molecular gas and star formation properties in 24 local luminous infrared galaxies, and found that the mm continuum emission is roughly twice more compact than the CO emission. They suggest that the mm extents are affected by flux loss at the outer regions of their sources, but note that a more detailed analysis is required to confirm it. Adopting a smaller $R_{\mathrm{mm}}$ will increase the star formation surface density, for both post-starburst candidates and the comparison galaxies, and thus does not change our conclusion.

\underline{To summarize:} if far-infrared SFRs are used, we find no evidence that post-starbursts are significantly offset from the KS relation. If infrared Neon lines are used to estimate the SFR, our analysis suggests that the median post-starburst is within $\sim 2\sigma$ of the mean relation observed in other galaxies, where $\sigma$ represents the scatter around the mean relation. Since this is based on six sources, and given that it is not clear whether the Neon lines are powered by star formation, we suggest that the claim by \citet{smercina22} that post-starbursts are significantly offset from the KS relation is premature. Additional observations of infrared lines in post-starburst candidates are required to check whether an offset truly exists. In particular, JWST observations can be used to spatially-resolve the Neon lines and disentangle [Ne II] and [Ne III] emission that originate from the AGN versus from star formation.

\section{Discussion}\label{s:disc}

\subsection{Far-infrared to SFR conversion factor}\label{s:FIR_to_SFR}

To convert from far-infrared luminosity to SFR, we used a conversion factor that assumes that the star-formation has been constant for the past 100 Myrs (see e.g., \citealt{kennicutt98, calzetti13}). This assumption may not be accurate for post-starburst candidates whose star formation is believed to be changing rapidly over time. In this section we use a simple model to estimate the far-infrared to SFR conversion factor for an exponentially-declining star formation history. We explore different burst properties and show that even under extreme assumptions about the burst strength, adopting such a conversion factor does not change our main conclusions.

We estimate the far-infrared to SFR conversion factor, $\mathrm{SFR} = C \times L_{\mathrm{FIR}}$, assuming an exponentially-declining star formation history. We use the {\sc starburst99} models and perform a similar calculation to that described in Section \ref{s:results:obscured_SF}. Our star formation history model includes a single exponential decay, with decay time scales in the range 25--200 Myrs, and burst ages in the range 50--500 Myrs. Under the extreme assumption that all post-starburst candidates are the descendants of (ultra)luminous infrared galaxies (e.g., \citealt{kaviraj07, french18}), we normalize each star-formation history so that the peak SFR during the burst (averaged over 100 Myrs) places the system two orders of magnitude above the star-forming main sequence. For each star formation history, we calculate the stellar population bolometric luminosity, and convert it to dust far-infrared emission using the \citet{charlot00} dust model. We define the instantaneous SFR as the average SFR over the past 10 Myrs, and the conversion factor $C$ is defined as the ratio between the instantaneous SFR and the dust far-infrared emission at present day. 

The dust model by \citet{charlot00} has been constrained using observations of star-forming and starburst galaxies, and has not been tested for post-starburst galaxies or galaxies that are below the star forming main sequence. Moreover, systems below the star-forming main sequence are expected to be dust poor, and thus the \citet{charlot00} model is expected to overestimate their dust far infrared emission following the quenching of the burst. Therefore, our simple model is incomplete and probably overestimates the differences between SFR(far-infrared) and the instantaneous SFR for systems below the main sequence. Using a more physical model for dust absorption in galaxies below the main sequence will probably reduce the difference between SFR(far-infrared) and the instantaneous SFR, and thus will only strengthen the conclusions of this section.

\begin{figure}
\includegraphics[width=3.5in]{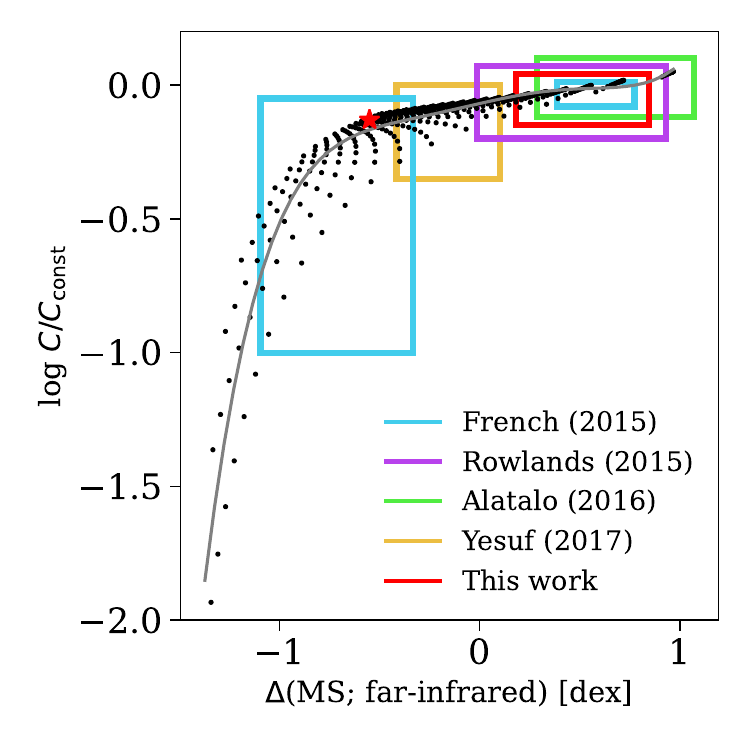}
\caption{\textbf{Far-infrared to SFR conversion factor, $\mathrm{SFR} = C \times L_{\mathrm{FIR}}$, for exponentially-decaying star formation histories.} Each black point represents an exponentially-decaying star formation history with a different burst age (50--500 Myr) and decay timescale (25--200 Myr). The far-infrared to SFR conversion factor $C$ is normalized by $C_{\mathrm{const}}$, the conversion factor assuming that star formation has been constant for the past 100 Myrs. $\Delta$(MS; far-infrared) is the distance of the galaxy from the star-forming main sequence, if $C_{\mathrm{const}}$ is used to convert from far-infrared luminosity to SFR (similar to the measured $\Delta$(MS) shown in Figure \ref{f:Delta_MS_versus_pb_age}). The grey solid line is a 5th-order polynomial fit to the models. The rectangles represent the different post-starburst candidates samples we consider (see additional details in the text). The red star represents the conversion factor of 1605-53062-0122 (see Figure \ref{f:SB99_visualization}).} \label{f:C_versus_delta_MS_FIR}
\end{figure}

In Figure \ref{f:C_versus_delta_MS_FIR} we show the conversion factor $C$ as a function of $\Delta$(MS; far-infrared) for the exponentially-declining star formation histories. We show $C$ normalized by $C_{\mathrm{const}}$, the conversion factor assuming that star formation has been constant for the past 100 Myrs. For example, $\log \big( C/C_{\mathrm{const}} \big) = -1$ represents a case where the constant conversion factor overestimates the instantaneous SFR by one order of magnitude. $\Delta$(MS; far-infrared) is the distance of the galaxy from the star-forming main sequence, if $C_{\mathrm{const}}$ is used to convert from far-infrared luminosity to SFR. This is similar to the measured distance from the main sequence, shown in Figure \ref{f:Delta_MS_versus_pb_age}. Since we normalized the models to have a similar peak SFR, $\Delta$(MS; far-infrared) also represents a sequence in post-burst age, such that $\Delta $(MS; far-infrared)$\sim 1$ represents a case where the burst is still ongoing, while $\Delta$(MS; far-infrared)$\sim -1$ represents older bursts.

Figure \ref{f:C_versus_delta_MS_FIR} shows that near the peak of the burst, $\log \big( C/C_{\mathrm{const}} \big) \sim 0$, suggesting that the constant conversion factor does not significantly overestimate the instantaneous SFR. Farther from the peak of the burst, $\log \big( C/C_{\mathrm{const}} \big)$ decreases as the galaxy transitions from being above the star-forming main sequence to below it. This is because as the burst ages, a larger fraction of the far-infrared emission is powered by older stars. According to this model, when far-infrared information places the galaxy one order of magnitude below the star-forming main sequence, $\Delta$(MS; far-infrared)$=-1$, the constant conversion factor overestimates the instantaneous SFR by one order of magnitude, and the galaxy is in fact two orders of magnitude below main sequence. This is in line with the trends found by \citet{hayward14} using hydrodynamical simulations of galaxy major mergers (see Figures 3 and 4 there).

We mark in Figure \ref{f:C_versus_delta_MS_FIR} the different post-starburst samples we consider. We use their distance from the star-forming main sequence, measured using a constant conversion factor for the far-infrared luminosity (e.g., Figure \ref{f:Delta_MS_versus_pb_age}), which we represent with rectangles. Each rectangle is centered around the median measured $\Delta$(MS; far-infrared), and its width represents two standard deviations. Since the \citet{french15} sample shows bimodal distribution in $\Delta$(MS; far-infrared), we use two rectangles to represent it. According to the model, for systems above the star-forming main sequence (\citealt{alatalo16a}, this work, \citealt{rowlands15}, and some of the \citealt{french15} sources), the far-infrared overestimates the instantaneous SFR by 10--20\%. For systems on the star-forming main sequence (the \citealt{yesuf17} sample), the far-infrared overestimates the instantaneous SFR by a factor of $\sim$2. For systems below the main sequence (some of the sources by \citealt{french15}), the far-infrared can overestimate the SFR by up to a factor of 10.

\citet{smercina18} and \citet{smercina22} found that far-infrared SFRs are a factor of a few larger than the SFRs derived from infrared lines ([CII] or [NeII]+[NeIII]) for their post-starburst candidates. They therefore suggested that the far-infrared overestimates the instantaneous SFR by a factor of a few (see also \citealt{hayward14}). Both far-infrared and infrared emission lines place these sources above the star-forming main sequence, where our simple model predicts a difference of only 10--20\% between SFR(far-infrared) and the instantaneous SFR. It is not clear what causes this discrepancy, and we leave a more detailed comparison to a future study.

\begin{figure*}
	\centering
\includegraphics[width=1\textwidth]{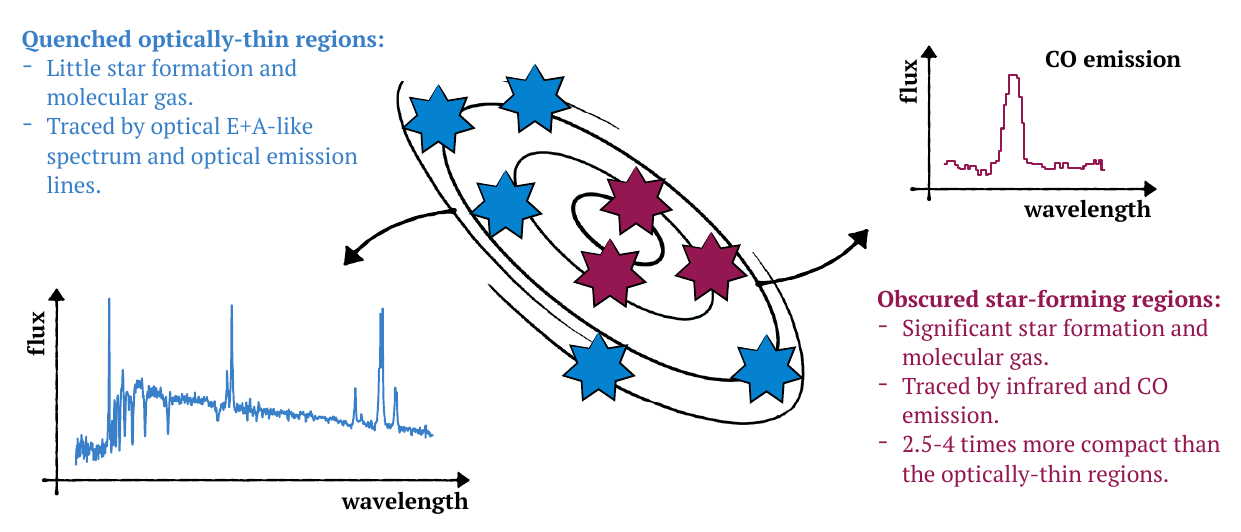}
\caption{\textbf{Emerging picture of optically-selected post-starburst candidates.} 
Such galaxies consist of two types of regions. Some regions have experienced a recent quenching of star formation, they are optically-thin, and emit an E+A-like optical spectrum. Their H$\alpha$ emission suggests little ongoing star formation. Other regions host significant obscured star formation, and emit primarily at mm and far-infrared wavelengths. Their far-infrared SFRs place them on or above the star-forming main sequence, and their CO emission is consistent with this SFR.
 }\label{f:system_geometry}
\end{figure*}

We corrected the far-infrared SFRs for all the post-starbursts we consider using the best-fitting line from Figure \ref{f:C_versus_delta_MS_FIR}, and verified that our conclusions from sections \ref{s:results:obscured_SF} and \ref{s:results:molecular_gas} do not change (see appendix \ref{a:SFR_FIR_checks}). In particular, if the corrected far-infrared SFRs are used instead (Figure \ref{f:SFR_versus_m}), sources that were above, on, and below the star-forming main sequence remain above, on, and below it respectively. The $\Delta$(MS)-post-burst age relation (Figure \ref{f:Delta_MS_versus_pb_age}) becomes steeper and with a larger scatter, but the Kendall's correlation coefficient and p-value remain the same since the rank-ordering of the sources does not change. The SFR-$M_{\mathrm{H_2}}$ relation of post-starburst candidates (Figure \ref{f:SFR_versus_MH2}) shows a negligibly steeper slope and a smaller scatter, bringing it even closer to the SFR-$M_{\mathrm{H_2}}$ relation of other sources. Finally, in the KS plane (Figure \ref{f:KS_relation}), the corrected far-infrared SFRs for our NOEMA sample are shifted by $\sim$0.04 dex downwards, leaving them on the KS relation of other galaxies.

\underline{To summarize:} our simple model of a single exponentially-decaying star formation rate qualitatively reproduces the main results presented by \citet{hayward14}. We find that using a far-infrared to SFR conversion factor that assumes that SFR has been constant for the past 100 Myrs may overestimate the instantaneous SFR in galaxies that have undergone a recent dramatic change in their SFR. For systems above the star-forming main sequence, the difference between SFR(far-infrared) and the instantaneous SFR is small, and cannot account for the 1--2 dex difference we found for some of the post-starburst candidates we consider. This, and our analysis in Section \ref{s:results:obscured_SF}, suggest that these sources host obscured starbursts. For systems on or below the star-forming main sequence, a constant conversion factor may overestimate the instantaneous SFR by a large factor. We verified that our conclusions do not change if we correct SFR(far-infrared) according to the best-fitting line from Figure \ref{f:C_versus_delta_MS_FIR}.

\subsection{Emerging picture of post-starbursts}\label{s:implications}

The combination of optical, infrared, and mm observations allows us to investigate the picture of post-starburst galaxy candidates, shown in Figure \ref{f:system_geometry}. In an optically-selected post-starburst galaxy, some of the regions have experienced a recent and abrupt quenching of their star formation. As a result, these regions became optically-thin, and their observed optical spectrum is dominated by A-type stars. The optical emission lines trace the warm ionized gas within these quenched regions, with H$\alpha$ emission that suggests little ongoing star formation. These regions are dust and molecular gas poor, resulting in low SFR. Other regions within the galaxy host significant star formation which is obscured in optical wavelengths. The dust in these regions emits in infrared wavelengths, and their far-infrared SFRs place them on or above the star-forming main sequence. These regions have significant molecular gas reservoirs, and their star formation efficiencies are similar to those observed in star-forming and starburst systems. With the current data, it is unclear whether the star formation in these regions is at its peak, or already began declining.

There is a weak correlation between the far-infrared SFR and the optically-derived post-burst age (figure \ref{f:Delta_MS_versus_pb_age}). Galaxies with a post-burst age of 200 Myrs in the quenched optically-thin regions are found between 1 dex above and 1 dex below the star-forming main sequence when considering their far-infrared SFR. In addition, many of the sources pass the PCA criteria by \citet{wild10} for quenched post-starburst galaxies\footnote{\citealt{wild10} performed PCA decomposition of SDSS spectra in the wavelength range 3175--4150 \AA\, and selected post-starburst candidates according to both H$\delta$ and Dn4000\AA.}, despite far-infrared observations placing them above the star-forming main sequence. All these suggest that optical information alone paint a partial picture of these complex transitioning sources. Since far-infrared bolometric luminosities can overestimate their instantaneous SFRs (section \ref{s:FIR_to_SFR}), a multi-wavelength approach is required to unveil their true nature.

In \citet{baron22} we studied the SFRs in post-starburst candidates, and found that their far-infrared properties strongly depend on their selection criteria. In particular, we found that post-starburst candidates selected with the traditional E+A selection criteria (weak emission lines; e.g., H$\alpha$ EW $<$ 3 \AA; \citealt{french18}) are the least probable to host obscured star formation, where only 4.5\% were detected by {\it IRAS} far-infrared observations. As explained in Section \ref{s:other_samples}, the subset of E+A galaxies with molecular gas measurements \citep{french15} no longer represents the general E+A population, as it shows IRAS far-infrared detection fraction of 40\%. Moreover, in Section \ref{s:results:obscured_SF} we found that 26\% of these sources host significant obscured star formation. Combining the results from \citet{baron22} and this work, we conclude that H$\delta$-strong, mid-infrared-faint galaxies with weak emission lines are the best candidates for truly-quenched post-starburst galaxies.

\section{Summary and Conclusions}\label{s:conclusions} 

We used NOEMA to observe the CO(1-0) line in 15 galaxies selected from our parent sample of post-starburst candidates with AGN and ionized outflows \citep{baron22}. We also collected all previous samples of post-starburst candidates with available CO observations. Combining the CO with archival far-infrared observations, we studied the star formation and molecular gas in post-starburst galaxy candidates with different emission line properties. Our main results are as follows.

\begin{enumerate}

\item The derived far-infrared SFRs are significantly larger than the optical, Dn4000\AA-based, SFRs for most of the post-starburst candidates we consider, in particular those selected with strong emission lines. While the optical SFRs place many of the systems on or below the star-forming main sequence, far-infrared observations place them above the main sequence, with some showing SFRs which are comparable to those of (ultra)luminous infrared galaxies. For post-starburst candidates with strong emission lines, we rule out the possibility that the far-infrared emission is driven by the post-burst population in most of the sources (e.g., A-type stars), and instead argue that it is powered by significant obscured star formation (consistent with \citealt{baron22}). For sources selected with the traditional E+A criteria, the observed far-infrared emission can be accounted for by the aging stellar population in $\sim$74\% of the cases.

\item In \citet{baron22} we found that samples selected according to the traditional E+A criteria (strong H$\delta$ absorption and weak H$\alpha$ emission) show a small contamination by obscured starbursts ($\sim$4.5\%). Here we concentrated on the subset of E+A galaxies with molecular gas measurements (\citealt{french15}), and found that 26\% (7 out of 26) host significant obscured star formation. We therefore conclude that this subset does not represent the star-formation and molecular gas properties of the general E+A population.

\item Using stellar population synthesis modeling, we derive the age of the recently-quenched starburst seen in optical, the "post-burst age" (\citealt{french18}). The relation between the far-infrared SFR and the post-burst age shows a weak downward trend ($\tau=-0.22$, p-value=0.002) with significant scatter, with, for example, systems with post-burst ages of 200 Myrs showing far-infrared SFRs 1 dex above and 1 dex below the main sequence.
	
\item Using far-infrared SFRs, post-starburst galaxy candidates show similar SFR-$M_{\mathrm{H_2}}$ relation to that observed in star-forming and starburst galaxies. In particular, systems reported to have exceptionally large molecular gas reservoirs host in fact significant obscured star formation. We show that the recently-discovered trends between star formation efficiency and the optical post-burst age (\citealt{french18, li19}) are the result of the tight SFR(far-infrared)-$M_{\mathrm{H_2}}$ correlation and the usage of the inaccurate optical SFRs. If far-infrared SFRs are used instead, the star formation efficiency observed in post-starburst candidates does not evolve significantly with the post-burst age, and is consistent with the efficiency observed in star-forming and starburst galaxies. Our results show no contradiction with the common galaxy evolution picture, where the decrease in SFR is due to the consumption of molecular gas by the starburst.

\item Using far-infrared luminosity to trace star formation, we find that post-starburst candidates are not offset from the KS relation. For previous studies that used optical SFRs, we show that the offset disappears if far-infrared information is used to account for the presence of obscured star formation. For previous studies that used infrared emission lines, we were unable to reproduce their reported factor $\sim$10 offset, and instead find a factor $\sim 2$ offset for 4 out of 6 sources. A more careful analysis, that includes disentangling line emission due to AGN photoionization versus star formation, of a larger sample of post-starburst galaxies is required to reach more robust conclusions.

\item Using a simple model of an exponentially-declining star formation rate, we examined the relation between SFR(far-infrared) and the instantaneous SFR in rapidly transitioning galaxies. According to the model, for systems above the star-forming main sequence, L(far-infrared) overestimates the instantaneous SFR by 10--20\%. For systems on the main sequence, L(far-infrared) overestimates the instantaneous SFR by a factor of $\sim$2. For systems below the main sequence, L(far-infrared) can overestimate the SFR by up to a factor of 10.
\end{enumerate}

The combination of optical, infrared, and mm observations suggests that optically-selected post-starburst galaxies are not necessarily post-starburst, in particular, when strong limits against ionized line emission are not employed. While some regions within these galaxies experienced a recent quenching of their star formation, other regions contain active star forming regions. These regions also contain significant amounts of molecular gas. All the above is in contradiction with the traditional classification of these sources as a single class of objects that are rapidly transitioning to quiescence. Spatially-resolved optical and mm observations are required to better constrain the geometry and physical properties of these systems.

\section{Data availability}\label{s:data_avail}

The properties of the 15 sources observed with NOEMA are given in Tables \ref{tab:CO_and_mm_properties} and \ref{tab:integ_properties}. The best-fitting SFH parameters for our sources are given in Table \ref{table:ages}. The best-fitting SFH parameters for the \citet{rowlands15} and \citet{alatalo16b} samples are given in \citet{french18}, while the parameters for the \citet{yesuf17} sample are given in Table \ref{table:ages_yesuf}. The optical and far-infrared data were extracted from public catalogues by the SDSS and {\it IRAS}.

\section*{Acknowledgments}
We thank our local NOEMA contact for this project, Jan Martin Winters, for his help in calibrating the observations and producing the uv tables.
We thank our anonymous referee for a constructive report. We thank I. Smail for useful comments and suggestions about this manuscript.
We appreciate the discussion with V. Wild about topics related to this publication.
D. Baron is supported by the Adams Fellowship Program of the Israel Academy of Sciences and Humanities.
This research made use of {\sc Astropy}\footnote{http://www.astropy.org}, a community-developed core Python package for Astronomy \citep{astropy2013, astropy2018}.

This work made use of SDSS-III\footnote{www.sdss3.org} data. Funding for SDSS-III has been provided by the Alfred P. Sloan Foundation, the Participating Institutions, the National Science Foundation, and the U.S. Department of Energy Office of Science. SDSS-III is managed by the Astrophysical Research Consortium for the Participating Institutions of the SDSS-III Collaboration including the University of Arizona, the Brazilian Participation Group, Brookhaven National Laboratory, Carnegie Mellon University, University of Florida, the French Participation Group, the German Participation Group, Harvard University, the Instituto de Astrofisica de Canarias, the Michigan State/Notre Dame/JINA Participation Group, Johns Hopkins University, Lawrence Berkeley National Laboratory, Max Planck Institute for Astrophysics, Max Planck Institute for Extraterrestrial Physics, New Mexico State University, New York University, Ohio State University, Pennsylvania State University, University of Portsmouth, Princeton University, the Spanish Participation Group, University of Tokyo, University of Utah, Vanderbilt University, University of Virginia, University of Washington, and Yale University. 

\bibliographystyle{mn2e}
\bibliography{ref_noema_sample}

\clearpage

\onecolumn

\appendix

\section{NOEMA and {\it IRAS} data}\label{a:CO_and_mm_data}	

\begin{figure*}
	\centering
\includegraphics[width=0.95\textwidth]{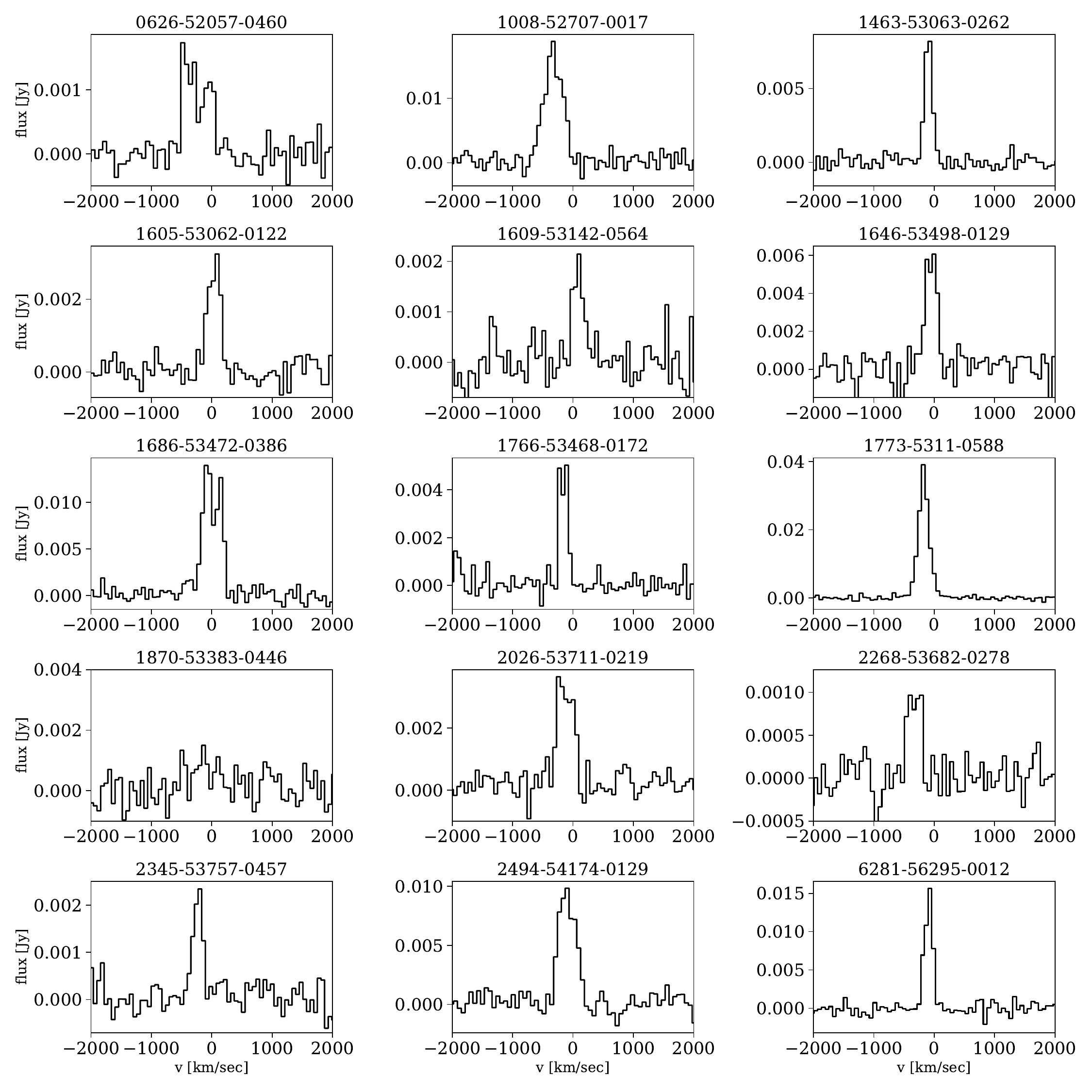}
\caption{Integrated CO(1-0) spectra obtained from our NOEMA observations. The SDSS IDs (PLATE-MJD-FIBER) are indicated at the top of each plot. 
 }\label{f:CO_spectra}
\end{figure*}

\begin{figure*}
	\centering
\includegraphics[width=0.95\textwidth]{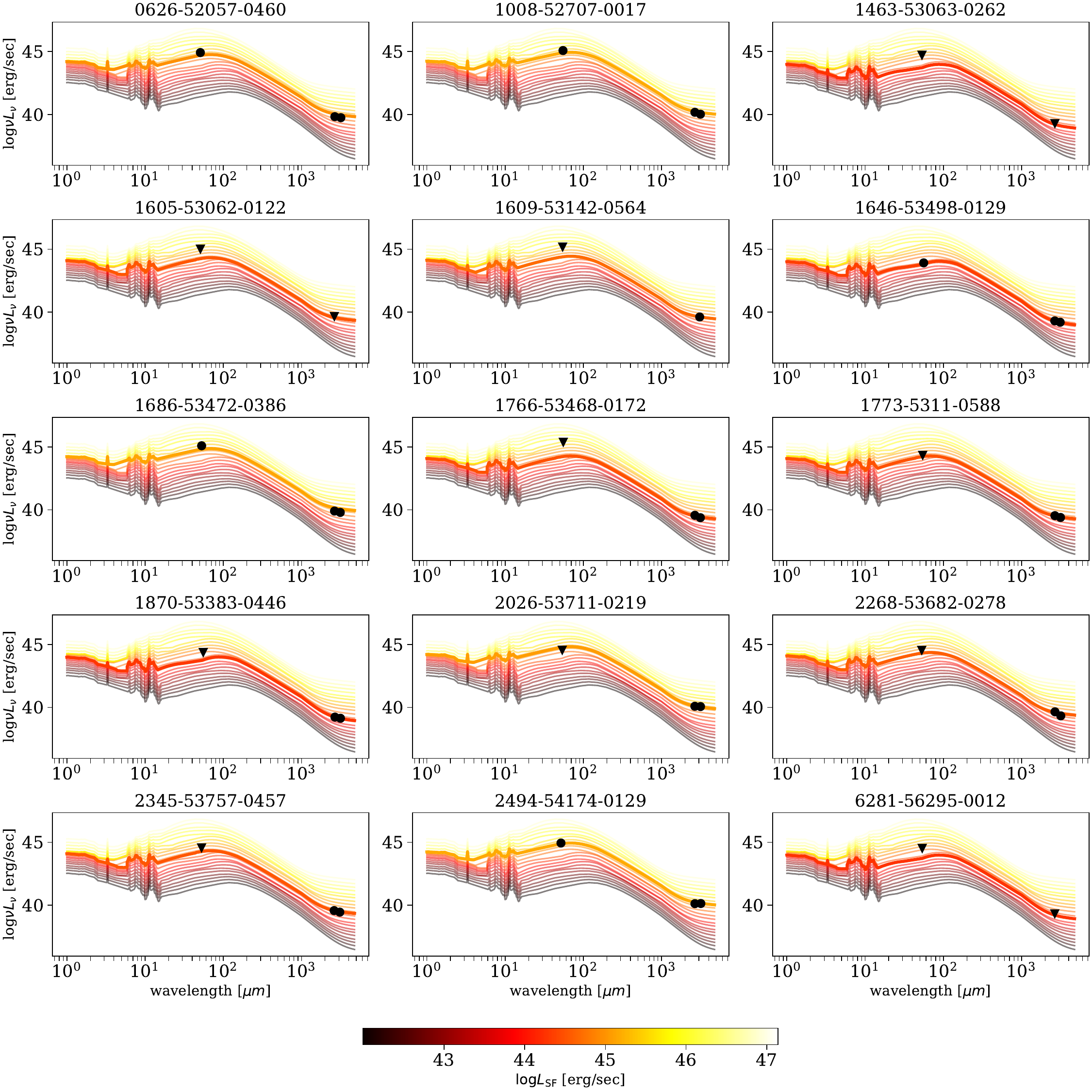}
\caption{Star formation template fits of the far-infrared and mm continuum emission. The SDSS IDs (PLATE-MJD-FIBER) are indicated at the top. The black circles represent measurements and triangles represent upper limits. The solid lines represent the \citet{chary01} star formation templates, color-coded by the star formation luminosity. The best-fitting star formation template is marked with a thicker line.
 }\label{f:FIR_and_mm_fit}
\end{figure*}

Figure \ref{f:CO_spectra} shows the integrated CO(1-0) spectra obtained with NOEMA. Figure \ref{f:FIR_and_mm_fit} shows the best-fitting \citet{chary01} star formation templates for each of the galaxies in our sample. 

\clearpage

\begin{landscape}
\begin{table*}
\caption{Age-dating best-fitting parameters: this work.}\label{table:ages}
\scriptsize
\centering
\begin{tabular}{ccccccccccccccccc}
\hline
\hline
Object ID &  SFH$^{a}$ & \multicolumn{3}{c}{Post-burst Age (Myr)} & \multicolumn{3}{c}{Age since Burst Start(Myr)} & \multicolumn{3}{c}{$\tau$ or $\Delta t$ $^b$ (Myr)} &  \multicolumn{2}{c}{Burst Light Fraction $y_f$ $^c$} & \multicolumn{3}{c}{Burst Mass Fraction $m_{\rm burst}$ $^{d}$} & $A_V$ $^e$ \\
 &  & (16\%) & (50\%) & (84\%) & (16\%) & (50\%) & (84\%) & (16\%) & (50\%) & (84\%) & (burst 1) & (burst 2) &(16\%) & (50\%) & (84\%) & (mag)\\
\hline
0626-52057-0460 & 2   &  90 & 118 & 142 & 230 & 258 & 282 & 100 & 100 & 100 & 0.29 & 0.41 & 0.06 & 0.08 & 0.10 & 0.8 \\
1008-52707-0017 & 2   &  79 & 118 & 147 & 219 & 258 & 287 & 100 & 100 & 100 & 0.28 & 0.41 & 0.06 & 0.08 & 0.11 & 1.2 \\
1463-53063-0262$^{*}$ &     &     &     &     &     &     &     &     &     &      &      &      &      &      &     \\
1605-53062-0122 & 1   & -88 & -18 &   7 & 141 & 211 & 237 &  25 & 100 & 100 & 0.79 & --  & 0.02 & 0.02 & 0.03 & 1.2 \\
1609-53142-0564 & 2   &  69 &  93 & 113 & 209 & 233 & 253 & 100 & 100 & 100 & 0.28 & 0.45 & 0.06 & 0.08 & 0.10 & 0.8 \\
1646-53498-0129 & 2   & 279 & 335 & 388 & 619 & 675 & 728 & 100 & 300 & 300 & 0.29 & 0.43 & 0.09 & 0.13 & 0.17 & 1.0 \\
1686-53472-0386 & 2   &  57 &  78 & 106 & 197 & 218 & 246 & 100 & 100 & 100 & 0.27 & 0.45 & 0.04 & 0.06 & 0.08 & 1.2 \\
1766-53468-0172 & 2   &  20 &  30 &  40 & 160 & 170 & 180 & 100 & 100 & 100 & 0.18 & 0.39 & 0.02 & 0.02 & 0.02 & 0.8 \\
1773-53112-0588$^{*}$ &   &    &    &    &    &    &    &    &    &    &   &      &   &   &   &   \\
1870-53383-0446 & 2   & 104 & 127 & 167 & 244 & 267 & 307 & 100 & 100 & 100 & 0.21 & 0.30 & 0.03 & 0.04 & 0.05 & 0.8 \\
2026-53711-0219 & 2   & 121 & 148 & 174 & 261 & 288 & 314 & 100 & 100 & 100 & 0.25 & 0.34 & 0.05 & 0.06 & 0.08 & 0.8 \\
2268-53682-0278 & 2   &  63 &  85 & 115 & 203 & 225 & 255 & 100 & 100 & 100 & 0.20 & 0.33 & 0.02 & 0.03 & 0.04 & 0.8 \\
2345-53757-0457 & 2   &   8 &  85 & 122 & 148 & 225 & 262 & 100 & 100 & 100 & 0.26 & 0.42 & 0.04 & 0.06 & 0.09 & 0.0 \\
2494-54174-0129 & 2   &  68 &  85 & 103 & 208 & 225 & 243 & 100 & 100 & 100 & 0.30 & 0.49 & 0.08 & 0.10 & 0.12 & 1.0 \\
6281-56295-0012$^{*}$ &   &     &     &     &     &     &     &     &     &     &      &      &      &      &      &     \\
\hline 
\vspace{-0.2in}
\tablenotetext{} {For additional details about the best-fitting parameters, see description in \citet{french18}. $^{*}$: Due to unphysical SFHs, we omit these three sources from the analysis (see Section \ref{s:data:optical-sdss} for details). $^{a}$: Number of recent bursts. $^{b}$:If SFH = 1, burst duration ($\tau$). If SFH = 2, separation between bursts ($\Delta t$). $^{c}$: If SFH = 2, light fraction for each burst is shown (burst mass fractions are the same for each burst, so the light fractions will be different). $^{d}$: If SFH = 2 recent bursts, burst mass fraction shown is combined from both recent bursts. $^{e}$: Includes Galactic foreground extinction.}

\end{tabular}
\end{table*}


\begin{table*}
\caption{Age-dating best-fitting parameters: \citet{yesuf17}.}\label{table:ages_yesuf}
\scriptsize
\centering
\begin{tabular}{cccccccccccccccccc}
\hline
\hline
RA (deg) & DEC (deg) &  SFH$^{a}$ & \multicolumn{3}{c}{Post-burst Age (Myr)} & \multicolumn{3}{c}{Age since Burst Start(Myr)} & \multicolumn{3}{c}{$\tau$ or $\Delta t$ $^b$ (Myr)} &  \multicolumn{2}{c}{Burst Light Fraction $y_f$ $^c$} & \multicolumn{3}{c}{Burst Mass Fraction $m_{\rm burst}$ $^{d}$} & $A_V$ $^e$ \\
 & &  & (16\%) & (50\%) & (84\%) & (16\%) & (50\%) & (84\%) & (16\%) & (50\%) & (84\%) & (burst 1) & (burst 2) &(16\%) & (50\%) & (84\%) & (mag)\\
\hline
126.015310 & 51.904355 & 2 &   83 &  109 &  186 &  223 &  249 &  326 &  100 &  100 &  100 & 0.15 & 0.23 & 0.02 & 0.03 & 0.03 & 1.2  \\
134.619160 &  0.023468 & 2 &  322 &  381 &  440 & 1362 & 1421 & 1480 & 1000 & 1000 & 1000 & 0.28 & 0.69 & 0.39 & 0.50 & 0.68 & 0.8  \\ 
139.499370 & 50.002183 & 2 &  185 &  276 &  390 & 1225 & 1316 & 1430 & 1000 & 1000 & 1000 & 0.14 & 0.40 & 0.03 & 0.05 & 0.08 & 0.6  \\
137.874860 & 45.468319 & 2 &  173 &  211 &  340 &  713 &  751 &  880 &  100 &  500 & 1000 & 0.11 & 0.28 & 0.02 & 0.03 & 0.04 & 1.2  \\
173.412840 & 52.674611$^{*}$ &     &     &     &     &     &     &     &     &     &      &      &      &      &      &     \\
173.167690 & 52.950379 & 1 &  273 &  318 &  363 &  331 &  375 &  420 &   25 &   25 &   25 & 0.32 & --  & 0.02 & 0.02 & 0.03 & 0.6  \\
179.028470 & 59.424927 & 1 & 1460 & 1720 & 2045 & 1517 & 1778 & 2102 &   25 &   25 &  200 & 0.79 & --  & 0.19 & 0.39 & 0.59 & 0.4  \\
182.019420 & 55.407672 & 2 &  136 &  183 &  217 & 1176 & 1223 & 1257 & 1000 & 1000 & 1000 & 0.15 & 0.55 & 0.04 & 0.05 & 0.08 & 0.6  \\
236.933940 & 41.402300 & 2 &   96 &  118 &  140 &  236 &  258 &  280 &  100 &  100 &  100 & 0.14 & 0.20 & 0.02 & 0.02 & 0.02 & 1.0  \\
247.636060 & 39.384201 & 2 &  199 &  241 &  281 & 1239 & 1281 & 1321 & 1000 & 1000 & 1000 & 0.15 & 0.51 & 0.05 & 0.06 & 0.08 & 0.6  \\
240.658050 & 41.293433 & 2 &  366 &  433 &  499 & 1406 & 1473 & 1539 & 1000 & 1000 & 1000 & 0.17 & 0.43 & 0.06 & 0.08 & 0.10 & 0.4  \\
200.951880 & 43.301187 & 2 &  152 &  183 &  215 &  292 &  323 &  355 &  100 &  100 &  100 & 0.24 & 0.31 & 0.05 & 0.06 & 0.08 & 1.8  \\
178.622550 & 42.980193 & 2 &  163 &  197 &  233 & 1203 & 1237 & 1273 & 1000 & 1000 & 1000 & 0.08 & 0.29 & 0.02 & 0.02 & 0.02 & 0.2  \\
189.517340 & 48.345096$^{*}$ &     &     &     &     &     &     &     &     &     &      &      &      &      &      &     \\
190.450280 & 47.708848 & 2 &  494 &  709 & 1743 & 1534 & 1749 & 2783 & 1000 & 1000 & 1000 & 0.20 & 0.43 & 0.09 & 0.16 & 1.00 & 0.6  \\
198.749300 & 51.272582 & 2 & 1403 & 1638 & 1950 & 1543 & 1778 & 2090 &  100 &  100 &  300 & 0.45 & 0.48 & 0.61 & 0.80 & 1.00 & 0.6  \\
212.016680 &  7.327645 & 2 &   36 &   49 &   61 &  176 &  189 &  201 &  100 &  100 &  100 & 0.20 & 0.40 & 0.02 & 0.03 & 0.04 & 1.6  \\
117.966200 & 49.814318 & 2 &  112 &  137 &  162 &  252 &  277 &  302 &  100 &  100 &  100 & 0.13 & 0.19 & 0.02 & 0.02 & 0.02 & 1.0  \\
203.561730 & 34.194146 & 2 &  366 &  433 &  499 & 1406 & 1473 & 1539 & 1000 & 1000 & 1000 & 0.29 & 0.69 & 0.45 & 0.80 & 1.00 & 1.0  \\
180.519230 & 35.321681 & 2 &  450 &  555 &  639 &  990 & 1095 & 1179 &  500 &  500 &  500 & 0.24 & 0.45 & 0.09 & 0.16 & 0.40 & 1.0  \\
170.945910 & 35.442308 & 2 &  161 &  258 &  313 &  301 &  398 &  453 &  100 &  100 &  100 & 0.11 & 0.14 & 0.02 & 0.04 & 0.05 & 0.8  \\
222.657720 & 22.734337 & 2 &   -9 &   -4 &    0 &  130 &  135 &  140 &  100 &  100 &  100 & 0.18 & 0.52 & 0.02 & 0.02 & 0.02 & 1.8  \\
172.082990 & 27.622098$^{*}$ &     &     &     &     &     &     &     &     &     &      &      &      &      &      &     \\
145.185430 & 21.234273$^{*}$ &     &     &     &     &     &     &     &     &     &      &      &      &      &      &     \\
\hline 	
\vspace{-0.2in}
\tablenotetext{} {For additional details about the best-fitting parameters, see description in \citet{french18}. $^{*}$: Due to unphysical SFHs, we omit these three sources from the analysis (see Section \ref{s:data:optical-sdss} for details). $^{a}$: Number of recent bursts. $^{b}$:If SFH = 1, burst duration ($\tau$). If SFH = 2, separation between bursts ($\Delta t$). $^{c}$: If SFH = 2, light fraction for each burst is shown (burst mass fractions are the same for each burst, so the light fractions will be different). $^{d}$: If SFH = 2 recent bursts, burst mass fraction shown is combined from both recent bursts. $^{e}$: Includes Galactic foreground extinction.}
\end{tabular}
\end{table*}

\end{landscape}

\section{Star formation and molecular gas properties of post-starburst candidates}\label{a:SF_and_mol_PSB_sample}

Tables \ref{table:french_meas}, \ref{table:rowlands_meas}, \ref{table:alatalo_meas}, and \ref{table:yesuf_meas} summarize the SFRs and molecular gas masses we used in this work, which are based on the studies: \citet{french15}, \citet{rowlands15}, \citet{alatalo16b}, and \citet{yesuf17}.

\begin{table*}
\caption{Star formation and molecular gas measurements of the \citet{french15} sample.}\label{table:french_meas}
\scriptsize
\centering
\begin{tabular}{cccccccc}
\hline
\hline
Object name & $\log \mathrm{M_{*}}$ [$\mathrm{M_{\odot}}$] & $\log \mathrm{SFR(optical)}$ [$\mathrm{M_{\odot}/yr}$] & $\log \mathrm{SFR(FIR)}$ [$\mathrm{M_{\odot}/yr}$] & $f(\mathrm{FIR})$ & $\log \mathrm{M_{H2}}$ [$\mathrm{M_{\odot}/yr}$] & $f(\mathrm{M_{H2}})$\\
(1) & (2) & (3) & (4) & (5) & (6) & (7) \\
\hline
EAH01 & 10.45 & -1.01 & 0.81 & 1 & $9.74 \pm 0.06$ & 1 \\
EAH02 & 9.96 & -1.02 & 0.48 & 1 & $9.56 \pm 0.09$ & 1 \\
EAH03 & 10.34 & -1.46 & 0.83 & 1 & $9.83 \pm 0.05$ & 1 \\
EAH04 & 10.18 & -0.60 & -0.15 & 1 & $8.60 \pm 0.09$ & 1 \\
EAH05 & 10.00 & -1.26 & 1.05 & 1 & $9.59 \pm 0.09$ & 1 \\
EAH06 & 10.53 & -1.19 & -0.07 & 1 & $9.03 \pm 0.10$ & 0 \\
EAH07 & 10.65 & -0.54 & -0.32 & 1 & $8.65 \pm 0.10$ & 0 \\
EAH08 & 10.41 & -1.11 & 1.15 & 1 & $9.19 \pm 0.15$ & 1 \\
EAH09 & 10.21 & -1.38 & -0.35 & 1 & $8.53 \pm 0.13$ & 1 \\
EAH10 & 10.24 & -1.03 & 0.86 & 1 & $9.89 \pm 0.12$ & 1 \\
EAH11 & 10.61 & -1.29 & -0.07 & 1 & $9.22 \pm 0.10$ & 0 \\
EAH12 & 10.55 & -1.47 & 0.20 & 1 & $9.45 \pm 0.10$ & 0 \\
EAH13 & 11.00 & 0.16 & 0.69 & 1 & $9.92 \pm 0.08$ & 1 \\
EAH14 & 10.04 & -1.24 & -0.34 & 1 & $9.29 \pm 0.10$ & 0 \\
EAH15 & 10.40 & -0.84 & -0.62 & 1 & $9.07 \pm 0.10$ & 0 \\
EAH16 & 10.74 & 0.92 & -0.05 & 1 & $9.83 \pm 0.10$ & 0 \\
EAH17 & 10.05 & -0.97 & -0.52 & 1 & $9.04 \pm 0.10$ & 0 \\
EAS01 & 10.24 & -0.89 & -1.12 & 1 & $8.43 \pm 0.10$ & 0 \\
EAS02 & 10.08 & -0.89 & -0.45 & 1 & $8.74 \pm 0.13$ & 1 \\
EAS03 & 10.86 & -1.24 & 0.10 & 1 & $9.79 \pm 0.06$ & 1 \\
EAS04 & 9.99 & -0.39 & -1.07 & 1 & $7.77 \pm 0.10$ & 0 \\
EAS05 & 10.57 & -0.75 & -0.33 & 1 & $9.12 \pm 0.14$ & 1 \\
EAS06 & 10.14 & -2.14 & 0.09 & 1 & $9.26 \pm 0.04$ & 1 \\
EAS07 & 10.54 & -0.43 & -1.26 & 1 & $8.67 \pm 0.10$ & 0 \\
EAS08 & 10.67 & 0.03 & -0.96 & 1 & $8.64 \pm 0.10$ & 0 \\
EAS09 & 10.56 & -1.29 & -0.20 & 1 & $9.16 \pm 0.06$ & 1 \\
EAS10 & 10.53 & -0.99 & -0.71 & 1 & $8.82 \pm 0.10$ & 0 \\
EAS11 & 10.74 & -0.41 & -0.16 & 1 & $8.87 \pm 0.10$ & 0 \\
EAS12 & 10.01 & -0.45 & -0.64 & 1 & $8.57 \pm 0.14$ & 1 \\
EAS13 & 10.95 & -1.13 & -0.31 & 1 & $9.15 \pm 0.10$ & 0 \\
EAS14 & 11.31 & -0.77 & 0.21 & 1 & $9.73 \pm 0.09$ & 1 \\
EAS15 & 10.83 & 0.37 & -0.29 & 1 & $9.11 \pm 0.14$ & 1 \\
\hline 
\vspace{-0.2in}
\tablenotetext{} {Columns: (1) The object name in the original study. (2) Stellar mass from the SDSS. (3) MPA/JHU optical SFR, extracted from SDSS. (4) IRAS far-infrared SFR. (5) A flag indicating whether the source was detected in far-infrared (1) or only upper limit is available (0). (6) Molecular gas mass and uncertainty. (7) A flag indicating whether the molecular gas mass is a measurement (1) or upper limit (0).}
\end{tabular}
\end{table*}

\begin{table*}
\caption{Star formation and molecular gas measurements of the \citet{rowlands15} sample.}\label{table:rowlands_meas}
\scriptsize
\centering
\begin{tabular}{cccccccc}
\hline
\hline
Object name & $\log \mathrm{M_{*}}$ [$\mathrm{M_{\odot}}$] & $\log \mathrm{SFR(optical)}$ [$\mathrm{M_{\odot}/yr}$] & $\log \mathrm{SFR(FIR)}$ [$\mathrm{M_{\odot}/yr}$] & $f(\mathrm{FIR})$ & $\log \mathrm{M_{H2}}$ [$\mathrm{M_{\odot}/yr}$] & $f(\mathrm{M_{H2}})$\\
(1) & (2) & (3) & (4) & (5) & (6) & (7) \\
\hline
PSB1 & 9.33 & 1.05 & 0.93 & 1 & $8.48 \pm 0.10$ & 0 \\
PSB2 & 10.48 & 1.10 & 1.32 & 1 & $9.56 \pm 0.04$ & 1 \\
PSB3 & 9.50 & 1.13 & 0.63 & 1 & $8.48 \pm 0.14$ & 1 \\
PSB4 & 9.81 & 0.76 & 0.81 & 1 & $9.32 \pm 0.04$ & 1 \\
PSB5 & 9.46 & 0.07 & 0.78 & 1 & $9.41 \pm 0.02$ & 1 \\
PSB6 & 10.47 & 0.46 & 0.55 & 1 & $9.61 \pm 0.05$ & 1 \\
PSB7 & 10.08 & 0.37 & 0.19 & 0 & $9.48 \pm 0.07$ & 1 \\
PSB8 & 9.89 & -0.93 & 0.18 & 1 & $8.90 \pm 0.11$ & 1 \\
PSB9 & 10.11 & 0.15 & 0.27 & 1 & $9.49 \pm 0.04$ & 1 \\
PSB10 & 10.04 & 0.14 & 0.56 & 1 & $9.28 \pm 0.05$ & 1 \\
PSB11 & 10.49 & -0.70 & 0.31 & 0 & $8.85 \pm 0.06$ & 1 \\
\hline 
\vspace{-0.2in}
\tablenotetext{} {Columns: (1) The object name in the original study. (2) Stellar mass from the SDSS. (3) MPA/JHU optical SFR, extracted from SDSS. (4) IRAS far-infrared SFR. (5) A flag indicating whether the source was detected in far-infrared (1) or only upper limit is available (0). (6) Molecular gas mass and uncertainty. (7) A flag indicating whether the molecular gas mass is a measurement (1) or upper limit (0).}
\end{tabular}
\end{table*}

\begin{table*}
\caption{Star formation and molecular gas measurements of the \citet{alatalo16b} sample.}\label{table:alatalo_meas}
\scriptsize
\centering
\begin{tabular}{cccccccc}
\hline
\hline
Object name & $\log \mathrm{M_{*}}$ [$\mathrm{M_{\odot}}$] & $\log \mathrm{SFR(optical)}$ [$\mathrm{M_{\odot}/yr}$] & $\log \mathrm{SFR(FIR)}$ [$\mathrm{M_{\odot}/yr}$] & $f(\mathrm{FIR})$ & $\log \mathrm{M_{H2}}$ [$\mathrm{M_{\odot}/yr}$] & $f(\mathrm{M_{H2}})$\\
(1) & (2) & (3) & (4) & (5) & (6) & (7) \\
\hline
J0003+0048 & 10.72 & 0.44 & 1.67 & 1 & $10.22 \pm 0.08$ & 1 \\
J0011-0054 & 10.34 & -1.42 & 0.38 & 0 & $9.14 \pm 0.08$ & 1 \\
J0029+1433 & 10.69 & 0.40 & 1.37 & 0 & $10.52 \pm 0.00$ & 0 \\
J0037+0024 & 10.18 & -1.09 & 0.74 & 0 & $9.85 \pm 0.00$ & 0 \\
J0119+1334 & 10.95 & 1.15 & 1.93 & 1 & $10.48 \pm 0.09$ & 1 \\
J0803+2530 & 10.93 & 0.10 & 1.65 & 0 & $10.03 \pm 0.10$ & 1 \\
J0807+2006 & 10.39 & -0.49 & 0.95 & 1 & $9.52 \pm 0.08$ & 1 \\
J0816+1936 & 10.40 & -0.49 & 1.08 & 0 & $9.72 \pm 0.10$ & 1 \\
J0845+2006 & 10.61 & 0.18 & 1.17 & 1 & $10.17 \pm 0.12$ & 1 \\
J0853+0310 & 10.88 & 0.49 & 1.19 & 1 & $10.27 \pm 0.06$ & 1 \\
J0859+1006 & 10.54 & -0.32 & 1.15 & 1 & $9.39 \pm 0.07$ & 1 \\
J0914+3753 & 10.30 & 0.26 & 1.01 & 1 & $9.89 \pm 0.01$ & 1 \\
J0918+4200 & 10.30 & -1.51 & 0.36 & 1 & $8.75 \pm 0.05$ & 1 \\
J0925+0623 & 10.51 & 0.91 & 0.83 & 1 & $9.58 \pm 0.01$ & 1 \\
J0928+0741 & 10.11 & 0.91 & 0.95 & 1 & $9.74 \pm 0.08$ & 1 \\
J0938+1819 & 10.65 & -0.96 & 1.52 & 1 & $10.22 \pm 0.01$ & 1 \\
J0957-0012 & 10.00 & -0.88 & 0.35 & 1 & $8.90 \pm 0.01$ & 1 \\
J1008+0936 & 9.97 & -0.37 & -0.12 & 1 & $8.43 \pm 0.02$ & 1 \\
J1008+1916 & 10.96 & 0.67 & 1.41 & 0 & $10.35 \pm 0.10$ & 1 \\
J1008+5123 & 10.60 & 0.71 & 1.37 & 0 & $10.29 \pm 0.09$ & 1 \\
J1018+1536 & 10.78 & -0.88 & 1.77 & 1 & $10.01 \pm 0.05$ & 1 \\
J1026+4340 & 10.25 & 0.61 & 1.52 & 1 & $9.75 \pm 0.02$ & 1 \\
J1028+5736 & 10.19 & 0.01 & 0.64 & 0 & $9.89 \pm 0.09$ & 1 \\
J1031+0540 & 10.72 & 0.74 & 1.53 & 0 & $10.41 \pm 0.11$ & 1 \\
J1046+2804 & 10.42 & 0.54 & 1.85 & 0 & $9.93 \pm 0.00$ & 0 \\
J1057+0554 & 10.06 & 0.00 & - & - & $9.37 \pm 0.09$ & 1 \\
J1126+1913 & 10.48 & -0.62 & 1.81 & 1 & $10.40 \pm 0.01$ & 1 \\
J1127+1256 & 10.87 & -0.48 & 1.48 & 0 & $9.71 \pm 0.11$ & 1 \\
J1136+2453 & 10.12 & -0.88 & 0.73 & 1 & $9.25 \pm 0.02$ & 1 \\
J1139+4631 & 11.05 & 0.79 & 1.81 & 1 & $10.29 \pm 0.01$ & 1 \\
J1153+0930 & 10.75 & 0.04 & 1.94 & 1 & $10.33 \pm 0.01$ & 1 \\
J1211+2936 & 10.59 & -1.30 & 1.54 & 1 & $9.73 \pm 0.08$ & 1 \\
J1216+1904 & 10.90 & -0.75 & 0.71 & 0 & $9.74 \pm 0.05$ & 1 \\
J1229+3224 & 11.09 & 0.48 & 2.11 & 1 & $10.64 \pm 0.01$ & 1 \\
J1248+5514 & 10.75 & -1.00 & 0.89 & 1 & $9.93 \pm 0.04$ & 1 \\
J1313+0207 & 10.42 & -0.05 & -0.13 & 0 & $8.99 \pm 0.09$ & 1 \\
J1314+2106 & 10.47 & -1.02 & 0.59 & 1 & $9.62 \pm 0.01$ & 1 \\
J1315+2437 & 10.11 & -1.57 & 1.17 & 1 & $9.40 \pm 0.01$ & 1 \\
J1326+1922 & 10.89 & -0.26 & 1.89 & 1 & $10.17 \pm 0.02$ & 1 \\
J1336+3008 & 9.75 & -1.51 & 0.05 & 1 & $8.72 \pm 0.02$ & 1 \\
J1339+4422 & 10.47 & -1.33 & 0.75 & 1 & $9.98 \pm 0.01$ & 1 \\
J1356+2816 & 10.79 & 0.87 & 1.39 & 1 & $10.61 \pm 0.00$ & 1 \\
J1409+1016 & 10.87 & 0.06 & 1.51 & 1 & $10.06 \pm 0.05$ & 1 \\
J1505+5847 & 10.80 & 0.53 & 1.39 & 1 & $10.24 \pm 0.10$ & 1 \\
J1506+0806 & 10.24 & 0.18 & 0.24 & 1 & $9.13 \pm 0.06$ & 1 \\
J1529+0601 & 10.69 & -1.22 & 1.35 & 1 & $9.80 \pm 0.11$ & 1 \\
J1529+0913 & 10.49 & 0.48 & 1.19 & 0 & $9.92 \pm 0.10$ & 1 \\
J1555+2955 & 10.32 & -0.67 & 1.04 & 1 & $9.70 \pm 0.09$ & 1 \\
J1611+0840 & 10.59 & 0.55 & 1.32 & 0 & $10.70 \pm 0.00$ & 0 \\
J1645+3048 & 9.98 & -0.78 & 0.69 & 1 & $9.36 \pm 0.12$ & 1 \\
J2245+1232 & 10.70 & -0.57 & 0.92 & 0 & $9.63 \pm 0.00$ & 0 \\
J2326-0114 & 10.83 & 0.36 & 1.76 & 0 & $10.57 \pm 0.12$ & 1 \\
\hline 
\vspace{-0.2in}
\tablenotetext{} {Columns: (1) The object name in the original study. (2) Stellar mass from the SDSS. (3) MPA/JHU optical SFR, extracted from SDSS. (4) IRAS far-infrared SFR. (5) A flag indicating whether the source was detected in far-infrared (1) or only upper limit is available (0). (6) Molecular gas mass and uncertainty. (7) A flag indicating whether the molecular gas mass is a measurement (1) or upper limit (0).}
\end{tabular}
\end{table*}

\begin{table*}
\caption{Star formation and molecular gas measurements of the \citet{yesuf17} sample.}\label{table:yesuf_meas}
\scriptsize
\centering
\begin{tabular}{cccccccc}
\hline
\hline
Object name & $\log \mathrm{M_{*}}$ [$\mathrm{M_{\odot}}$] & $\log \mathrm{SFR(optical)}$ [$\mathrm{M_{\odot}/yr}$] & $\log \mathrm{SFR(FIR)}$ [$\mathrm{M_{\odot}/yr}$] & $f(\mathrm{FIR})$ & $\log \mathrm{M_{H2}}$ [$\mathrm{M_{\odot}/yr}$] & $f(\mathrm{M_{H2}})$\\
(1) & (2) & (3) & (4) & (5) & (6) & (7) \\
\hline
TPSB1 & 10.04 & -0.19 & -0.17 & 1 & $8.79 \pm 0.20$ & 1 \\
TPSB2 & 10.40 & -0.20 & 0.22 & 1 & $8.64 \pm 0.20$ & 1 \\
TPSB4 & 10.30 & -0.13 & 0.21 & 1 & $9.19 \pm 0.20$ & 0 \\
TPSB5 & 10.19 & -0.89 & 0.17 & 0 & $9.10 \pm 0.20$ & 0 \\
TPSB6 & 10.29 & -0.64 & 0.18 & 0 & $8.97 \pm 0.20$ & 0 \\
TPSB7 & 10.15 & -0.87 & -0.38 & 0 & $8.62 \pm 0.20$ & 1 \\
TPSB8 & 10.21 & -1.11 & -0.02 & 0 & $8.81 \pm 0.20$ & 0 \\
TPSB9 & 10.54 & -1.57 & 0.19 & 1 & $8.90 \pm 0.20$ & 0 \\
TPSB10 & 10.45 & -0.24 & -0.13 & 1 & $9.29 \pm 0.20$ & 1 \\
TPSB10b & 10.12 & -0.52 & -0.05 & 1 & $9.12 \pm 0.20$ & 0 \\
TPSB11 & 10.70 & 0.35 & -0.05 & 1 & $8.93 \pm 0.20$ & 1 \\
TPSB12 & 10.33 & -0.30 & -0.12 & 0 & $8.94 \pm 0.20$ & 0 \\
TPSB13 & 10.54 & -0.52 & 0.42 & 1 & $8.68 \pm 0.20$ & 0 \\
TPSB14 & 10.05 & -0.90 & -0.41 & 0 & $8.71 \pm 0.20$ & 0 \\
TPSB15 & 10.49 & -0.41 & 0.01 & 0 & $8.85 \pm 0.20$ & 0 \\
TPSB16 & 10.25 & -0.37 & 0.39 & 0 & $8.81 \pm 0.20$ & 0 \\
TPSB17 & 10.02 & -0.93 & -0.36 & 0 & $8.66 \pm 0.20$ & 0 \\
TPSB18 & 10.68 & -0.22 & 0.70 & 1 & $9.21 \pm 0.20$ & 1 \\
TPSB19 & 10.47 & -0.07 & 0.31 & 1 & $9.00 \pm 0.20$ & 0 \\
TPSB20 & 10.59 & -0.77 & -0.15 & 0 & $8.97 \pm 0.20$ & 0 \\
TPSB21 & 10.54 & -0.21 & 0.52 & 1 & $8.82 \pm 0.20$ & 0 \\
TPSB24 & 10.33 & -0.32 & -0.31 & 0 & $8.60 \pm 0.20$ & 0 \\
TPSB26 & 10.31 & -1.05 & - & - & $8.89 \pm 0.20$ & 0 \\
TPSB28 & 10.09 & -0.76 & -0.62 & 0 & $8.47 \pm 0.20$ & 0 \\
\hline 
\vspace{-0.2in}
\tablenotetext{} {Columns: (1) The object name in the original study. (2) Stellar mass from the SDSS. (3) MPA/JHU optical SFR, extracted from SDSS. (4) IRAS far-infrared SFR. (5) A flag indicating whether the source was detected in far-infrared (1) or only upper limit is available (0). (6) Molecular gas mass and uncertainty. (7) A flag indicating whether the molecular gas mass is a measurement (1) or upper limit (0).}
\end{tabular}
\end{table*}

\section{Star formation and molecular gas properties of comparison galaxies}\label{a:SF_and_mol_comp_sample}

\begin{figure}
\includegraphics[width=3.5in]{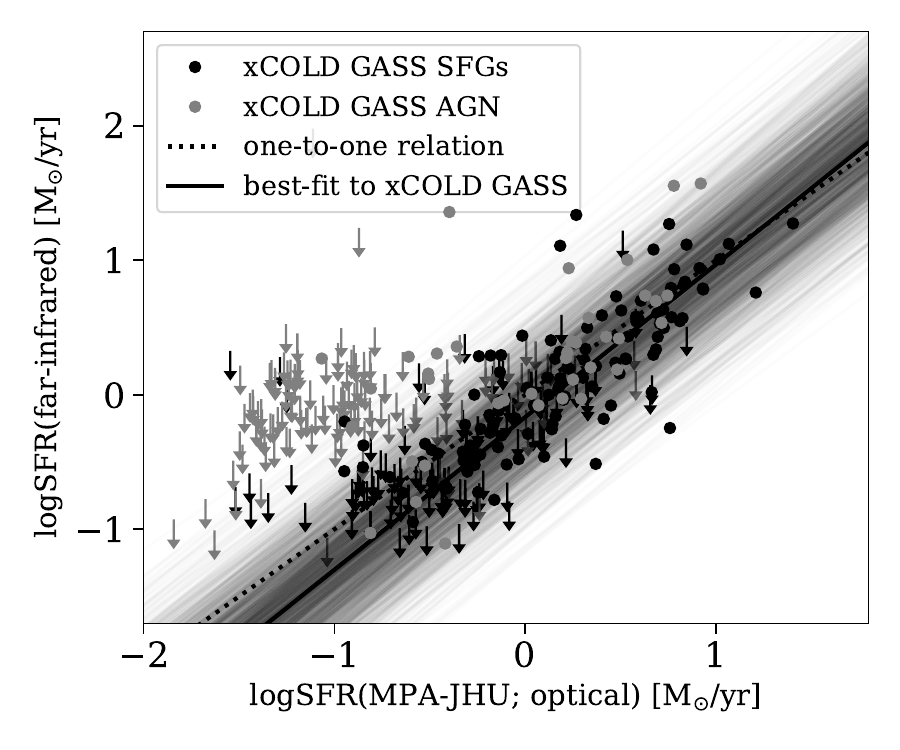}
\caption{\textbf{Comparison of far-infrared and optical SFRs for the xCOLD GASS sample.} The points represent measurements and the arrows upper limits. We include all emission line galaxies by xCOLD GASS, where black markers represent star-forming galaxies with emission line ratios consistent with star formation, and grey markers represent composites, LINERs, and Seyferts. The black solid line shows the best-fitting linear relation, including upper limits. The shaded lines represent the best-fitting scatter to the relation, as described in section \ref{s:data:stat}.} \label{f:SFR_FIR_versus_optical_xcold_gass}
\end{figure}

\begin{figure}
\includegraphics[width=3.5in]{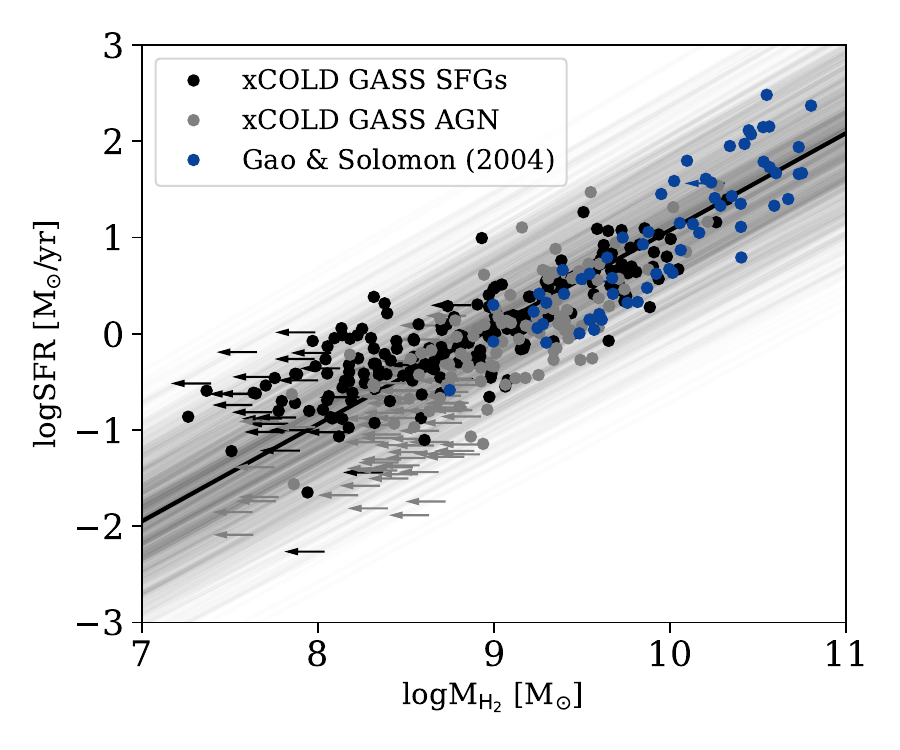}
\caption{\textbf{SFR versus molecular gas mass for the comparison galaxies.} The points represent measurements and the arrows upper limits. We include all emission line galaxies by xCOLD GASS, where black markers represent star-forming galaxies with emission line ratios consistent with star formation, and grey markers represent composites, LINERs, and Seyferts. The (ultra)luminous infrared galaxies by \citet{gao04} are marked with dark blue. The black solid line shows the best-fitting linear relation, including upper limits. The shaded lines represent the best-fitting scatter to the relation, as described in section \ref{s:data:stat}.} \label{f:SFR_versus_MH2_comparison_samples}
\end{figure}

Figure \ref{f:SFR_FIR_versus_optical_xcold_gass} shows the far-infrared versus Dn4000\AA-based SFR relation for our comparison sample, xCOLD GASS. The best-fitting relation includes the observed upper limits, and the shaded lines around the relation represent the full posterior distribution. Figure \ref{f:SFR_versus_MH2_comparison_samples} shows the SFR versus $M_{\mathrm{H_2}}$ relation for the comparison galaxies, which include the xCOLD GASS sample and the (ultra)luminous infrared galaxies by \citet{gao04}. 

\section{Re-analysis of the Li et al. (2019) sample}\label{a:li_et_al_reanalysis}

\citet{li19} combined the post-starburst galaxies from \citet{rowlands15}, \citet{alatalo16a}, and \citet{french18} for which \emph{Herschel} observations were publicly-available. They used the dust-corrected H$\alpha$ emission line to estimate the SFR in their sources. Some of the sources in their combined sample were observed in mm wavelengths, and their CO emission lines were used to estimate the molecular gas masses. For the rest, \citet{li19} used SED-based dust masses to estimate $M_{\mathrm{H_2}}$. We collected the H$\alpha$-based SFRs and $M_{\mathrm{H_2}}$ measurements of the sources in their sample. We used {\sc scanpi} to extract their 60 $\mathrm{\mu m}$ fluxes, and estimated their far-infrared SFRs. 

\begin{figure*}
	\centering
\includegraphics[width=1\textwidth]{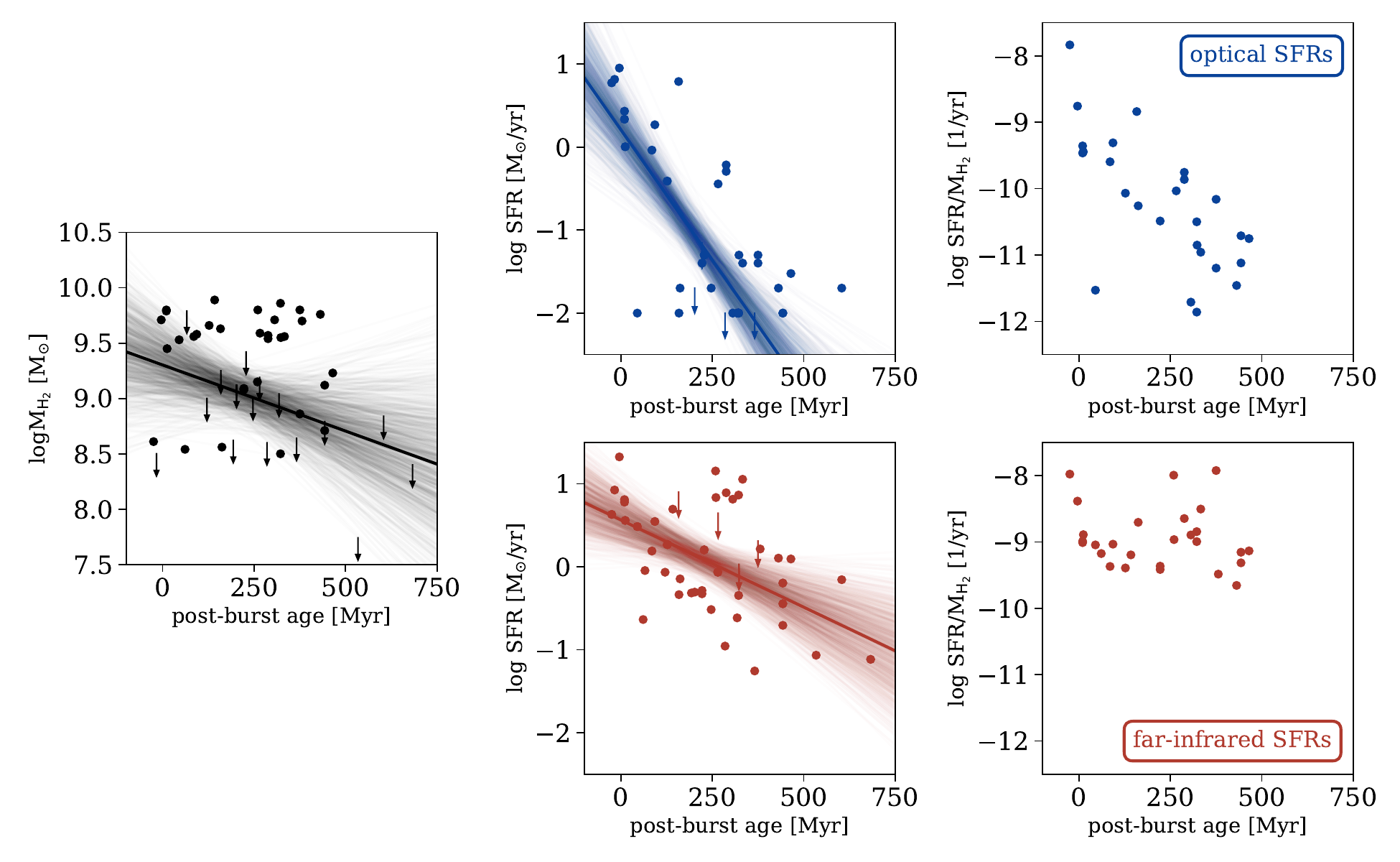}
\caption{\textbf{SFR and $M_{\mathrm{H_2}}$ versus post-burst age for the \citet{li19} sample.} 
First column: $M_{\mathrm{H_2}}$ versus the post-burst age, where measurements are marked with points and upper limits with arrows. The solid line represents the best-fitting relation obtained with {\sc linmix}, and the lighter bands represent the uncertainty of the best fit. 
Second column: optical (top) and far-infrared (bottom) SFR versus the post-burst age. 
Third column: optical (top) and far-infrared (bottom) star formation efficiency. Only galaxies with measured SFR and $M_{\mathrm{H_2}}$ are shown. 
 }\label{f:reanalysis_of_SFE_li_et_al}
\end{figure*}

Figure \ref{f:reanalysis_of_SFE_li_et_al} shows the SFR and $M_{\mathrm{H_2}}$ versus the post-burst age for the \citet{li19} sample, where the top row shows optical SFRs and the bottom row far-infrared SFRs. Using the optical SFRs, we reproduce the declining trend of the star formation efficiency as a function of the post-burst age reported by \citet{li19}. However, when using far-infrared SFRs, the resulting star formation efficiency is practically independent of the post-burst age. The Figure therefore suggests that the trend reported by \citet{li19} is driven by the inaccurate optical SFRs, and that there is no significant evolution of the star formation efficiency as the starburst ages.

\begin{figure*}
	\centering
\includegraphics[width=1\textwidth]{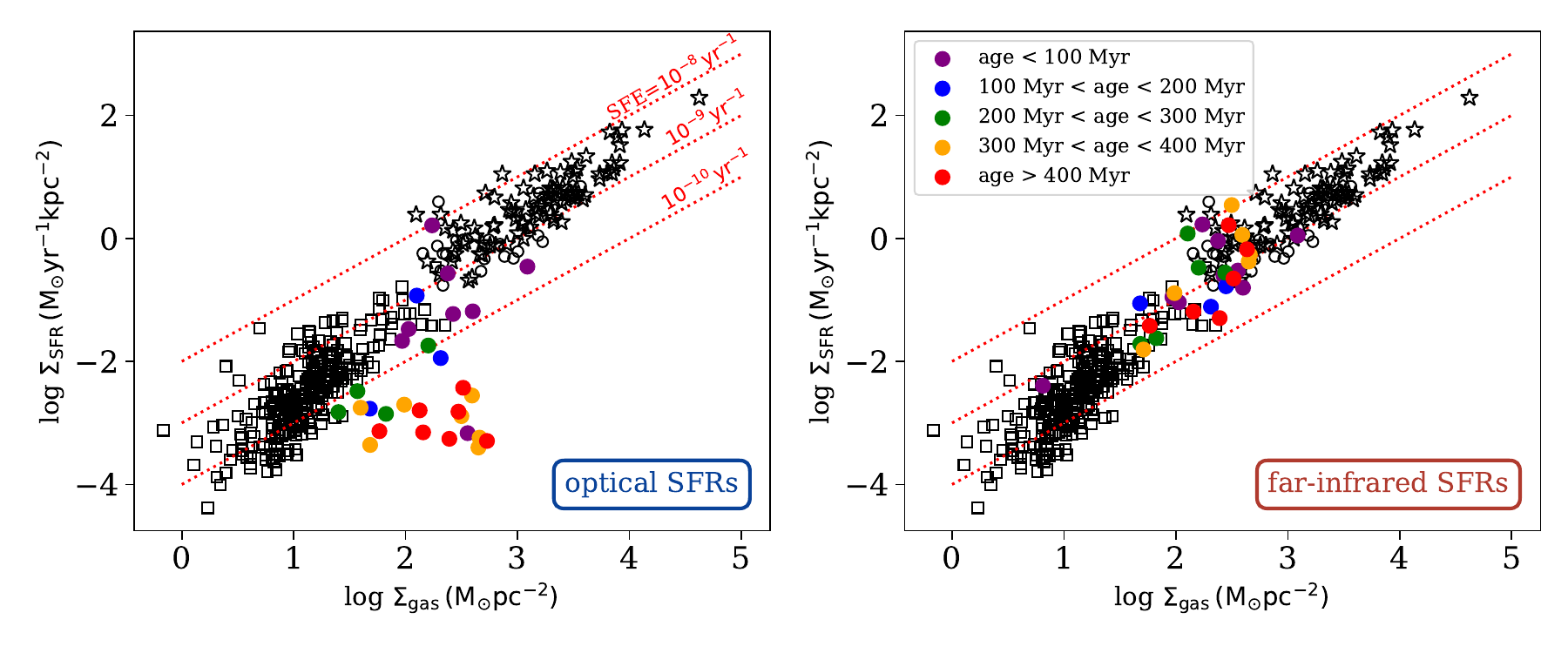}
\caption{\textbf{Location of the \citet{li19} post-starbursts in the Kennicutt--Schmidt plane.} 
The SFR surface density versus the molecular gas surface density, using optical (left) and far-infrared (right) SFRs. The \citet{li19} galaxies are color-coded according to their post-burst age, similarly to their color-coding in \citet{li19}. The comparison galaxies include the dwarfs, spirals, circumnuclear disks, and luminous infrared galaxies from \citet{de_los_reyes19} and \citet{kennicutt21}, and are marked with black empty symbols. 
}\label{f:reanalysis_of_KS_li_et_al}
\end{figure*}

Figure \ref{f:reanalysis_of_KS_li_et_al} shows the location of the \citet{li19} post-starburst galaxies in the Kennicutt--Schmidt (KS) plane. To reproduce their results, we used the SDSS Petrosian radii to convert the SFR and $M_{\mathrm{H_2}}$ into surface densities. Using optical SFRs, we reproduce their main result that post-starbursts are offset from the KS relation of other galaxies. When far-infrared SFRs are used instead, post-starburst galaxies lie on the same KS relation as other types of galaxies. This suggests that the offset reported by \citet{li19} is due to their usage of optical SFRs. 

\section{SFR(far-infrared) assuming constant and exponentially-decaying SFH}\label{a:SFR_FIR_checks}

\begin{figure}
	\centering
\includegraphics[width=3.25in]{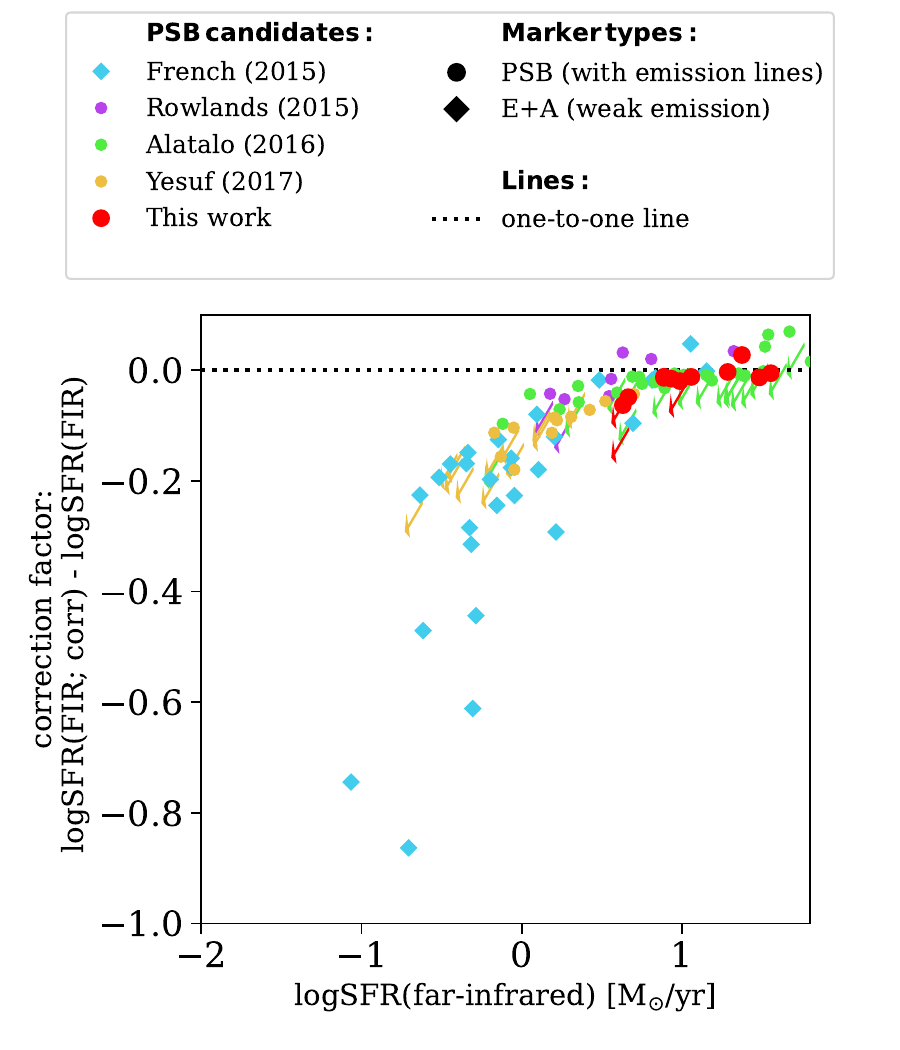}
\caption{\textbf{Far-infrared SFR correction factor as a function of SFR(far-infrared).} The correction factor applied to every galaxy we consider $\log$SFR(far-infrared corrected) - $\log$SFR(far-infrared), where $\log$SFR(far-infrared) is the SFR derived from the far-infrared assuming a constant conversion factor, while $\log$SFR(far-infrared corrected) is the SFR after the correction derived in section \ref{s:FIR_to_SFR}. }\label{f:SFR_FIR_before_and_after_corr}
\end{figure}

\begin{figure}
	\centering
\includegraphics[width=3.25in]{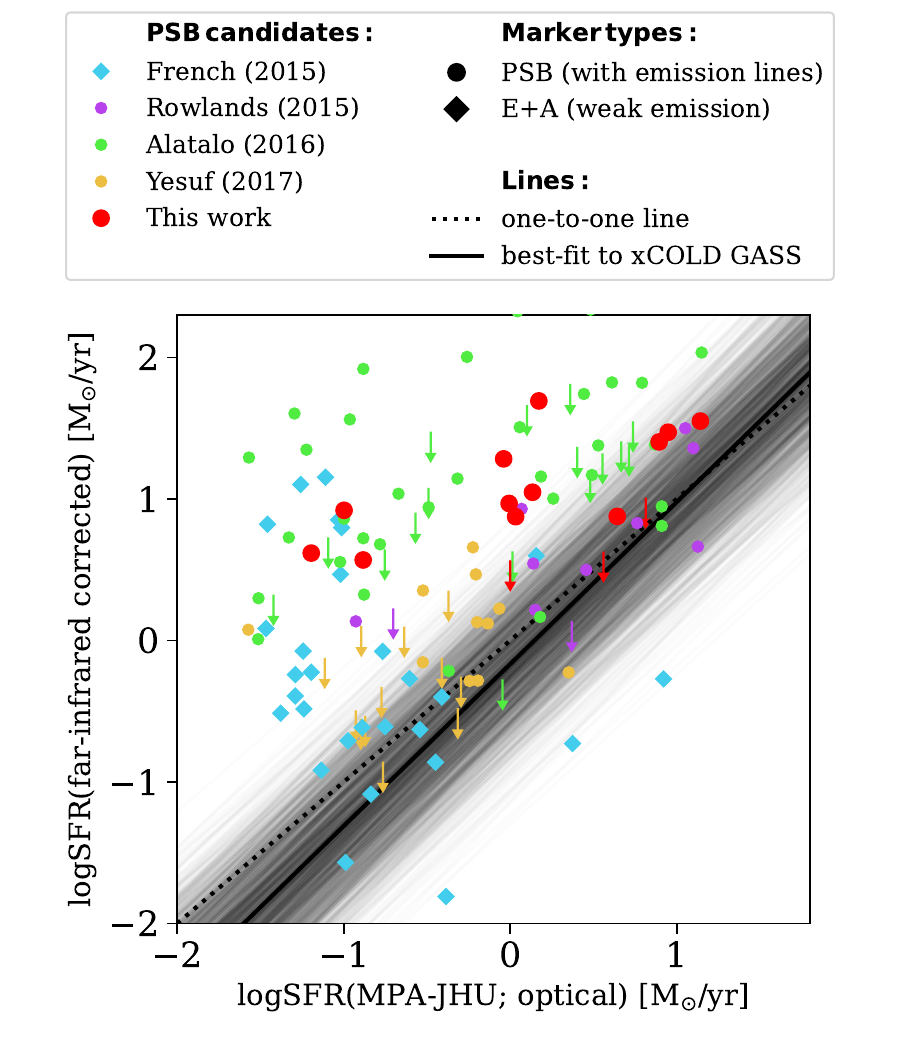}
\caption{\textbf{Comparison of far-infrared and optical SFRs in post-starburst candidates.} 60 $\mathrm{\mu m}$-based SFR versus the Dn4000\AA-based SFR for different post-starburst galaxy samples, where points represent measurements and arrows represent upper limits. Each color corresponds to a different sample, as indicated in the legend at the top. The dotted black line is a 1:1 relation. The black solid line is the best-fitting relation obtained for the comparison galaxies from xCOLD GASS. In contrast to figure \ref{f:SFR_FIR_versus_Dn4000} where we used far-infrared to SFR conversion factor that assumes constant SFR over the past 100 Myrs, here the SFR was corrected according to the best-fitting relation from section \ref{s:FIR_to_SFR}. }\label{f:SFR_FIR_versus_Dn4000_conv}
\end{figure}

Figure \ref{f:SFR_FIR_before_and_after_corr} shows the far-infrared correction factor applied to the galaxies we consider according to the relation derived in section \ref{s:FIR_to_SFR}. Figure \ref{f:SFR_FIR_versus_Dn4000_conv} shows the far-infrared versus optical SFR for the post-starburst galaxies we consider, assuming an exponentially-decaying SFH to estimate the conversion factor between far-infrared and SFR (see section \ref{s:FIR_to_SFR}). This is compared to figure \ref{f:SFR_FIR_versus_Dn4000} where we used a standard conversion factor between far-infrared and SFR that assumes that the SFR has been constant over the past 100 Myrs.

\begin{figure*}
	\centering
\includegraphics[width=1\textwidth]{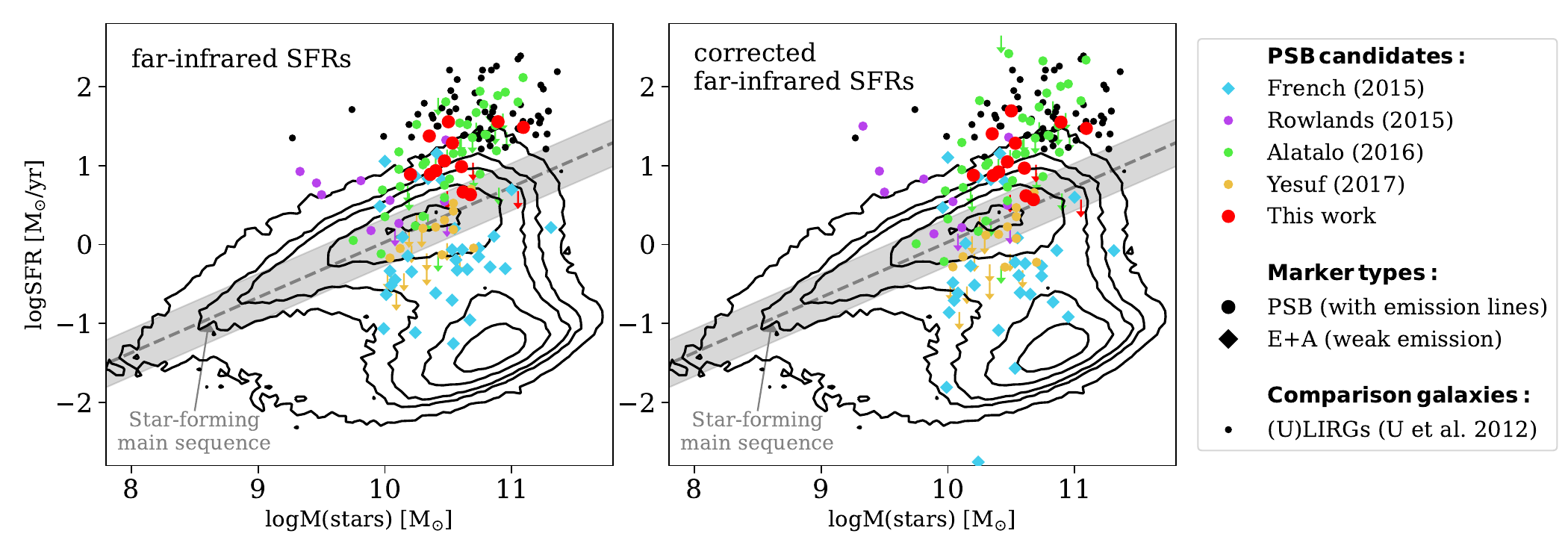}
\caption{\textbf{SFR versus stellar mass using different far-infrared to SFR conversion factors.} 
The points represent measurements and arrows represent upper limits, where each color corresponds to a different sample, as indicated in the legend on the right. The grey dotted line represents the star-forming main sequence at $z=0$ (\citealt{whitaker12}), and the light-grey band a $\pm$0.3 dex interval around it. For the post-starburst galaxy candidates, the left panel shows SFR(far-infrared) when using a far-infrared to SFR conversion factor that assumes that SFR has been constant over the past 100 Myrs. The right panel shows SFR(far-infrared) when assuming an exponentially-declining SFH (see section \ref{s:FIR_to_SFR}).}\label{f:SFR_versus_m_conv}
\end{figure*}

\begin{figure}
\includegraphics[width=3.5in]{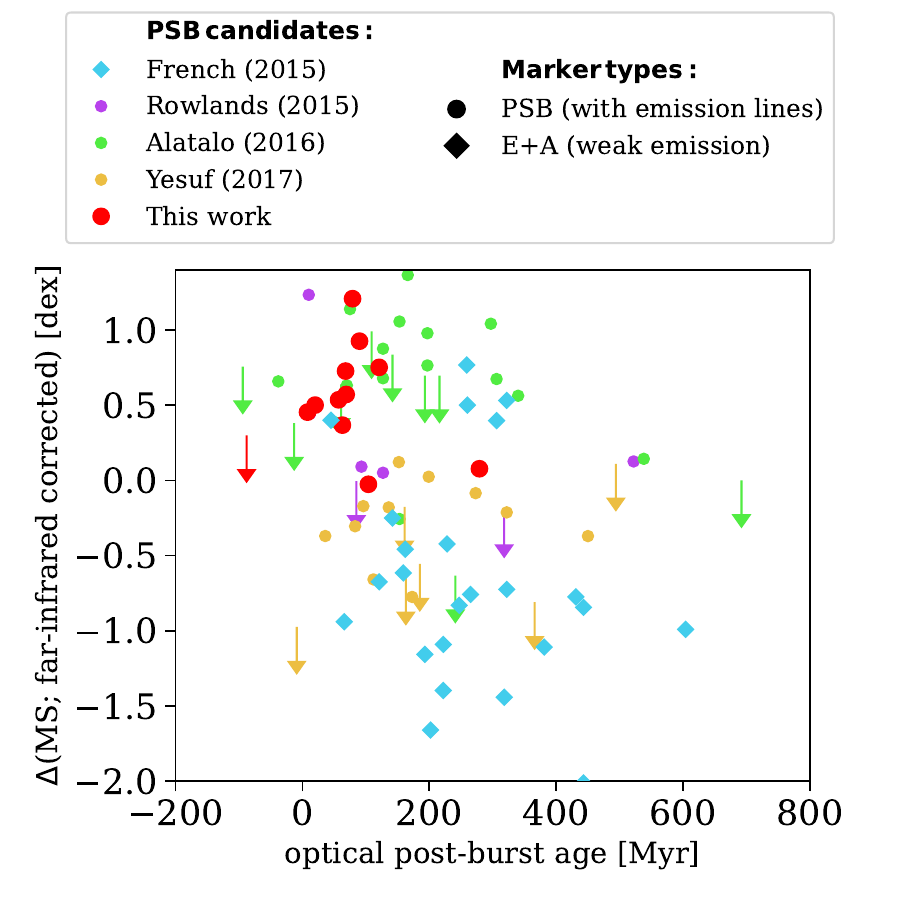}
\caption{\textbf{The location of a galaxy with respect to the star-forming main sequence versus the post-burst age.} The points represent measurements and arrows represent upper limits, where each color corresponds to a different sample, as indicated in the legend at the top. $\mathrm{\Delta(MS)}$ is estimated using the far-infrared-based SFR, assuming the exponentially-declining SFH presented in section \ref{s:FIR_to_SFR}.} \label{f:Delta_MS_versus_pb_age_conv}
\end{figure}

\begin{figure*}
	\centering
\includegraphics[width=1\textwidth]{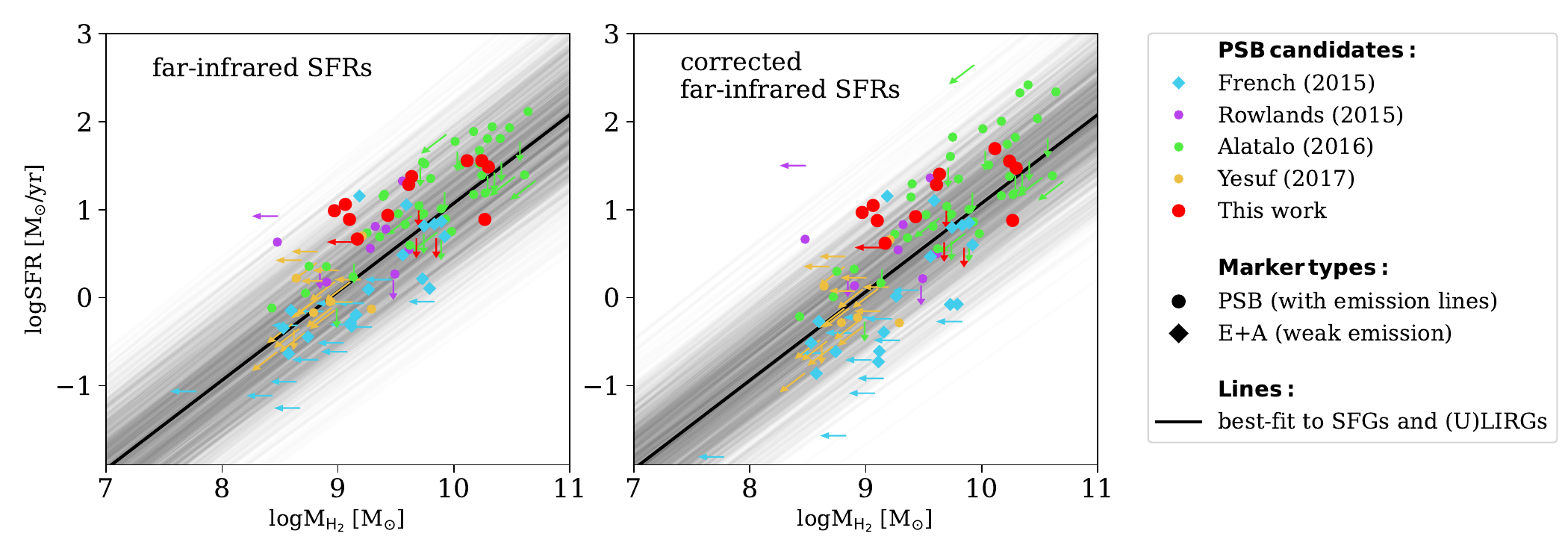}
\caption{\textbf{SFR versus $M_{\mathrm{H_2}}$ using different far-infrared to SFR conversion factors.} 
The points represent measurements and arrows represent upper limits, where arrows pointing down are upper limits on SFR, arrows to the left on $M_{\mathrm{H_2}}$, and diagonal arrows to the bottom left on both SFR and $M_{\mathrm{H_2}}$. Each color corresponds to a different sample, as indicated in the legend on the right. For the post-starburst galaxy candidates, the left panel shows SFR(far-infrared) when using a far-infrared to SFR conversion factor that assumes that SFR has been constant over the past 100 Myrs. The right panel shows SFR(far-infrared) when assuming an exponentially-declining SFH (see section \ref{s:FIR_to_SFR}).}\label{f:SFR_versus_MH2_conv}
\end{figure*}

Figure \ref{f:SFR_versus_m_conv} compares between the location of the post-starburst galaxies we consider on the SFR-stellar mass plane when using the standard conversion factor and when assuming an exponentially-decaying SFH. Figure \ref{f:Delta_MS_versus_pb_age_conv} presents the location of the galaxy with respect to the star-forming main sequence versus the post-burst age, using far-infrared and assuming an exponentially-decaying SFH to estimate the SFR. Figure \ref{f:SFR_versus_MH2_conv}  compares between the SFR-$M_{\mathrm{H_2}}$ relation of post-starburst galaxies obtained when using the standard conversion factor, versus the conversion factor obtained when assuming an exponentially-decaying SFH.

\end{document}